\documentclass[twocolumn,preprintnumbers,prb,aps,amssymb,amsmath,showpacs,floatfix,superscriptaddress,longbibliography]{revtex4-1}

\usepackage{graphicx}
\usepackage{epsfig}
\usepackage{natbib}
\usepackage{dcolumn}
\usepackage{bm} 
\usepackage{soul}
\usepackage{sidecap}
\usepackage[usenames,dvipsnames]{color}
\usepackage{setspace}
\usepackage[colorlinks=true,linkcolor=blue,urlcolor=blue,citecolor=blue]{hyperref}
\usepackage[usenames,dvipsnames,svgnames,table]{xcolor}

\begin{document}

\newcommand{\lsco}{La$_{2-x}$Sr$_x$CuO$_4$}
\newcommand{\lbco}{La$_{2-x}$Ba$_x$CuO$_4$}
\newcommand{\lnsco}{La$_{1.6-x}$Nd$_{0.4}$Sr$_x$CuO$_4$}
\newcommand{\lesco}{La$_{1.8-x}$Eu$_{0.2}$Sr$_x$CuO$_4$}
\newcommand{\lresco}{La$_{2-x-y}$RE$_{y}$Sr$_x$CuO$_4$}

\newcommand{\lnscoxy}{La$_{2-x-y}$Nd$_{y}$Sr$_x$CuO$_4$}
\newcommand{\lnco}{La$_{1.6}$Nd$_{0.4}$CuO$_4$}
\newcommand{\pcco}{Pr$_{2-x}$Ce$_{x}$CuO$_{4}$}
\newcommand{\ybco}{YBa$_{2}$Cu$_{3}$O$_{y}$}
\newcommand{\tltwotwoone}{Tl$_{2}$Ba$_{2}$CuO$_{6+\delta}$}
\newcommand{\hbco}{HgBa$_{2}$CuO$_{4+\delta}$}
\newcommand{\bsco}{Bi$_2$Sr$_{2}$CuO$_{6+\delta}$}
\newcommand{\bscco}{Bi$_2$Sr$_{2}$CaCu$_{2}$O$_{8+\delta}$}

\newcommand{\muohmcm}[1]{#1\,$\mu\Omega$\,cm}

\newcommand{\TN}{$T_{\rm N}$}
\newcommand{\Tc}{$T_{\rm c}$}
\newcommand{\Tstar}{$T^{\star}$}
\newcommand{\TCDW}{$T_{\rm CDW}$}
\newcommand{\Hc}{$H_{\rm c2}$}

\newcommand{\ie}{{\it i.e.}}
\newcommand{\eg}{{\it e.g.}}
\newcommand{\etal}{{\it et al.}}


\title{Sensitivity of \texorpdfstring{$T_{\rm c}$}{\textit{T}c} to pressure and magnetic field in the cuprate superconductor YBa\texorpdfstring{$_2$}{2}Cu\texorpdfstring{$_3$}{3}O\texorpdfstring{$_y$}{y}: evidence of charge order suppression by pressure}


\author{O. Cyr-Choini\`{e}re}
\altaffiliation{Present address: Yale School of Engineering and Applied Science, Yale University, New Haven, Connecticut 06511}
\affiliation{Institut quantique, D\'{e}partement de physique  \&  RQMP, Universit\'{e} de Sherbrooke, Sherbrooke,  Qu\'{e}bec J1K 2R1, Canada}

\author{D. LeBoeuf}
\altaffiliation{Present address: Laboratoire National des Champs Magn\'{e}tiques Intenses, UPR 3228, (CNRS-INSA-UJF-UPS), Grenoble 38042, France}
\affiliation{Institut quantique, D\'{e}partement de physique  \&  RQMP, Universit\'{e} de Sherbrooke, Sherbrooke,  Qu\'{e}bec J1K 2R1, Canada}

\author{S. Badoux}
\affiliation{Institut quantique, D\'{e}partement de physique  \&  RQMP, Universit\'{e} de Sherbrooke, Sherbrooke,  Qu\'{e}bec J1K 2R1, Canada}

\author{S. Dufour-Beaus\'{e}jour}
\affiliation{Institut quantique, D\'{e}partement de physique  \&  RQMP, Universit\'{e} de Sherbrooke, Sherbrooke,  Qu\'{e}bec J1K 2R1, Canada}

\author{D.~A.~Bonn}
\affiliation{Department of Physics and Astronomy, University of British Columbia, Vancouver, British Columbia V6T 1Z4, Canada}
\affiliation{Canadian Institute for Advanced Research, Toronto, Ontario M5G 1Z8, Canada}

\author{W. N. Hardy}
\affiliation{Department of Physics and Astronomy, University of British Columbia, Vancouver, British Columbia V6T 1Z4, Canada}
\affiliation{Canadian Institute for Advanced Research, Toronto, Ontario M5G 1Z8, Canada}

\author{R. Liang}
\affiliation{Department of Physics and Astronomy, University of British Columbia, Vancouver, British Columbia V6T 1Z4, Canada}
\affiliation{Canadian Institute for Advanced Research, Toronto, Ontario M5G 1Z8, Canada}

\author{D. Graf}
\affiliation{National High Magnetic Field Laboratory, Florida State University, Tallahassee, FL 32306, USA}

\author{N. Doiron-Leyraud}
\email{nicolas.doiron-leyraud@usherbrooke.ca}
\affiliation{Institut quantique, D\'{e}partement de physique  \&  RQMP, Universit\'{e} de Sherbrooke, Sherbrooke,  Qu\'{e}bec J1K 2R1, Canada}

\author{Louis~Taillefer}
\email{louis.taillefer@usherbrooke.ca}
\affiliation{Institut quantique, D\'{e}partement de physique  \&  RQMP, Universit\'{e} de Sherbrooke, Sherbrooke,  Qu\'{e}bec J1K 2R1, Canada}
\affiliation{Canadian Institute for Advanced Research, Toronto, Ontario M5G 1Z8, Canada}

\date{\today}


\begin{abstract}

Cuprate superconductors have a universal tendency to form charge density-wave (CDW) order which competes with superconductivity and is strongest at a doping $p$\,$\simeq$\,$0.12$.
Here we show that in the archetypal cuprate YBa$_{2}$Cu$_{3}$O$_{y}$ (YBCO) pressure suppresses charge order, but does not affect the pseudogap phase.
This is based on transport measurements under pressure, which reveal that the onset of the pseudogap at $T^*$ is independent of pressure, while the negative Hall effect, a clear signature of CDW order in YBCO, is suppressed by pressure.
We also find that pressure and magnetic field shift the superconducting transition temperature $T_{\rm c}$ of YBCO in the same way as a function of doping -- but in opposite directions -- and most effectively at $p$\,$\simeq$\,$0.12$.
This shows that the competition between superconductivity and CDW order can be tuned in two ways, either by suppressing superconductivity with field or suppressing CDW order by pressure. 
Based on existing high-pressure data and our own work, we observe that when CDW order is fully suppressed at high pressure, the so-called ``1/8 anomaly'' in the superconducting dome vanishes, revealing a smooth $T_{\rm c}$ dome which now peaks at $p$\,$\simeq$\,$0.13$. 
We propose that this $T_{\rm c}$ dome is shaped by the competing effects of the pseudogap phase below its critical point $p^{\star}$\,$\sim$\,$0.19$ and spin order at low doping.
\end{abstract}




\maketitle


\section{Introduction}
\label{sec:Intro}

\begin{figure}[t!]
\centering
\includegraphics[width=0.46\textwidth]{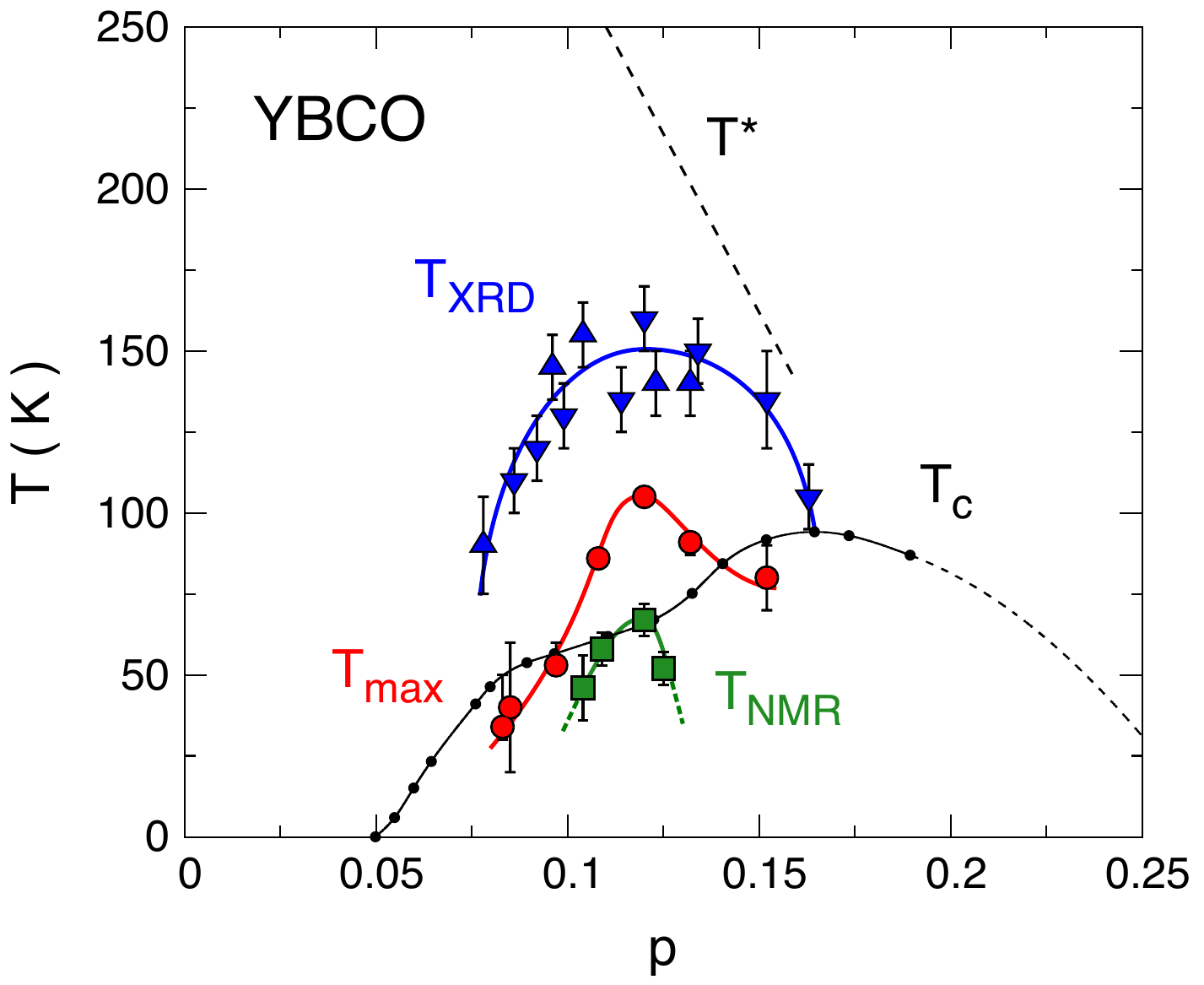}
\caption{
Temperature-doping phase diagram of YBCO, showing the superconducting phase below $T_{\rm c}$ (black dots~[\onlinecite{Liang2006}]) 
and the onset of charge order seen by NMR, above a threshold magnetic field, below $T_{\rm NMR}$ (green squares~[\onlinecite{Wu2013}]).
CDW modulations are detected by x-ray diffraction below $T_{\rm XRD}$ 
(up triangles~[\onlinecite{Hucker2014}]; down triangles~[\onlinecite{Blanco-Canosa2014}]).
The Fermi surface undergoes a reconstruction seen as a downturn in the Hall coefficient 
below $T_{\rm max}$ (red dots~[\onlinecite{LeBoeuf2011}]).
\Tstar{} marks the onset of the pseudogap phase (dashed line [\onlinecite{Ando2004},\onlinecite{Daou2010}]). Full lines are guides to the eye.
}
\label{Phasediag-CDW}
\end{figure}

The recent observation of charge density modulations in \ybco~(YBCO){} [\onlinecite{Wu2011,Ghiringhelli2012,Chang2012a,Achkar2012}], 
\lsco{} (LSCO)~[\onlinecite{Croft2014}], \hbco~[\onlinecite{Tabis2014}], \bsco~[\onlinecite{Comin2014}] and \bscco~[\onlinecite{daSilvaNeto2014}] 
shows that charge density-wave (CDW) order is a generic tendency of cuprates, 
not specific to materials such as \lbco{} (LBCO), where it has long been known to exist~[\onlinecite{Tranquada1995}]. 
In Fig.~\ref{Phasediag-CDW}, the onset temperature of CDW modulations seen in YBCO by x-ray diffraction, $T_{\rm XRD}$, 
is plotted as a function of doping~[\onlinecite{Hucker2014},\onlinecite{Blanco-Canosa2014}]. 
It forms a dome peaked at $p$\,=\,$0.12$, as does the onset temperature of CDW order seen by nuclear magnetic resonance (NMR) (above a threshold magnetic field),  $T_{\rm NMR}$~[\onlinecite{Wu2013}]. 
The Fermi surface of YBCO undergoes a reconstruction (FSR), attributed to CDW order, into small electron~[\onlinecite{Doiron-Leyraud2007}] and hole~[\onlinecite{Doiron-Leyraud2015}] 
pockets at low temperature. This process is detected as a downturn in the Hall coefficient $R_{\rm H}(T)$ towards negative values~[\onlinecite{LeBoeuf2007}],
characterized by a maximum in $R_{\rm H}(T)$ at a temperature $T_{\rm max}$~[\onlinecite{LeBoeuf2011}]. 
As seen in Fig.~\ref{Phasediag-CDW}, $T_{\rm max}$ also peaks at $p$\,=\,$0.12$. 
CDW and FSR also both peak at $p$\,=\,$0.12$ in \lesco~[\onlinecite{Fink2011},\onlinecite{Laliberte2011}].


That the CDW phase in cuprates is universally peaked at $p$\,=\,$0.12$ is a striking experimental fact which naturally begs an understanding.
Prior explanations in terms of a commensurate match of the CDW period with either the lattice or the hole density are no longer viable. 
Indeed, while in LBCO or LSCO-based materials the CDW incommensurability tracks $p$ and the period becomes nearly commensurate with the lattice 
at $p$\,$\simeq$\,$0.12$, neither of these facts are true for YBCO~[\onlinecite{Hucker2014},\onlinecite{Blanco-Canosa2014}]. 
For some as yet unknown reason, the conditions for CDW formation in cuprates are most favourable  at $p$\,=\,$0.12$.


\begin{figure*}[t]
\centering
\includegraphics[width=1.0\textwidth]{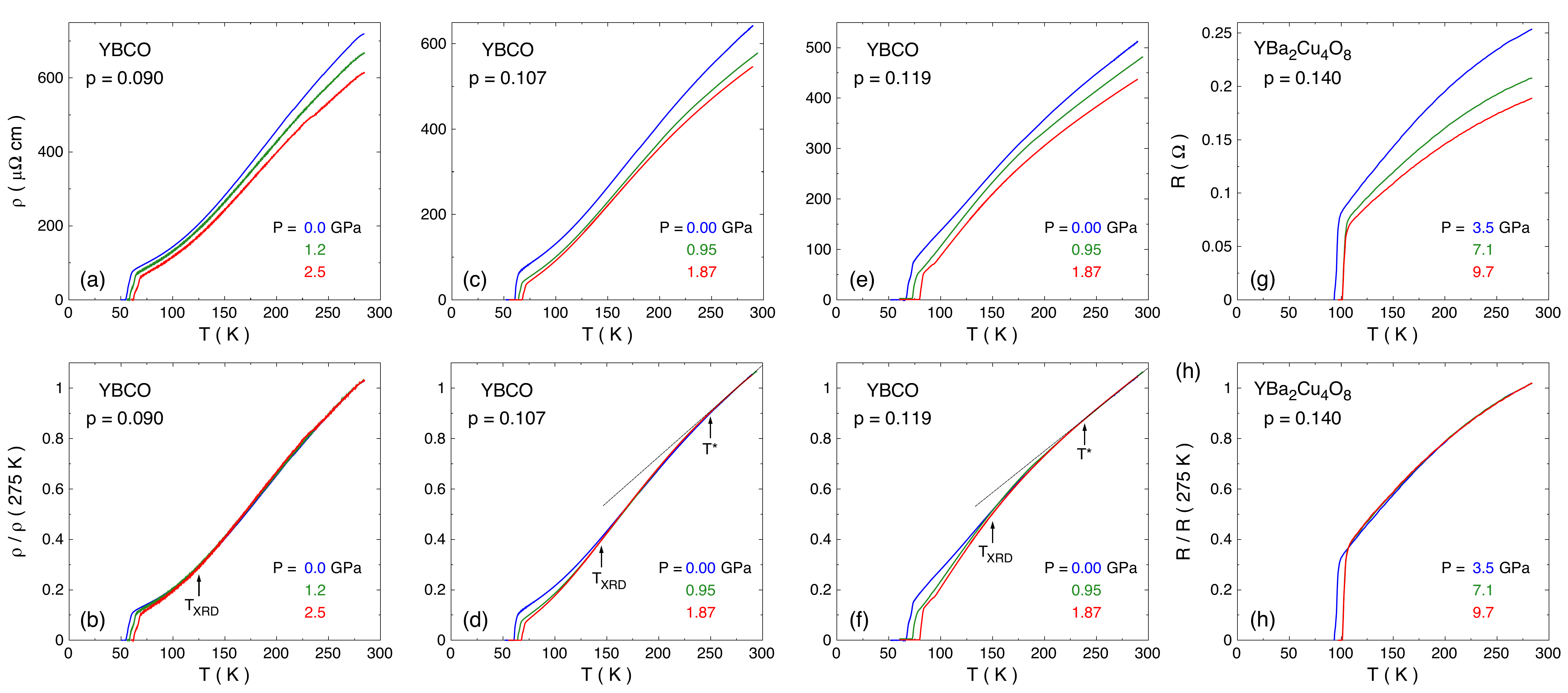}
\caption{
	$a$-axis electrical resistivity $\rho_{\rm a}(T)$ of YBCO as a function of temperature at dopings and pressures as indicated. 
	The top panels show the $\rho_{\rm a}(T)$ (or the resistance $R(T)$ for panel (g)). The bottom panels show $\rho_{\rm a}(T)$ ($R(T)$) normalized by its value at $T$\,=\,$275$\,K. 
	The data in panels (g) and (h) are reproduced from ref.~\onlinecite{Nakayama2014}, on YBa$_2$Cu$_4$O$_8$ with $p = 0.14$ (\Tc\,=\,80\,K), which is stoichiometric with perfect oxygen order.  
	In panels (d) and (f), the straight line is a linear fit to the data at high temperature. 
	The pseudogap temperature \Tstar (arrow) is defined as the temperature below which $\rho_{\rm a}(T)$ 
	deviates from its linear dependence at high temperature~[\onlinecite{Ando2004},\onlinecite{Daou2010}]. 
	Note that \Tstar\,$>$\,$300$\,K for $p = 0.090$~[\onlinecite{Ando2004}]. 
	The absence of linearity in YBa$_2$Cu$_4$O$_8$ (panels (g) and (h)) may come from measurements on a twinned sample with randomly oriented domains.
	The important aspect is that there is no change in the functional form of $R(T)$ in YBa$_2$Cu$_4$O$_8$ up to 10\,GPa. 
	$T_{\rm XRD}$ marks the onset of CDW modulations seen in x-ray diffraction at the corresponding doping (see Fig.~\ref{Phasediag-CDW}). 
 	The normalized resistivity is affected by pressure only below $T_{\rm XRD}$. 
}
\label{rho-vs-T-P-YBCO}
\end{figure*}


CDW order and superconductivity are competing phases. 
The x-ray intensity drops sharply below \Tc~[\onlinecite{Ghiringhelli2012},\onlinecite{Chang2012a}], showing that superconductivity weakens CDW order in YBCO. 
The absence of NMR splitting under an in-plane magnetic field ($H$\,$\parallel$\,$ab$), as opposed to an out-of-plane field ($H$\,$\parallel$\,$c$), 
is another evidence of the phase competition between charge order and superconductivity~[\onlinecite{Wu2011}]. 
Conversely, CDW order weakens superconductivity. 
This shows up in the doping dependence of the superconducting critical temperature \Tc{} and upper critical field \Hc,
as a dip in the former (Fig.~\ref{Phasediag-CDW})~[\onlinecite{Liang2006}] and a local minimum in the latter~[\onlinecite{Grissonnanche2014}],
both centred at $p$\,=\,$0.12$, where CDW order is strongest.
The dip in \Tc{} was shown to scale with the onset of FSR, closely linking the two~[\onlinecite{LeBoeuf2011}].
Application of a magnetic field $H$ restores the CDW amplitude below \Tc, while it has no effect above \Tc~[\onlinecite{Chang2012a}].
This shows that one can tune the competition between CDW order and superconductivity by applying a magnetic field.

Here we show that pressure is a second, independent tuning parameter for this competition, 
shifting $T_{\rm c}$ in the same way as the magnetic field, as a function of doping, but in opposite direction. 
Therefore, pressure is seen as a tuning parameter that weakens CDW order, with little direct effect on superconductivity or on the pseudogap phase.
Recent x-ray studies observe that pressure suppresses CDW order in YBCO~[\onlinecite{Souliou2018},\onlinecite{Huang2018}].
As a result, the increase of $T_{\rm c}$ with pressure is a consequence of competition between superconductivity and charge order. 
By applying sufficiently large pressures one can fully suppress CDW order and obtain the superconducting phase diagram free of competition which, based on existing data, displays a \Tc{} dome peaked at $p$\,$\simeq$\,$0.13$ and not at $p$\,=\,$0.16$.
The fact that both CDW order and superconductivity peak around the same doping in the absence of mutual competition 
suggests that some competing mechanism from another origin acts to suppress both at low doping. 
Identifying this mechanism will be key to understanding the cuprate phase diagram.


\section{Methods}
\label{sec:Methods}

Single crystals of \ybco{} were prepared as described elsewhere~[\onlinecite{Liang2012}], with oxygen content $y$ ranging from $y$\,=\,$6.35$ to $y$\,=\,$6.998$. 
The hole concentration (doping) $p$ of each sample is given by its superconducting critical temperature \Tc~[\onlinecite{Liang2006}].
\Tc{} was determined from measurements of the electrical resistivity $\rho(T)$ or Nernst signal $N(T)$ in $H$\,=\,$0$ and $15$\,T (applied along the $c$ axis of the orthorhombic structure), giving \Tc($H$\,=\,$0$) and \Tc($H$\,=\,$15$\,T) as the temperature below which $\rho$ and $N$ are zero. The values of $y$, $p$, \Tc($H$\,=\,$0$) and \Tc($H$\,=\,$15$\,T) for our 13 YBCO samples are listed in Table~\ref{table:Table-dTcdH} of the Appendix. This includes a sample with 1.4~\% of Ca substitution (at $y$\,$\simeq$\,$7$), for which $p$\,=\,$0.19$.

The $a$-axis electrical resistivity $\rho_{\rm a}(T)${} at ambient and high pressure, in $H$\,=\,$0$ and $15$\,T, was measured at Sherbrooke on three single crystals with a high degree of oxygen order: 1)\,$y$\,=\,$6.50$, \Tc(0)\,=\,54.5\,K, $p$\,=\,$0.090$ (ortho-II); 2)\,$y$\,=\,$6.54$, \Tc(0)\,=\,60.2\,K, $p$\,=\,$0.107$ (ortho-II); and 3)\,$y$\,=\,$6.67$, \Tc(0)\,=\,65.3\,K, $p$\,=\,$0.119$ (ortho-VIII).
The Hall coefficient $R_{\rm H}(T) = \rho_{\rm ab}(T)/H${} at ambient and high pressure, and magnetic fields up to $H$\,=\,$35$\,T, was measured at the National High Magnetic Field Laboratory in Tallahassee on two single crystals with high oxygen order: 1)\,$y$\,=\,$6.54$, \Tc(0)\,=\,61.3\,K, $p$\,=\,$0.109$ (ortho-II) and 2)\,$y$\,=\,$6.67$, \Tc(0)\,=\,66.0\,K, $p$\,=\,$0.120$ (ortho-VIII).
The samples were pressurized using a nonmagnetic piston-cylinder clamp cell with either a 1/1 mixture of pentane and 3-methyl-1-butanol or 7373 Daphne oil as the pressure medium, ensuring a hydrostatic pressure during pressurization. The pressure was determined from the superconducting transition of a lead gauge or from the fluorescence of a ruby chip. Note that pressure can enhance oxygen order in YBCO, and oxygen ordering increases the doping in the CuO$_2$ planes (see Sec.~\ref{subsec:Doping-pressure}). To avoid this, one should apply pressure at temperatures below $\sim$\,$200$\,K~[\onlinecite{Sadewasser1997},\onlinecite{Sadewasser2000}]. Another (simpler) way is to start with oxygen-ordered samples, apply the pressure at room temperature and rapidly cool the samples ( $<$~2 hours at 300~K) to avoid relaxation effects. This is the approach we used.


\section{Effect of pressure}
\label{sec:Sensitivity-pressure}


\subsection{Electrical resistivity}
\label{subsec:Resistivity}

In Figs.~\ref{rho-vs-T-P-YBCO}(a), (c) and (e), the electrical resistivity $\rho_{\rm a}(T)$ of our 3 oxygen-ordered YBCO samples is plotted as a function temperature, for different values of the applied pressure. 
In Figs.~\ref{rho-vs-T-P-YBCO}(b), (d) and (f), we show the resistivity normalized at $T$\,=\,$275$\,K. 
We see that above $150$\,K and for all three dopings, pressure has essentially no effect on the functional form of $\rho_{\rm a}(T)$ and only induces a slight reduction in amplitude. At $p$\,=\,$0.107$ and $0.119$, the data exhibit a linear-$T$ regime at high temperature, followed by a drop below linearity at \Tstar, a clear signature of the pseudogap phase [\onlinecite{Cyr-Choiniere2018}]. At $p$\,=\,$0.090$, \Tstar~is above 300~K ~[\onlinecite{Ando2004}] which is too high to lend a clear linear-$T$ regime within our experimental range, but the normalized curves all fall on top of each other. The fact that both \Tstar~and the subsequent drop are insensitive to pressure shows that in YBCO the pseudogap itself is not affected by pressure in this doping range. As shown in Figs.~\ref{rho-vs-T-P-YBCO}(g) and (h), the same holds true in YBa$_2$Cu$_4$O$_8$ ($p$\,=\,$0.14$), measured up to 10\,GPa~[\onlinecite{Nakayama2014}], suggesting that this conclusion is valid up to such high pressures. In the related system \lnsco (Nd-LSCO) where a departure from linear-$T$ resistivity [\onlinecite{Collignon2017}] is unambiguously connected to the pseudogap opening as seen in Angle Resolved Photoemission Spectroscopy data [\onlinecite{Matt2015}], the pseudogap was also found to be independent of pressure close to $p$\,=\,$0.15$ [\onlinecite{Doiron-Leyraud2017}]. (In Nd-LSCO, pressure tunes the pseudogap critical point $p^{\star}$ down but does not affect \Tstar~at dopings well below $p^{\star}$~[\onlinecite{Doiron-Leyraud2017}].)
Note that NMR data on the cuprate (Ca$_x$La$_{1-x}$)(Ba$_{1.75-x}$La$_{0.25+x}$)Cu$_3$O$_y$, where Ca doping modifies the lattice parameters and acts as internal pressure while keeping hole doping constant, showed that this internal pressure does not affect the pseudogap temperature \Tstar~but changes $T_{\rm c}$~[\onlinecite{Cvitanic2014}], consistent with our interpretation. Going to lower temperatures below $T_{\rm XRD}$ (Figs.~\ref{rho-vs-T-P-YBCO}(b), (d) and (f)), our data now reveal changes in $\rho_{\rm a}(T)$ which we attribute to the CDW and which we discuss below in the light of normal-state Hall effect measurements.


\subsection{Hall effect}
\label{subsec:Hall}

\begin{figure}[t]
\centering
\includegraphics[width=0.49\textwidth]{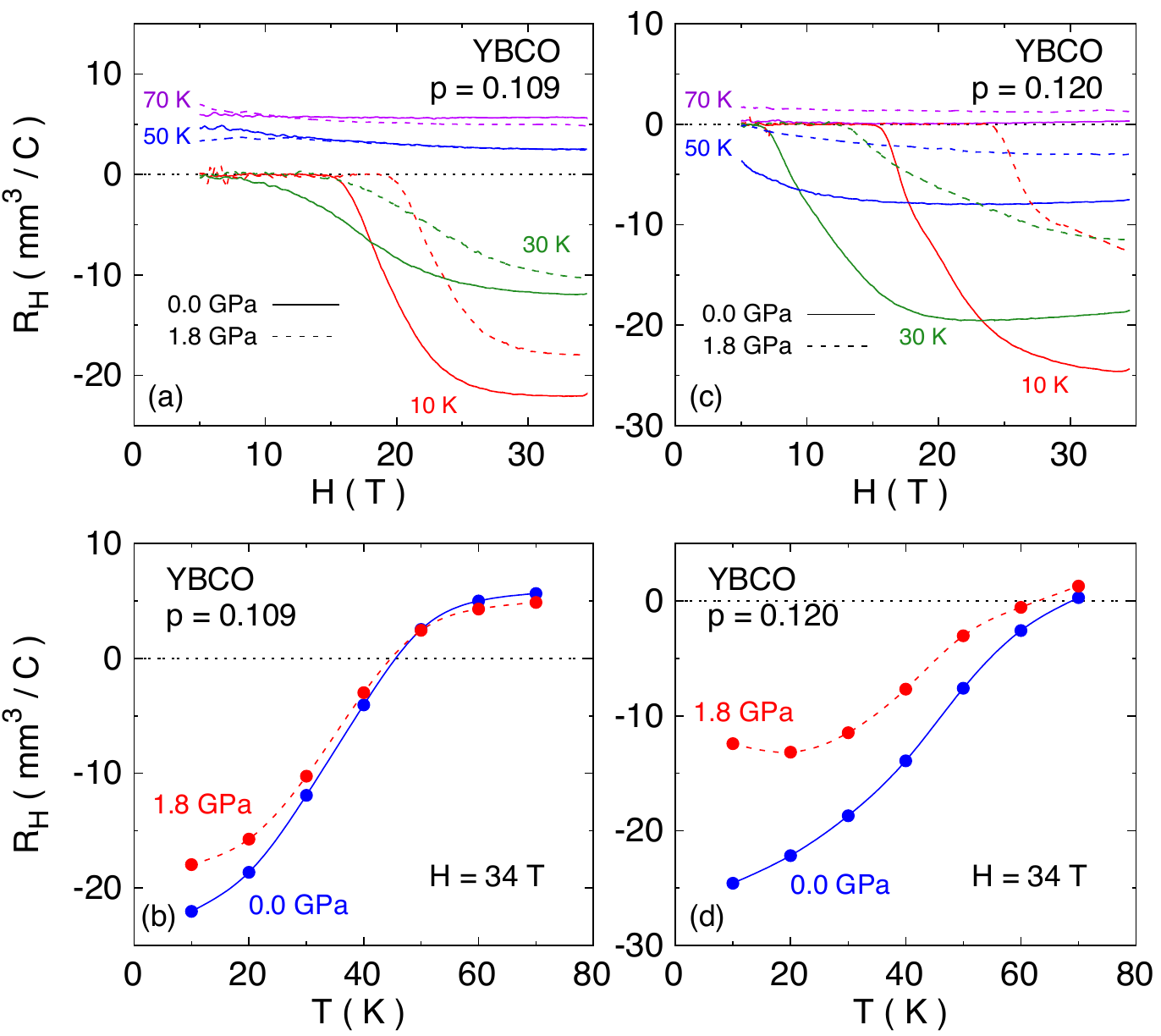}
\caption{Hall coefficient $R_{\rm H}${} of YBCO at dopings and pressures as indicated. In panels (a,c) we show $R_{\rm H}${} as a function of field at temperatures as indicated. In panels (b,d) we show $R_{\rm H}${} as a function of temperature from our data at $H$\,=\,$34$\,T. In all panels the full lines are data at ambient pressure and dashed lines are data at 1.8 GPa. Note that the slight upturn at $10$\,K and $p$\,=\,$0.120$ (d) is an artefact caused by superconductivity, as $R_{\rm H}${} has not fully transitioned (saturated) to the normal state value at $34$\,T (red dashed line in (c)).
}
\label{RH-vs-H-T}
\end{figure}

In YBCO, a clear consequence of the CDW is the fact that the Fermi surface is reconstructed at low temperatures. This FSR was first established through quantum oscillation measurements [\onlinecite{Doiron-Leyraud2007},\onlinecite{Yelland2008},\onlinecite{Bangura2008}], which revealed a small Fermi surface, and Hall effect data, which showed that the Hall coefficient $R_{\rm H}${} is negative [\onlinecite{LeBoeuf2007}] and therefore that the Fermi surface is electron-like. This is shown in Figs.~\ref{RH-vs-H-T}(b) and (d) where we plot $R_{\rm H}${} at low temperature for YBCO at $p$\,=\,$0.109$ and $0.120$. As a function of temperature, the Hall effect is positive at high temperature, reaches a maximum at a temperature $T_{\rm max}${}, and then falls rapidly to negative values because of FSR by the CDW [\onlinecite{LeBoeuf2011}]. In Fig.~\ref{Phasediag-CDW} we reproduce $T_{\rm max}${} as a function of doping from ref.~[\onlinecite{LeBoeuf2011}] and see that it forms a dome that peaks near 1/8 and correlates with the presence of the CDW. A negative Hall signal is only observed over the doping range where CDW order is seen by x-rays [\onlinecite{Hucker2014},\onlinecite{Blanco-Canosa2014}], above $p$\,=\,$0.08$ [\onlinecite{LeBoeuf2011}] and below $p$\,=\,$0.16$ [\onlinecite{Badoux2016}]. The amplitude of the negative $R_{\rm H}${} was also shown to be maximal where CDW is strongest [\onlinecite{Hucker2014},\onlinecite{LeBoeuf2011}] and where the dip in \Tc~is more pronounced [\onlinecite{LeBoeuf2011}]. Consequently, in YBCO $R_{\rm H}${}  is a reliable marker of the CDW phase. In Fig.~\ref{RH-vs-H-T} we show its evolution as a function of pressure for $p$\,=\,$0.109$ and $0.120$. In Figs.~\ref{RH-vs-H-T}(a) and (c), we plot a set of representative isotherms at ambient pressure and 1.8 GPa, down to $10$\,K and at fields up to $34$\,T, which is sufficient to reach the normal state value~[\onlinecite{Grissonnanche2014}] (at ambient pressure). In the bottom panels we show the normal-state $R_{\rm H}${} at $34$\,T as a function of temperature. At $p$\,=\,$0.120$ we see that pressure has a large effect on both the amplitude of $R_{\rm H}${} and the temperature at which it changes sign, suppressing both quantities significantly. The same is observed at $p$\,=\,$0.109$ albeit in a much reduced fashion.

We interpret this reduction in the amplitude of the negative $R_{\rm H}${} as a clear signature of the suppression of the CDW by pressure. As discussed below, doping is also affected by pressure, but the change in $R_{\rm H}${} seen here cannot be explained by that alone. At $p$\,=\,$0.120$ and $20$\,K, we see a relative change $\Delta R_{\rm H}/R_{\rm H}$\,=\,$(R_{\rm H} (1.8 {\rm GPa}) - R_{\rm H} (0 {\rm GPa}))/R_{\rm H} (0 {\rm GPa})$ of about 40\%. According to $R_{\rm H}${} data as a function of doping in YBCO [\onlinecite{LeBoeuf2011}], this would require a change in doping larger than $\Delta p = 0.012$ if that was the sole effect. This is much larger than the actual change in doping induced by 1.8 GPa, estimated to be $\Delta p = 0.002$ at $p$\,=\,$0.120$ (see Sec.~\ref{subsec:Doping-pressure} below). We therefore conclude that pressure suppresses the CDW phase. This is consistent with direct observations of the CDW by x-ray measurements on YBCO [\onlinecite{Souliou2018},\onlinecite{Huang2018}], which also found that the CDW disappears under pressure. We note that 1.0~GPa appears sufficient to fully suppress the CDW signature in x-ray~[\onlinecite{Souliou2018}] at $p \sim 0.105$ while $R_{\rm H}${} remains negative up to at least 1.8~GPa, presumably because CDW fluctuations survive up to much higher pressure and can still cause a FSR.

Recent Hall effect measurements on YBCO at $p$\,=\,$0.11$ up to 2.6 GPa revealed a very weak effect of pressure on $R_{\rm H}${}~[\onlinecite{Putzke2018}], which led the authors to conclude that the CDW is only weakly affected by pressure. This is consistent with our observation that the effect of pressure is much weaker at $p$\,=\,$0.109$ than at $p$\,=\,$0.120$, which we explain as follows. Pressure suppresses the CDW dome at all dopings across the phase diagram, which should suppress the negative $R_{\rm H}${}. At the same time, pressure increases the doping (see Sec.~\ref{subsec:Doping-pressure}) and since the CDW dome is peaked at $p$\,=\,$0.12$, a slight increase in doping from $p$\,=\,$0.11$ should make $R_{\rm H}${} more negative. The two effects therefore balance each other at $p$\,=\,$0.11$. At $p$\,=\,$0.12$, however, they reinforce each other, hence the much greater sensitivity of $R_{\rm H}${} under pressure. We note that our $p$\,=\,$0.109$ sample has a \Tc = $61.3$\,K, which is slightly higher than the \Tc = $60.7$\,K reported in ref.~[\onlinecite{Putzke2018}], consistent with the fact that they observe an even weaker effect of pressure on $R_{\rm H}${}. A simple test would be to apply more pressure to their sample with $p$\,=\,$0.11$, thereby tuning $p$ beyond 0.12: $R_{\rm H}${} should rapidly become less negative, as we find in our sample with $p$\,=\,$0.120$.


\subsection{Superconducting \Tc}
\label{subsec:Tc}

In contrast to \Tstar, pressure has a large effect on \Tc. 
In Fig.~\ref{dTcdP-raw}(a), we plot the pressure dependence of \Tc~for our samples whose resistivity data are shown in Fig.~\ref{rho-vs-T-P-YBCO}.
The slope ${dT_{\rm {c}}}/{dP}$ has a positive value, of magnitude 
$3.0$\,$\pm$\,$0.2$, 
$3.5$\,$\pm$\,$0.4$, and 
$7.5$\,$\pm$\,$0.5$\,K\,/\,GPa
for $p$\,=\,$0.090$, $0.107$, and $0.119$, respectively. 
Fig.~\ref{dTcdP-raw}(b) displays these measured values of ${dT_{\rm {c}}}/{dP}$ (open red circles) as a function of doping, 
and they show a good agreement with published data obtained by applying pressure at low temperature (open blue circles). 
This confirms that our sample are negligibly affected by oxygen-ordering effects, but still raises the relevance of discussing the effect of pressure on doping in YBCO.


\subsection{Hole doping}
\label{subsec:Doping-pressure}

There are two mechanisms by which pressure increases doping in YBCO. 
The first, previously mentioned, has to do with the re-arrangement of oxygen atoms in the CuO chains. 
Pressure improves the degree of oxygen order.
With oxygen ordering comes an enhanced charge transfer between CuO chains and CuO$_2$ planes, and hence an increased doping of holes into the planes. 
This has been studied in detail (\eg{} ref.~\onlinecite{Sadewasser2000}), and there are two approaches to eliminate this ordering process: 
1) apply pressure at a temperature of 200\,K or lower (to freeze oxygen movement in the chains); 
2) use samples that already have a high degree of oxygen order. 

\begin{figure}[t]
\centering
\includegraphics[width=0.46\textwidth]{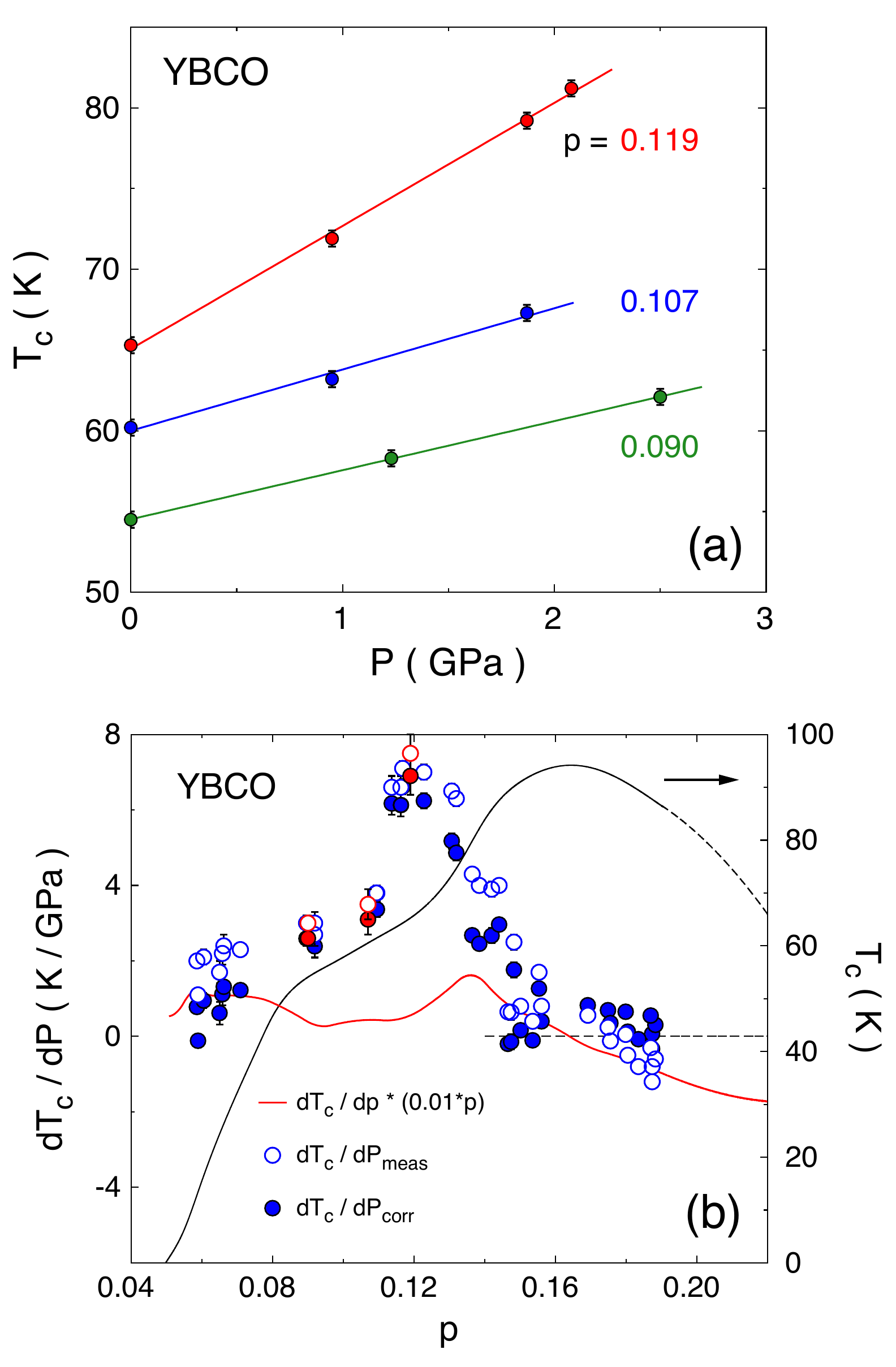}
\caption{(a) Superconducting \Tc{} of our YBCO samples with dopings as indicated as a function of pressure (applied at room temperature). \Tc{} is obtained from the resistivity data in Fig.~\ref{rho-vs-T-P-YBCO}. (b) Sensitivity of \Tc{} to pressure $P$ in YBCO, defined as ${dT_{\rm {c}}}/{dP}$, the initial slope in the \Tc{} vs $P$ dependence, as a function of doping $p$. The data points come from the literature (blue circles and dots; see Table~\ref{table:Table-dTcdP} and references therein) and from our own measurements (red circles and dots; data from panel (a)). Two quantities are plotted: 1) the measured values of ${dT_{\rm {c}}}/{dP}$ (labelled $({dT_{\rm {c}}}/{dP})_{\rm meas}$; open circles); 2) the corrected values (labelled $({dT_{\rm {c}}}/{dP})_{\rm corr}$; full dots). The corrected data are obtained via $({dT_{\rm {c}}}/{dP})_{\rm corr} = \left(\partial T_{\rm {c}}/\partial P\right)_{p} = ({dT_{\rm {c}}}/{dP})_{\rm meas} - 0.01\, p\, ({dT_{\rm {c}}}/{dp})$ (see Table~\ref{table:Table-dTcdP} and Sec.~\ref{subsec:Doping-pressure}). The red line is the magnitude of the correction, with $\left(\partial T_{\rm {c}}/\partial p\right)_{P}$ being the derivative of the \Tc{} vs $p$ curve (black line; right axis). The horizontal dashed line marks $({dT_{\rm {c}}}/{dP})_{\rm corr} = 0$. 
}
\label{dTcdP-raw}
\end{figure}

In Table~\ref{table:Table-dTcdP}, we collect published data on the suppression of \Tc{} with pressure $P$, specifically the rate ${dT_{\rm {c}}}/{dP}$, 
only for those studies that have used one or the other of these approaches to ensure that no oxygen relaxation effects take place. 
For dopings higher than $p $\,=\,$0.15$, in the range where no superstructure is favoured, these precautions are not necessary. 
To the long list of published data in Table~\ref{table:Table-dTcdP}, we add our three data points on oxygen-ordered samples (from Fig.~\ref{dTcdP-raw}(a)) 
and they fit very well with the other published data (Fig.~\ref{dTcdP-raw}(b)).


The second mechanism by which pressure increases doping is one that cannot be avoided. 
By bringing the chains closer to the planes, charge transfer is improved and hole doping is increased. 
The variation of \Tc, with pressure therefore includes two terms~[\onlinecite{Almasan1992},\onlinecite{Neumeier1993}]:
\begin{equation}
\frac{\partial T_{\rm {c}}}{\partial P} = {\frac{\partial p}{\partial P}}\left({\frac{\partial T_{\rm {c}}}{\partial p}}\right)_{P} + \left({\frac{\partial T_{\rm {c}}}{\partial P}}\right)_{p}.
\label{eq:dTcdP}
\end{equation}
The term on the left of the equation is the raw ``sensitivity to pressure'', as measured in the experiment. 
The first term on the right represents the doping effect of pressure and the second term is the direct (intrinsic) dependence of \Tc{} on $P$, of interest here. 
Fortunately, the first term is small. 
In Fig.~\ref{dTcdP-raw}(b), we plot that raw $\partial T_{\rm {c}}/ \partial P$ (referred as ${dT_{\rm {c}}}/{dP}_{\rm meas}$)  
vs doping $p$ (open circles) (see Table~\ref{table:Table-dTcdP}). We immediately see a sharp peak at $p$\,=\,$0.12$. 
To investigate the effect of pressure-induced doping on these data, we plot a corrected set of data, 
obtained by subtracting the product of $\partial p/\partial P$ and $\left(\partial T_{\rm {c}}/\partial p\right)_{P}$ (first term on the right of Eq.~\ref{eq:dTcdP}). 
The term $\partial p/\partial P$ represents the charge transfer (doping) rate, from chains to planes, as pressure is increased. 
Since this rate should depend on the initial doping $p$ (\eg{} charge transfer will be greater for highly doped chains than lightly doped), 
we assume that $\partial p/\partial P$ varies linearly with $p$. 
We then fix the prefactor by requiring that $\left(\partial T_{\rm {c}}/\partial P\right)_{p}$\,=\,$0$ for $p \geq 0.16$, in the overdoped region. 
This is based on our assumption that the small negative values of $\partial T_{\rm {c}}/\partial P$ 
measured at $p > 0.16$ are purely due to the doping effect, since $\left(\partial T_{\rm {c}}/\partial p\right)_{P} < 0$ on the downward sloping side of the \Tc\, vs $p$ curve. 
This yields  $\partial p/\partial P$\,=\,$0.01$\,$*$\,$p$ hole\,/\,GPa. 
As for  $\left(\partial T_{\rm {c}}/\partial p\right)_{P}$, it is simply the derivative of the \Tc\, vs $p$ curve (black curve in Fig.~\ref{Phasediag-CDW}). 
The resulting product term $\partial p/\partial P * \left(\partial T_{\rm {c}}/\partial p\right)_{P}$ is plotted as the red line in Fig.~\ref{dTcdP-raw}(b) and its subtraction from the measured $\partial T_{\rm {c}}/ \partial P$ corresponds to the corrected data $\left(\partial T_{\rm {c}}/\partial P\right)_{p}$, plotted as full dots. 
We see that the correction is small everywhere. 
The doping dependence of the corrected $\left(\partial T_{\rm {c}}/\partial P\right)_{p}$ is also plotted in Fig.~\ref{dTcdP-dTcdH} as blue circles. 
Since these doping effects of pressure directly come from the presence of chains and their doping role in the plane, 
they should be absent in chainless cuprates~[\onlinecite{Yamada1992,Chu1993,Crawford2005,Hucker2010}]. 

In summary, both the measured ${dT_{\rm {c}}}/{dP}$ and the corrected $\left({\partial T_{\rm {c}}} / {\partial P}\right)_{p}$ 
peak sharply at $p$\,=\,$0.12$ (Fig.~\ref{dTcdP-raw}(b)), as noted earlier~[\onlinecite{Benischke1992}]. 
The dramatic increase in ${dT_{\rm {c}}}/{dP}$ between $p$\,$\simeq$\,$0.11$ and $p$\,$\simeq$\,$0.12$ (Fig.~\ref{dTcdP-raw}(a)), 
previously detected in thermal expansion measurements at ambient pressure~[\onlinecite{Kraut1993}], signals a rapid change in the properties of YBCO near $p$\,=\,$0.12$.
Note also that \Tc{} is enhanced by pressure only below $p$\,$\simeq$\,$0.16$. 
Given that CDW order (as detected by x-ray diffraction) onsets below $p$\,$\simeq$\,$0.16$~[\onlinecite{Blanco-Canosa2014}] 
and peaks at $p$\,=\,$0.12$ (Fig.~\ref{Phasediag-CDW}), we attribute the pressure enhancement of  \Tc{} to a suppression of the competing CDW order. 
This interpretation is confirmed by looking at the effect of a magnetic field, 
an established tuning parameter for this phase competition~[\onlinecite{Chang2012a},\onlinecite{Hucker2014},\onlinecite{Blanco-Canosa2014}].


\section{Effect of magnetic field}
\label{sec:Sensitivity-field}

\begin{figure}[t!]
\centering
\includegraphics[width=0.44\textwidth]{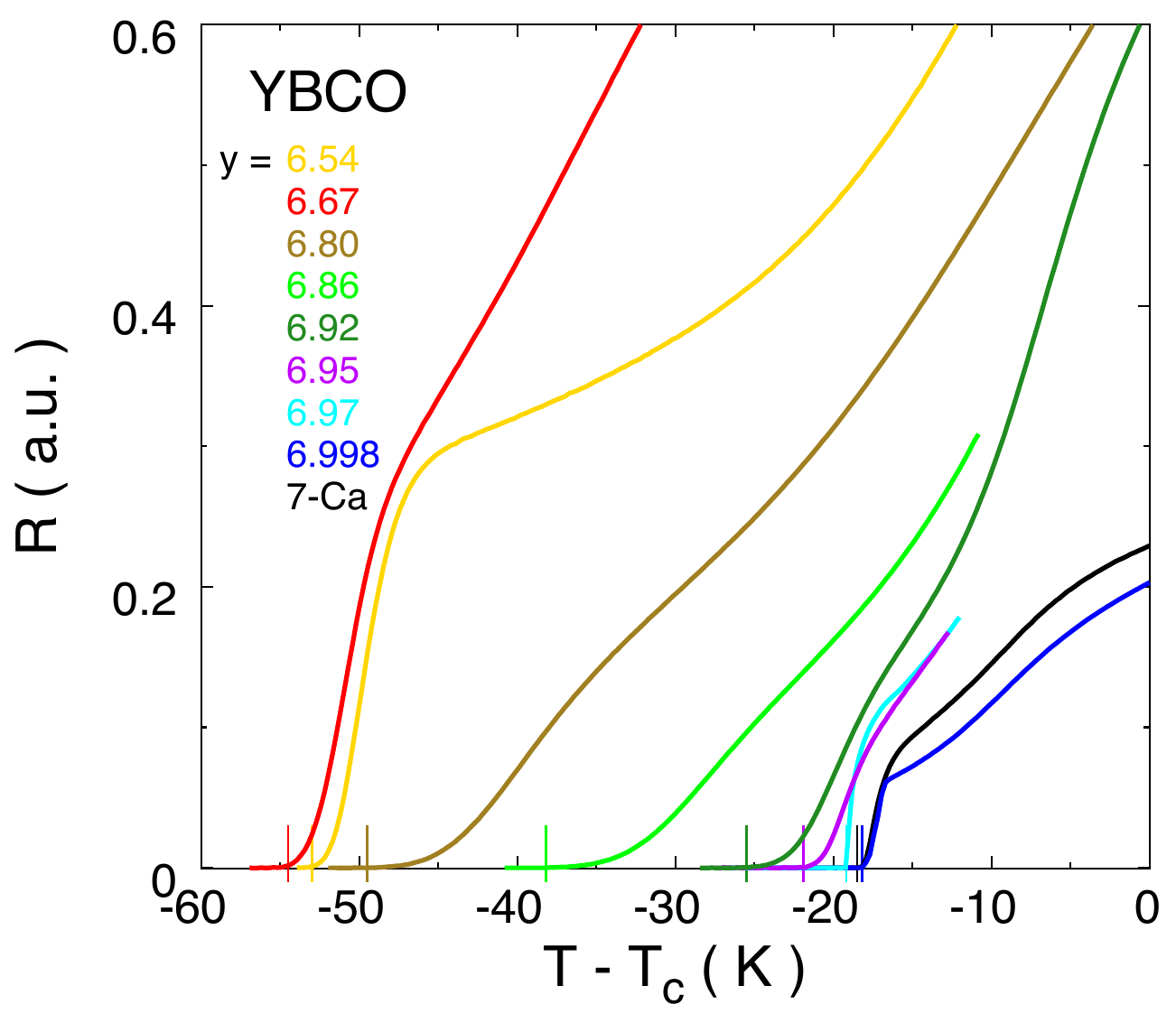}
\caption{Electrical resistance of YBCO with oxygen contents $y$ as indicated, in $H = 15$~T, plotted versus $T-$\Tc, where \Tc{}\,=\,\Tc(0) is the zero-field superconducting transition temperature.
	The shift in \Tc{} caused by the field, $\Delta$\Tc\,$=$\,\Tc(0)\,$-$\,\Tc(15\,T), is marked by a short vertical line and is a measure of the sensitivity to field, defined as $-$\,$d$\Tc\,/\,$dH$\,=\,$\Delta$\Tc\,/\,15\,T.
	All values of \Tc(0), \Tc(15\,T), and $\Delta$\Tc\,/\,15\,T are listed in Table~\ref{table:Table-dTcdH}. 
}
\label{Resistance-Tc}
\end{figure}

\begin{figure}[t!]
\centering
\includegraphics[width=0.44\textwidth]{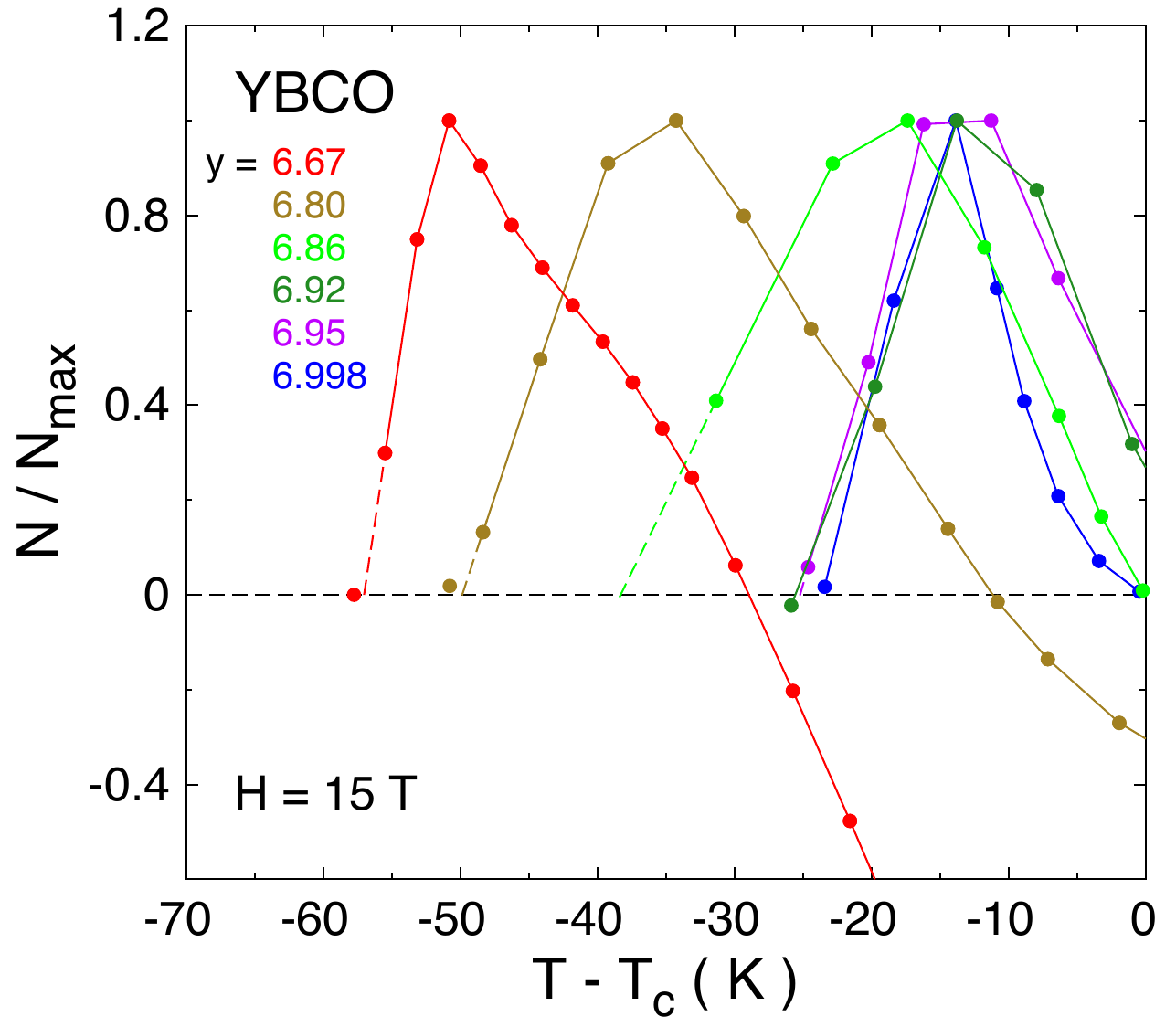}
\caption{Nernst effect of YBCO for six of the samples whose resistance data are shown in Fig.~\ref{Resistance-Tc}, in $H = 15$~T, plotted versus $T-$\Tc.
	\Tc(15\,T) is defined as the point where $N$\,=\,$0$, indicated by the linear extrapolations (dashed lines). The corresponding values are listed in Table~\ref{table:Table-dTcdH}. 
	Details on Nernst effect measurements can be found in ref.~\onlinecite{Daou2010}.
}
\label{Nernst-Tc}
\end{figure}

The sensitivity of the superconducting transition temperature \Tc{} to a magnetic field $H$ applied along the $c$ axis 
was studied across the doping range of YBCO, from $p$\,$\simeq$\,$0.06$ to $p$\,$\simeq$\,$0.19$. 
Using either resistivity (Fig.~\ref{Resistance-Tc}) or Nernst effect data (Fig.~\ref{Nernst-Tc}), the amount $d$\Tc{} by which \Tc{} is reduced when a field 
of 15\,T is applied was measured on 13 different single crystals (see Table~\ref{table:Table-dTcdH}). 
The sensitivity to field, defined as $-$\,$d$\Tc\,/\,$dH$\,=\,[\Tc$(H$\,=\,$0$)\,$-$\,\Tc($H$\,=\,$15$\,T)]\,/\,15\,T, is plotted in Fig.~\ref{dTcdP-dTcdH}. 

We see that it is small and flat above $p$\,=\,$0.16$, it rises rapidly below $p$\,=\,$0.16$, to reach a maximum at $p$\,=\,$0.12$, and then decreases at lower $p$. 
So $-$\,$d$\Tc/$dH$ vs $p$ peaks at $p$\,=\,$0.12$. 
This is not surprising, since we know that \Hc{} vs $p$ has a local minimum at $p$\,$\simeq$\,$0.12$~[\onlinecite{LeBoeuf2011},\onlinecite{Grissonnanche2014}]. 
Note also that $T_{\rm XRD}$ intersects \Tc{} at $p$\,$\simeq$\,$0.16$ (Fig.~\ref{Phasediag-CDW}), thereby explaining the low sensitivity at $p$\,$>$\,$0.16$. 
All this confirms that superconductivity is weakened when CDW order grows, consistent with the scenario of phase competition discussed above.


\section{Discussion}
\label{sec:Discussion}

\begin{figure}[t]
\centering
\includegraphics[width=0.46\textwidth]{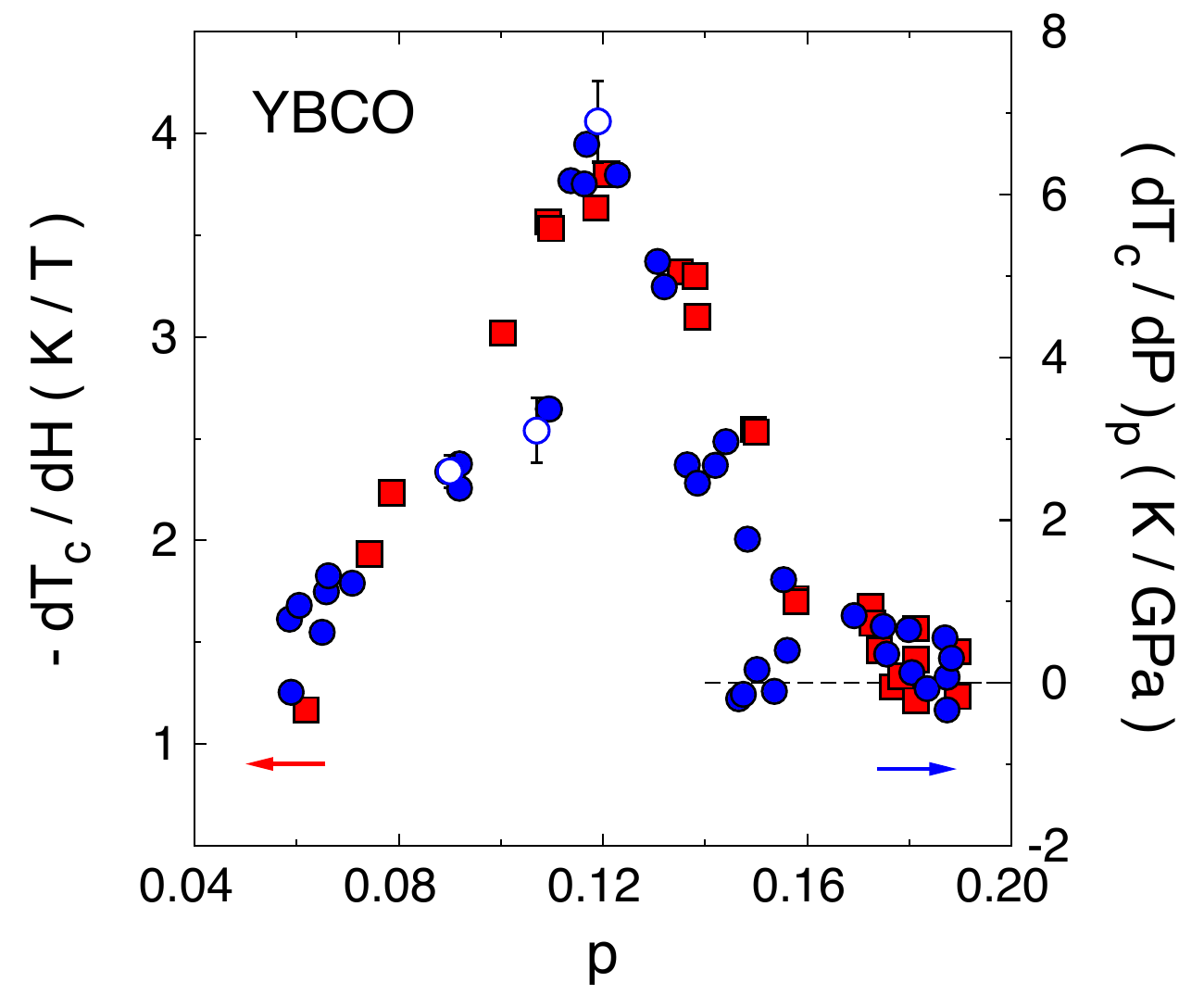}
\caption{
	Sensitivity of \Tc{} to pressure and magnetic field in YBCO as a function of doping. 
	The sensitivity to pressure is defined as the positive change $d$\Tc{} induced by a small pressure $dP$ (at constant $p$), 
	plotted as $(\partial T_{\rm {c}}$\,/\,$\partial P)_{p}$ vs $p$ (blue dots, right axis; see Table~\ref{table:Table-dTcdP} in the Appendix).
	Open circles with error bars are from our own data (see Fig.~\ref{dTcdP-raw}(a)); 
	full circles are from published data (Fig.~\ref{dTcdP-raw} and Table~\ref{table:Table-dTcdP} in the Appendix). 
	To get $(\partial T_{\rm {c}}$\,/\,$\partial P)_{p}$, a small doping-dependent correction is applied to the measured $d$\Tc{}\,/\,$dP$, 
	that accounts for the increase in doping, and hence in \Tc, due to pressure (see Sec.~\ref{subsec:Doping-pressure}). 
	The dashed line marks $(\partial T_{\rm {c}}$\,/\,$\partial P)_{p}$\,=\,$0$.
	The sensitivity to field is defined as the negative shift $d$\Tc{} in \Tc{} in 15\,T: 
	$-$\,$d$\Tc\,/\,$dH$\,=\,[\Tc$(H$\,=\,$0)$\,$-$\,\Tc$(H$\,=\,$15$\,T)]\,/\,15\,T (red squares, left axis; Table~\ref{table:Table-dTcdH} in the Appendix). 
}
\label{dTcdP-dTcdH}
\end{figure}

In Fig.~\ref{dTcdP-dTcdH}, we see that $(\partial T_{\rm {c}}$\,/\,$\partial P)_{p}$ and $-$\,$d$\Tc\,/\,$dH$ track each other : 
both are small and flat for $p$\,$>$\,$0.16$, both rise rapidly below $p$\,=\,$0.16$, and both peak at $p$\,=\,$0.12$. 
The two sensitivities look so identical as a function of doping because $H$ and $P$ are pure tuning parameters: 
$H$ does not directly couple to CDW order (which is independent of $H$ above $T_{\rm {c}}$~[\onlinecite{Chang2012a}]) 
and $P$ does not directly couple to superconductivity ($\partial T_{\rm {c}}$\,/\,$\partial P$\,$\sim$\,$0$ at $p$\,$>$\,$0.16$). 
Field and pressure are two complementary parameters with which to tune phase competition between CDW order and superconductivity in YBCO, in opposite directions. 
This is consistent with our interpretation that pressure suppresses CDW order in YBCO, as shown by x-ray studies in underdoped YBCO~[\onlinecite{Souliou2018},\onlinecite{Huang2018}]. We mention that the same is seen in LBCO at $p$\,=\,$0.125$, with pressure suppressing CDW order and raising \Tc{}~[\onlinecite{Hucker2010}].


\begin{figure}[t]
\centering
\includegraphics[width=0.46\textwidth]{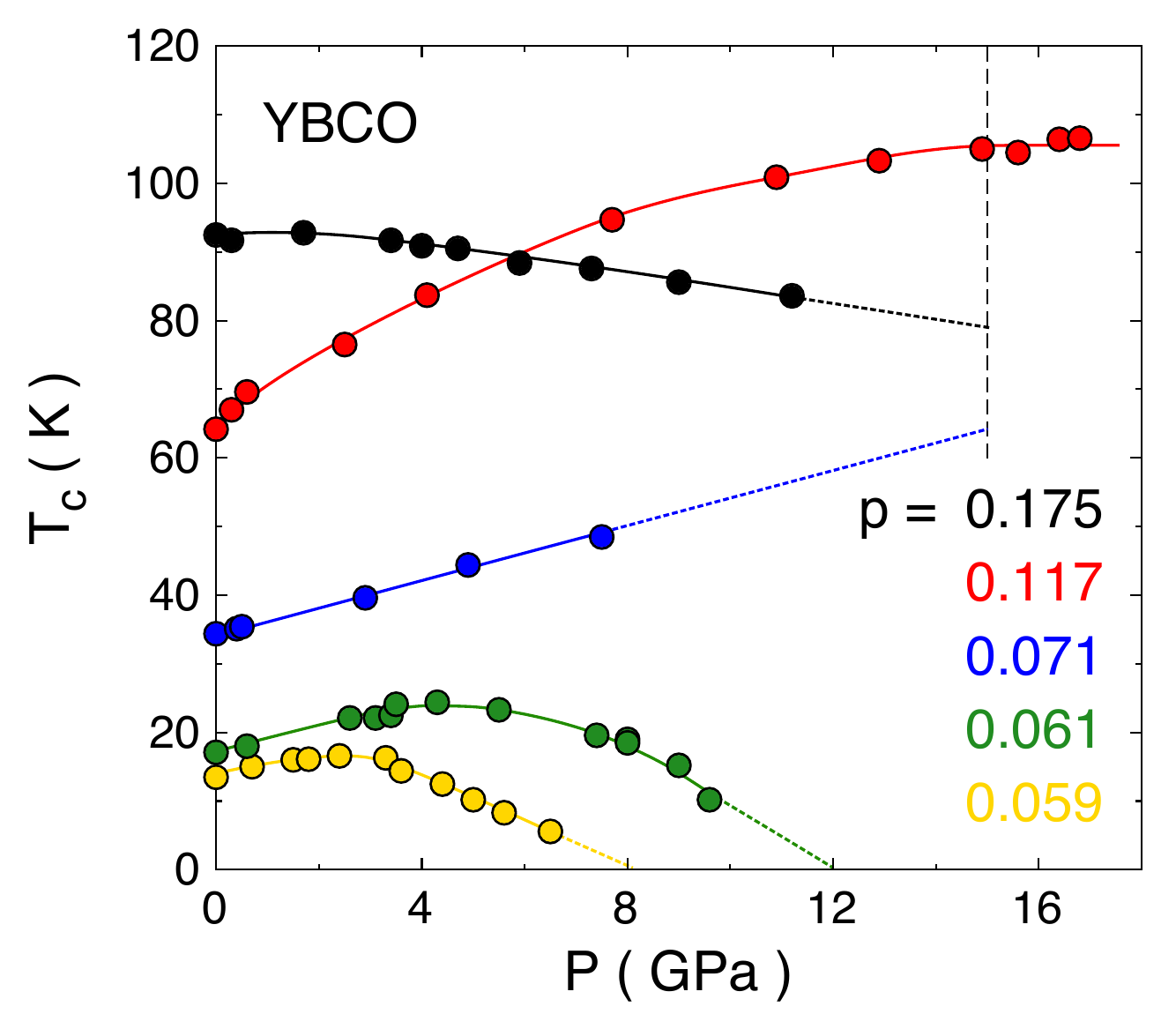}
\caption{
Superconducting \Tc{} versus pressure in YBCO for dopings as indicated, reproduced from ref.~\onlinecite{Sadewasser2000}.
Note that to avoid relaxation effects due to oxygen ordering, the pressure was applied at low temperature. Full lines are a guide to the eye. 
We extract the values of \Tc{} at $P$ = 2 and 7.5~GPa by taking cuts, and at $P = 15$~GPa (vertical dashed line) by extrapolating the data (dotted lines). 
The values of \Tc{} thus obtained are plotted in Fig.~\ref{Phasediag-15GPa}, with the doping adjusted to account for pressure effects.
The zero-pressure values of \Tc{} are (from bottom to top): \Tc(0)~= 14.2 (yellow), 17.5 (green), 34.1 (blue), 63.7 (red) and 92.3 K (black)~[\onlinecite{Sadewasser2000}]. 
}
\label{Tc-vs-P-Sadewasser}
\end{figure}

\begin{figure}[t]
\centering
\includegraphics[width=0.46\textwidth]{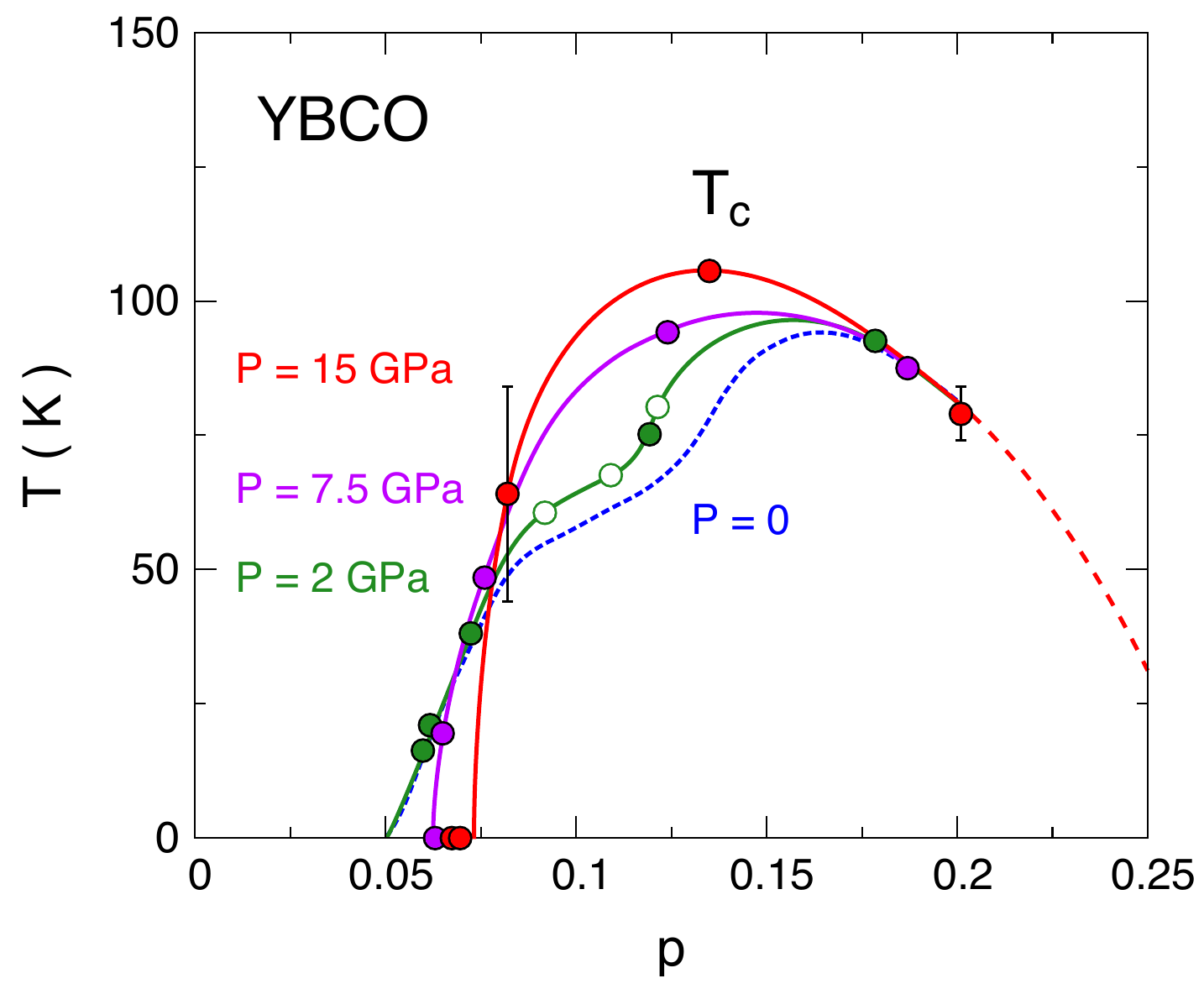}
\caption{
Superconducting phase diagram of YBCO, showing \Tc{} at $P = 0$ (dashed blue line; same as Fig.~\ref{Phasediag-CDW}~[\onlinecite{Liang2006}]), $P = 2$~GPa (green open circles, from our data in Fig.~\ref{dTcdP-raw}(a); green dots, from ref.~\onlinecite{Sadewasser2000} (see Fig.~\ref{Tc-vs-P-Sadewasser})) and $P = 7.5$~GPa (purple dots, from ref.~\onlinecite{Sadewasser2000} (see Fig.~\ref{Tc-vs-P-Sadewasser})).
\Tc{} measured at, or extrapolated to, $P = 15$~GPa is shown as red dots (from ref.~\onlinecite{Sadewasser2000} (see Fig.~\ref{Tc-vs-P-Sadewasser})).
The doping values have been adjusted to include the effect of pressure (see Sec.~\ref{subsec:Doping-pressure}). The green, purple, and red lines are guides to the eye.
}
\label{Phasediag-15GPa}
\end{figure}

We therefore expect that a sufficiently strong pressure will suppress CDW order entirely and reveal the superconducting phase diagram of YBCO free of competition. 
To examine this scenario, we reproduce in Fig.~\ref{Tc-vs-P-Sadewasser} the data for \Tc{} in YBCO as a function of pressure up to 17~GPa and over a wide range of doping as measured by Sadewasser \textit{et al.}~[\onlinecite{Sadewasser2000}].
Taking the measured \Tc{} at $P$ = 2 and 7.5~GPa, we show in Fig.~\ref{Phasediag-15GPa} the evolution of the \Tc{} dome with pressure. To obtain the \Tc{} dome in the high pressure limit, we take the measured value at 15~GPa for $p$\,=\,$0.117$ where \Tc{} saturates, and linearly extrapolate for the four other dopings shown in Fig.~\ref{Tc-vs-P-Sadewasser}. For the highest doping (black; $p$\,=\,$0.175$), the linear extrapolation is reasonable and the uncertainty is small, so that \Tc($15$\,GPa)\,=\,$80$\,$\pm$\,$5$\,K. 
Note that a recent study reports a linear decrease of \Tc{} with pressure in overdoped YBa$_{2}$Cu$_{3}$O$_{7}$ 
leading to a complete suppression above $\sim$\,$10$\,GPa~[\onlinecite{Alireza2017}].
For the lowest two dopings, it is clear that \Tc($15$\,GPa)\,=\,0 regardless of how one extrapolates to 15\,GPa. 
The only significant uncertainty is on the sample with $p$\,=\,$0.071$ (blue), whose data stop at 8\,GPa. 
The dependence of \Tc{} between 8\,GPa and 15\,GPa could be quite different from the linear extrapolation shown in Fig.~\ref{Tc-vs-P-Sadewasser}. 
To reflect that uncertainty, we assign a large error bar to that point, namely \Tc($15$\,GPa)\,=\,$63$\,$\pm$\,$20$\,K. 
This large uncertainty has little impact on the superconducting dome displayed in Fig.~\ref{Phasediag-15GPa}. 
In particular, the position of the peak in the dome of \Tc{} vs $p$ at $15$\,GPa must necessarily be in the interval $0.08$\,$<$\,$p$\,$<$\,$0.13$, most likely close to 0.13.


In Fig.~\ref{Phasediag-15GPa}, we summarize those \Tc{} data at $P$ = 2, 7.5 and $15$\,GPa and plot them as a function of doping.
Note that in order to obtain the doping values under pressure, we use the following formula: $p(P)$\,=\,$p(0)$\,+\,$0.01*p*P$ (see Sec.~\ref{subsec:Doping-pressure}), where $p(0)$ is the doping value at ambient pressure.
For instance, for $p(0)$\,=\,$0.117$, $p(15$\,GPa)\,=\,$0.135$. 
Therefore, as a primary effect (even without considering superconductivity), pressure changes the phase diagram of YBCO by increasing doping. 
More significantly, we see that when CDW order is removed, the superconducting phase in the temperature-doping diagram of YBCO is transformed in two important ways. 
First, the dip at $p$\,$\simeq$\,$0.12$ gradually goes away, so that by $P$\,=\,$15$\,GPa, \Tc{} forms a dome peaked at $p$\,$\simeq$\,$0.13$. 
Indeed, the fact that \Tc{} for $p(0)$\,=\,$0.117$ becomes flat above $15$\,GPa (Fig.~\ref{Tc-vs-P-Sadewasser}) 
implies that it has reached its maximal value of \Tc\,=\,$105$\,K and 
shows that the peak in the dome of \Tc{} vs $p$ at $15$\,GPa must be at $p(15$\,GPa)\,=\,$0.135$ or lower.
(We note that 15~GPa far exceeds the pressure at which the CDW signal in x-ray is suppressed ($\sim 1.0$~GPa) at $p \sim 0.105$~[\onlinecite{Souliou2018}]. Nevertheless, the fact that \Tc{} at $p(0)$\,=\,$0.117$ keeps evolving above 1.0~GPa is consistent with the fact that $R_{\rm H}${} is negative at 1.8~GPa: both show that the effects of the CDW persist to pressures well above 1.0~GPa.)
Secondly, the foot of the dome at low doping moves up, from $p$\,=\,$0.05$ to $p$\,$\simeq$\,$0.075$. 
As a result, the fall of \Tc{} with decreasing $p$ is much faster than it was at ambient pressure (Fig.~\ref{Phasediag-15GPa}). 
At $P$\,=\,$0$, \Tc{} falls below $p$\,$\simeq$\,$0.16$ because a competing phase of CDW order sets in below a $T$\,=\,$0$ critical point 
at $p$\,$\simeq$\,$0.16$~[\onlinecite{Badoux2016}]. 
At $P$\,=\,$15$\,GPa, this CDW critical point is removed (possibly), yet \Tc{} is still a dome, now falling below $p$\,$\simeq$\,$0.13$. 
What competing phase, resistant to pressure, is causing that fall?
Let us mention two possible scenarios.

The first scenario is spin order. 
In YBCO at ambient pressure, long-range antiferromagnetic order exists up to $p = 0.05$ and short-range incommensurate spin-density-wave (SDW) 
correlations extend up to $p \simeq 0.08$~[\onlinecite{Haug2010}]. 
CDW, SDW and superconducting phases all compete with each other~[\onlinecite{Blanco-Canosa2013}]. 
By suppressing CDW order, pressure could strengthen SDW order, extend its range up to higher $p$, and stiffen its competing effect on superconductivity at low doping. 
Adding Zn impurities in YBCO has shown to suppress superconductivity, \eg{} at $p$\,$\simeq$\,$0.12$, 
but to also suppress CDW order, and to nucleate SDW order~[\onlinecite{Blanco-Canosa2013}]. 
Note however that muon spin rotation studies in LBCO showed that hydrostatic pressure suppressed magnetic order (SDW) 
while enhancing superconducting fraction~[\onlinecite{Guguchia2013}]. 

A second scenario for a competition that persists at high pressure is the pseudogap phase.
In particular, the pseudogap due to strong correlations associated with the Mott insulator is known 
to compete with superconductivity~[\onlinecite{Georges1996},\onlinecite{Tremblay2013}], 
and to produce a dome of \Tc{} vs $p$~[\onlinecite{Gull2013}]. 
It remains to be seen where in doping the peak in \Tc{} lies with respect to the \Tstar{} line and the underlying 
critical point for the transition from Fermi-liquid phase at high $p$ to pseudogap phase at low $p$~[\onlinecite{Werner2009},\onlinecite{Sordi2012}]. 
We propose that both competing effects shape the CDW-free \Tc{} dome: the pseudogap phase below its critical point $p^{\star}$\,$\sim$\,$0.19$ and spin order at low doping.

In any scenario, two questions must be addressed: 
Why a dome of CDW order peaked at $p$\,$\simeq$\,$0.12$? 
Why does pressure have such a strong detrimental effect on CDW order, while it has little direct effect on either superconductivity or the pseudogap phase?
Note that in the present manuscript we focused on the short-range 2D CDW order present in zero (and low) magnetic field, which was shown to cause the FSR~[\onlinecite{Cyr-Choiniere2017}]. In future investigations, the pressure dependence of the long-range 3D CDW order seen in high fields~[\onlinecite{LeBoeuf2013},\onlinecite{Gerber2015}] should also be examined separately.


\section{Summary}
\label{sec:Summary}

Over the years, the numerous studies of the effect of pressure on \Tc{} in YBCO have collectively revealed a complex behavior that has remained a mystery. 
A crucial piece of information that had been missing until recently to make sense of the apparent complexity is the existence of a dome of CDW order in YBCO. 
Here we showed that magnetic field and pressure shift \Tc{} in the same way as a function of doping, but in opposite directions, 
and conclude that they are two independent parameters with which to tune the competition between superconductivity and CDW order in YBCO.
This is likely the reason why the record \Tc{} in cuprate superconductors was reached by applying pressure~[\onlinecite{Gao1994}]. 
In YBCO at high pressures, when CDW order is removed, the superconducting dome of \Tc{} vs $p$ is seen to peak at $p \simeq 0.13$, 
revealing that another competing mechanism is at play at low doping. 
We propose that this \Tc{} dome is shaped by the competing effects of the pseudogap phase below $p^{\star}$\,$\sim$\,$0.19$ and spin order at low doping.


\section*{ACKNOWLEDGEMENTS }

We thank J.~Corbin, M.-\`{E}.~Delage, G.~Grissonnanche, E.~Hassinger, F.~Lalibert\'{e}, S.~Ren\'{e} de Cotret, F~F.~Tafti for their assistance with the experiments, 
and B.~J.~Ramshaw for his assistance with sample preparation.
A portion of this work was performed at the National High Magnetic Field Laboratory, which is supported by National Science Foundation Cooperative Agreement No. DMR-1157490 and the State of Florida.
O.C.C. was supported by a fellowship from the Natural Sciences and Engineering Research Council of Canada (NSERC). 
L.T. thanks ESPCI-ParisTech, Universit\'{e} Paris-Sud, CEA-Saclay and the Coll\`{e}ge de France for their hospitality and support, 
and the European Research Council (Grant ERC-319286 QMAC)  
and LABEX PALM (ANR-10-LABX-0039-PALM) for their support, while this article was written. 
D.A.B., W.N.H. and R.L. acknowledge funding from the Natural Sciences and Engineering Research Council of Canada (NSERC). 
L.T. acknowledges support from the Canadian Institute for Advanced Research (CIFAR) and funding from 
the Natural Sciences and Engineering Research Council of Canada (NSERC; PIN:123817), 
the Fonds de recherche du Qu\'{e}bec - Nature et Technologies (FRQNT), 
the Canada Foundation for Innovation (CFI),
and a Canada Research Chair. 
Part of this work was funded by the Gordon and Betty Moore Foundation's EPiQS Initiative (Grant GBMF5306 to L.T.).


\section{APPENDIX}
\label{sec:Appendix}

Tables~\ref{table:Table-dTcdH} and~\ref{table:Table-dTcdP} below present the raw values of the data points plotted in Figs.~\ref{dTcdP-raw}(b) and \ref{dTcdP-dTcdH}.


\renewcommand{\arraystretch}{1.1}
\setlength{\tabcolsep}{7pt}

\begin{table}[t!]
\caption{
	Characteristics of the 13 YBCO samples whose sensitivity of \Tc\,to field is plotted in Fig.~\ref{dTcdP-dTcdH}, labelled by their oxygen content $y$, 
	doping $p$, zero-resistance \Tc\, at $H$\,=\,$0$ and $H$\,=\,$15$\,T, from resistivity (Fig.~\ref{Resistance-Tc}) or Nernst (Fig.~\ref{Nernst-Tc}) data. 
	$\Delta$\Tc\ = \Tc$(H$\,=\,$0)$\,$-$\,\Tc$(H$\,=\,$15$\,T); $-$\,$d$\Tc/$dH$\,=\,$\Delta$\Tc /15\,T. 
	The numbers in the last column are plotted as red squares in Fig.~\ref{dTcdP-dTcdH}. 
}
\centering
\begin{tabular}{cccccccccc}
\\
\hline
$p$ & $y$ & $T_{\rm c}$ (0) & $T_{\rm c}$ (15T) & $\Delta T_{\rm c}$ / 15T \\
 & & (K) & (K) & (K\,/\,T) \\
\hline
\hline
0.062 & 6.35 & 19.5 & 2.0 & 1.17 \\
0.074 & 6.45 & 39.5 & 10.5 & 1.93 \\
0.079 & 6.45 & 45.0 & 11.5 & 2.23 \\
0.100 & 6.49 & 57.8 & 12.5 & 3.02 \\
0.109 & 6.54 & 61.3 & 7.8 & 3.57 \\
0.110 & 6.54 & 61.5 & 8.5 & 3.53 \\
0.119 & 6.67 & 65.0 & 10.5 & 3.63 \\
0.121 & 6.67 & 66.0 & 9.0 & 3.80 \\
0.135 & 6.80 & 78.5 & 28.7 & 3.32 \\
0.138 & 6.80 & 82.0 & 32.5 & 3.30 \\
0.139 & 6.80 & 82.5 & 36.0 & 3.10 \\
0.150 & 6.86 & 90.8 & 52.6 & 2.55 \\
0.150 & 6.86 & 91.0 & 53.0 & 2.53 \\
0.158 & 6.92 & 93.5 & 68.0 & 1.70 \\
0.158 & 6.92 & 93.5 & 67.8 & 1.71 \\
0.172 & 6.95 & 93.1 & 68.0 & 1.67 \\
0.173 & 6.99 & 93.0 & 69.1 & 1.59 \\
0.174 & 6.95 & 92.7 & 70.8 & 1.46 \\
0.177 & 6.97 & 92.0 & 72.8 & 1.28 \\
0.178 & 6.97 & 91.5 & 71.5 & 1.33 \\
0.181 & 6.998 & 90.5 & 72.3 & 1.21 \\
0.181 & 6.998 & 90.5 & 69.3 & 1.41 \\
0.181 & 6.998 & 90.5 & 67.0 & 1.57 \\
0.190 & Ca1.4\% & 87.0 & 65.2 & 1.45 \\
0.190 & Ca1.4\% & 87.0 & 68.5 & 1.23 \\
\hline
\end{tabular}
\label{table:Table-dTcdH}
\end{table}



\renewcommand{\arraystretch}{1.1}
\setlength{\tabcolsep}{4pt}

\begin{table}[h!]
\caption{
	Characteristics of all samples whose sensitivity of \Tc{} to pressure is plotted in Figs.~\ref{dTcdP-raw}(b) and \ref{dTcdP-dTcdH}. 
	Zero-resistance \Tc{} at ambient pressure $(P$\,=\,$0)$, doping $p$, measured initial slope ${dT_{\rm {c}}}/{dP}$.
	$\left(\partial T_{\rm {c}}/\partial P\right)_{p}$ corresponds to the measured ${dT_{\rm {c}}}/{dP}$ 
	corrected for the doping effect of pressure (see section~\ref{subsec:Doping-pressure}). 
	In the 4th column, we list the term $\left(\partial T_{\rm {c}}/\partial p\right)_{P}$ that goes into this correction.
	The last column gives the reference for the data. 
	Our own three samples are identified as ``Own''.
}
\centering
\begin{tabular}{cccccccccc}
\\
\hline
$T_{\rm c}$(0) & $p$ & $dT_{\rm c}/dP$ & $\left(\partial T_{\rm {c}}/\partial p\right)_{P}$ & $\left(\partial T_{\rm {c}}/\partial P\right)_{p}$ & Ref. \\
(K) & & (K\,/\,GPa) & (K\,/\,hole) & (K\,/\,GPa) & \\
\hline
\hline
14.0 & 0.059 & 1.1 & 2064.32 & -0.1\,$\pm$\,0.1 &  [\onlinecite{Tissen1999}] \\
13.5 & 0.059 & 2.0 & 2079.08 & 0.8\,$\pm$\,0.1 & [\onlinecite{Sadewasser2000}] \\
17.1 & 0.061 & 2.1 & 1901.22 & 1.0\,$\pm$\,0.2 & [\onlinecite{Sadewasser2000}] \\
25.0 & 0.065 & 1.7 & 1661.15 & 0.6\,$\pm$\,0.3 & [\onlinecite{Fietz1996}] \\
26.4 & 0.066 & 2.2 & 1644.39 & 1.1\,$\pm$\,0.3 & [\onlinecite{Fietz1996}] \\
27.1 & 0.066 & 2.4 & 1636.17 & 1.3\,$\pm$\,0.3 & [\onlinecite{Fietz1996}] \\
34.4 & 0.071 & 2.3 & 1514.45 & 1.2\,$\pm$\,0.1 & [\onlinecite{Sadewasser2000}] \\
54.2 & 0.090 & 3.0 & 453.84 & 2.6\,$\pm$\,0.2 & [\onlinecite{Benischke1992}] \\
54.5 & 0.090 & 3.0 & 428.50 & 2.6\,$\pm$\,0.2 & Own \\ 
55.1 & 0.092 & 2.7 & 337.67 & 2.4\,$\pm$\,0.3 & [\onlinecite{Fietz1996}] \\
55.1 & 0.092 & 3.0 & 337.67 & 2.7\,$\pm$\,0.3 & [\onlinecite{Fietz1996}] \\
60.2 & 0.107 & 3.5 & 406.50 & 3.1\,$\pm$\,0.4 & Own \\ 
61.3 & 0.109 & 3.8 & 398.10 & 3.4\,$\pm$\,0.2 & [\onlinecite{Benischke1992}] \\
61.4 & 0.110 & 3.8 & 396.10 & 3.4\,$\pm$\,0.2 & [\onlinecite{Benischke1992}] \\
63.0 & 0.114 & 6.6 & 377.22 & 6.2\,$\pm$\,0.3 & [\onlinecite{Fietz1996}] \\
64.0 & 0.116 & 6.6 & 403.97 & 6.1\,$\pm$\,0.3 & [\onlinecite{Fietz1996}] \\
64.2 & 0.117 & 7.1 & 412.42 & 6.6\,$\pm$\,0.2 & [\onlinecite{Sadewasser2000}] \\
65.3 & 0.119 & 7.2 & 467.50 & 6.9\,$\pm$\,0.5 & Own \\ 
67.2 & 0.123 & 7.0 & 616.06 & 6.2\,$\pm$\,0.2 & [\onlinecite{Benischke1992}] \\
73.6 & 0.131 & 6.5 & 1012.29 & 5.2\,$\pm$\,0.2 & [\onlinecite{Benischke1992}] \\
75.0 & 0.132 & 6.3 & 1086.33 & 4.9\,$\pm$\,0.2 & [\onlinecite{Benischke1992}] \\
80.2 & 0.137 & 4.3 & 1187.43 & 2.7\,$\pm$\,0.1 & [\onlinecite{Almasan1992}] \\
82.5 & 0.139 & 4.0 & 1115.16 & 2.5\,$\pm$\,0.1 & [\onlinecite{Almasan1992}] \\
86.0 & 0.142 & 3.9 & 864.92 & 2.7\,$\pm$\,0.2 & [\onlinecite{Benischke1992}] \\
87.7 & 0.144 & 4.0 & 718.94 & 3.0\,$\pm$\,0.1 & [\onlinecite{Almasan1992}] \\
89.3 & 0.147 & 0.7 & 576.82 & -0.2$\pm$ 0.2 & [\onlinecite{Klotz1991}] \\
89.8 & 0.148 & 0.6 & 531.21 & -0.1\,$\pm$\,0.2 & [\onlinecite{Tozer1993}] \\
90.2 & 0.148 & 2.5 & 495.32 & 1.8\,$\pm$\,0.2 & [\onlinecite{Benischke1992}] \\
91.1 & 0.150 & 0.8 & 423.91 & 0.2\,$\pm$\,0.1 & [\onlinecite{Fietz1994}] \\
92.4 & 0.154 & 0.4 & 328.33 & -0.10\,$\pm$\,0.08 & [\onlinecite{Benischke1992}] \\
92.9 & 0.155 & 1.7 & 277.42 & 1.3\,$\pm$\,0.1 & [\onlinecite{Almasan1992}] \\
93.1 & 0.156 & 0.8 & 257.46 & 0.4\,$\pm$\,0.1 & [\onlinecite{Yoshida1999}] \\
93.7 & 0.169 & 0.6 & -157.27 & 0.83\,$\pm$\,0.06 & [\onlinecite{Almasan1992}] \\
92.5 & 0.175 & 0.2 & -259.58 & 0.69\,$\pm$\,0.06 & [\onlinecite{Sadewasser2000}] \\
92.3 & 0.176 & -0.1 & -269.86 & 0.35\,$\pm$\,0.08 & [\onlinecite{Fietz1996}] \\
91.0 & 0.180 & 0.05 & -335.27 & 0.65\,$\pm$\,0.09 & [\onlinecite{Fietz1994}] \\
90.8 & 0.181 & -0.5 & -346.04 & 0.1\,$\pm$\,0.1 & [\onlinecite{Fietz1994}] \\
89.7 & 0.184 & -0.8 & -396.45 & -0.1\,$\pm$\,0.1 & [\onlinecite{Fietz1994}] \\
88.2 & 0.187 & -0.3 & -455.77 & 0.6\,$\pm$\,0.1 & [\onlinecite{Yoshida1999}] \\
88.0 & 0.187 & -0.8 & -462.91 & 0.1\,$\pm$\,0.1 & [\onlinecite{Fietz1994}] \\
88.0 & 0.187 & -1.2 & -462.91 & -0.33\,$\pm$\,0.03 & [\onlinecite{Lortz2006}] \\
87.6 & 0.188 & -0.6 & -479.67 & 0.3\,$\pm$\,0.1 & [\onlinecite{Fietz1994}] \\
\hline	
\end{tabular}
\label{table:Table-dTcdP}
\end{table}


\clearpage


\bibliographystyle{apsrev4-1}

\begin{thebibliography}{65}%
\makeatletter
\providecommand \@ifxundefined [1]{%
 \@ifx{#1\undefined}
}%
\providecommand \@ifnum [1]{%
 \ifnum #1\expandafter \@firstoftwo
 \else \expandafter \@secondoftwo
 \fi
}%
\providecommand \@ifx [1]{%
 \ifx #1\expandafter \@firstoftwo
 \else \expandafter \@secondoftwo
 \fi
}%
\providecommand \natexlab [1]{#1}%
\providecommand \enquote  [1]{``#1''}%
\providecommand \bibnamefont  [1]{#1}%
\providecommand \bibfnamefont [1]{#1}%
\providecommand \citenamefont [1]{#1}%
\providecommand \href@noop [0]{\@secondoftwo}%
\providecommand \href [0]{\begingroup \@sanitize@url \@href}%
\providecommand \@href[1]{\@@startlink{#1}\@@href}%
\providecommand \@@href[1]{\endgroup#1\@@endlink}%
\providecommand \@sanitize@url [0]{\catcode `\\12\catcode `\$12\catcode
  `\&12\catcode `\#12\catcode `\^12\catcode `\_12\catcode `\%12\relax}%
\providecommand \@@startlink[1]{}%
\providecommand \@@endlink[0]{}%
\providecommand \url  [0]{\begingroup\@sanitize@url \@url }%
\providecommand \@url [1]{\endgroup\@href {#1}{\urlprefix }}%
\providecommand \urlprefix  [0]{URL }%
\providecommand \Eprint [0]{\href }%
\providecommand \doibase [0]{http://dx.doi.org/}%
\providecommand \selectlanguage [0]{\@gobble}%
\providecommand \bibinfo  [0]{\@secondoftwo}%
\providecommand \bibfield  [0]{\@secondoftwo}%
\providecommand \translation [1]{[#1]}%
\providecommand \BibitemOpen [0]{}%
\providecommand \bibitemStop [0]{}%
\providecommand \bibitemNoStop [0]{.\EOS\space}%
\providecommand \EOS [0]{\spacefactor3000\relax}%
\providecommand \BibitemShut  [1]{\csname bibitem#1\endcsname}%
\let\auto@bib@innerbib\@empty
\bibitem [{\citenamefont {Wu}\ \emph {et~al.}(2011)\citenamefont {Wu},
  \citenamefont {Mayaffre}, \citenamefont {Kr\"amer}, \citenamefont
  {Horvati\`c}, \citenamefont {Berthier}, \citenamefont {Hardy}, \citenamefont
  {Liang}, \citenamefont {Bonn},\ and\ \citenamefont {Julien}}]{Wu2011}%
  \BibitemOpen
  \bibfield  {author} {\bibinfo {author} {\bibfnamefont {T.}~\bibnamefont
  {Wu}}, \bibinfo {author} {\bibfnamefont {H.}~\bibnamefont {Mayaffre}},
  \bibinfo {author} {\bibfnamefont {S.}~\bibnamefont {Kr\"amer}}, \bibinfo
  {author} {\bibfnamefont {M.}~\bibnamefont {Horvati\`c}}, \bibinfo {author}
  {\bibfnamefont {C.}~\bibnamefont {Berthier}}, \bibinfo {author}
  {\bibfnamefont {W.~N.}\ \bibnamefont {Hardy}}, \bibinfo {author}
  {\bibfnamefont {R.}~\bibnamefont {Liang}}, \bibinfo {author} {\bibfnamefont
  {D.~A.}\ \bibnamefont {Bonn}}, \ and\ \bibinfo {author} {\bibfnamefont
  {M.-H.}\ \bibnamefont {Julien}},\ }\href
  {http://dx.doi.org/10.1038/nature10345} {\bibfield  {journal} {\bibinfo
  {journal} {Nature}\ }\textbf {\bibinfo {volume} {477}},\ \bibinfo {pages}
  {191} (\bibinfo {year} {2011})}\BibitemShut {NoStop}%
\bibitem [{\citenamefont {Ghiringhelli}\ \emph {et~al.}(2012)\citenamefont
  {Ghiringhelli}, \citenamefont {Le~Tacon}, \citenamefont {Minola},
  \citenamefont {Blanco-Canosa}, \citenamefont {Mazzoli}, \citenamefont
  {Brookes}, \citenamefont {De~Luca}, \citenamefont {Frano}, \citenamefont
  {Hawthorn}, \citenamefont {He}, \citenamefont {Loew}, \citenamefont {Sala},
  \citenamefont {Peets}, \citenamefont {Salluzzo}, \citenamefont {Schierle},
  \citenamefont {Sutarto}, \citenamefont {Sawatzky}, \citenamefont {Weschke},
  \citenamefont {Keimer},\ and\ \citenamefont {Braicovich}}]{Ghiringhelli2012}%
  \BibitemOpen
  \bibfield  {author} {\bibinfo {author} {\bibfnamefont {G.}~\bibnamefont
  {Ghiringhelli}}, \bibinfo {author} {\bibfnamefont {M.}~\bibnamefont
  {Le~Tacon}}, \bibinfo {author} {\bibfnamefont {M.}~\bibnamefont {Minola}},
  \bibinfo {author} {\bibfnamefont {S.}~\bibnamefont {Blanco-Canosa}}, \bibinfo
  {author} {\bibfnamefont {C.}~\bibnamefont {Mazzoli}}, \bibinfo {author}
  {\bibfnamefont {N.~B.}\ \bibnamefont {Brookes}}, \bibinfo {author}
  {\bibfnamefont {G.~M.}\ \bibnamefont {De~Luca}}, \bibinfo {author}
  {\bibfnamefont {A.}~\bibnamefont {Frano}}, \bibinfo {author} {\bibfnamefont
  {D.~G.}\ \bibnamefont {Hawthorn}}, \bibinfo {author} {\bibfnamefont
  {F.}~\bibnamefont {He}}, \bibinfo {author} {\bibfnamefont {T.}~\bibnamefont
  {Loew}}, \bibinfo {author} {\bibfnamefont {M.~M.}\ \bibnamefont {Sala}},
  \bibinfo {author} {\bibfnamefont {D.~C.}\ \bibnamefont {Peets}}, \bibinfo
  {author} {\bibfnamefont {M.}~\bibnamefont {Salluzzo}}, \bibinfo {author}
  {\bibfnamefont {E.}~\bibnamefont {Schierle}}, \bibinfo {author}
  {\bibfnamefont {R.}~\bibnamefont {Sutarto}}, \bibinfo {author} {\bibfnamefont
  {G.~A.}\ \bibnamefont {Sawatzky}}, \bibinfo {author} {\bibfnamefont
  {E.}~\bibnamefont {Weschke}}, \bibinfo {author} {\bibfnamefont
  {B.}~\bibnamefont {Keimer}}, \ and\ \bibinfo {author} {\bibfnamefont
  {L.}~\bibnamefont {Braicovich}},\ }\href {\doibase 10.1126/science.1223532}
  {\bibfield  {journal} {\bibinfo  {journal} {Science}\ }\textbf {\bibinfo
  {volume} {337}},\ \bibinfo {pages} {821} (\bibinfo {year}
  {2012})}\BibitemShut {NoStop}%
\bibitem [{\citenamefont {Chang}\ \emph {et~al.}(2012)\citenamefont {Chang},
  \citenamefont {Blackburn}, \citenamefont {Holmes}, \citenamefont
  {Christensen}, \citenamefont {Larsen}, \citenamefont {Mesot}, \citenamefont
  {Liang}, \citenamefont {Bonn}, \citenamefont {Hardy}, \citenamefont
  {Watenphul}, \citenamefont {Zimmermann}, \citenamefont {Forgan},\ and\
  \citenamefont {Hayden}}]{Chang2012a}%
  \BibitemOpen
  \bibfield  {author} {\bibinfo {author} {\bibfnamefont {J.}~\bibnamefont
  {Chang}}, \bibinfo {author} {\bibfnamefont {E.}~\bibnamefont {Blackburn}},
  \bibinfo {author} {\bibfnamefont {A.~T.}\ \bibnamefont {Holmes}}, \bibinfo
  {author} {\bibfnamefont {N.~B.}\ \bibnamefont {Christensen}}, \bibinfo
  {author} {\bibfnamefont {J.}~\bibnamefont {Larsen}}, \bibinfo {author}
  {\bibfnamefont {J.}~\bibnamefont {Mesot}}, \bibinfo {author} {\bibfnamefont
  {R.}~\bibnamefont {Liang}}, \bibinfo {author} {\bibfnamefont {D.~A.}\
  \bibnamefont {Bonn}}, \bibinfo {author} {\bibfnamefont {W.~N.}\ \bibnamefont
  {Hardy}}, \bibinfo {author} {\bibfnamefont {A.}~\bibnamefont {Watenphul}},
  \bibinfo {author} {\bibfnamefont {M.~v.}\ \bibnamefont {Zimmermann}},
  \bibinfo {author} {\bibfnamefont {E.~M.}\ \bibnamefont {Forgan}}, \ and\
  \bibinfo {author} {\bibfnamefont {S.~M.}\ \bibnamefont {Hayden}},\ }\href
  {\doibase 10.1038/nphys2456} {\bibfield  {journal} {\bibinfo  {journal} {Nat.
  Phys.}\ }\textbf {\bibinfo {volume} {8}},\ \bibinfo {pages} {871} (\bibinfo
  {year} {2012})}\BibitemShut {NoStop}%
\bibitem [{\citenamefont {Achkar}\ \emph {et~al.}(2012)\citenamefont {Achkar},
  \citenamefont {Sutarto}, \citenamefont {Mao}, \citenamefont {He},
  \citenamefont {Frano}, \citenamefont {Blanco-Canosa}, \citenamefont
  {Le~Tacon}, \citenamefont {Ghiringhelli}, \citenamefont {Braicovich},
  \citenamefont {Minola}, \citenamefont {Moretti~Sala}, \citenamefont
  {Mazzoli}, \citenamefont {Liang}, \citenamefont {Bonn}, \citenamefont
  {Hardy}, \citenamefont {Keimer}, \citenamefont {Sawatzky},\ and\
  \citenamefont {Hawthorn}}]{Achkar2012}%
  \BibitemOpen
  \bibfield  {author} {\bibinfo {author} {\bibfnamefont {A.~J.}\ \bibnamefont
  {Achkar}}, \bibinfo {author} {\bibfnamefont {R.}~\bibnamefont {Sutarto}},
  \bibinfo {author} {\bibfnamefont {X.}~\bibnamefont {Mao}}, \bibinfo {author}
  {\bibfnamefont {F.}~\bibnamefont {He}}, \bibinfo {author} {\bibfnamefont
  {A.}~\bibnamefont {Frano}}, \bibinfo {author} {\bibfnamefont
  {S.}~\bibnamefont {Blanco-Canosa}}, \bibinfo {author} {\bibfnamefont
  {M.}~\bibnamefont {Le~Tacon}}, \bibinfo {author} {\bibfnamefont
  {G.}~\bibnamefont {Ghiringhelli}}, \bibinfo {author} {\bibfnamefont
  {L.}~\bibnamefont {Braicovich}}, \bibinfo {author} {\bibfnamefont
  {M.}~\bibnamefont {Minola}}, \bibinfo {author} {\bibfnamefont
  {M.}~\bibnamefont {Moretti~Sala}}, \bibinfo {author} {\bibfnamefont
  {C.}~\bibnamefont {Mazzoli}}, \bibinfo {author} {\bibfnamefont
  {R.}~\bibnamefont {Liang}}, \bibinfo {author} {\bibfnamefont {D.~A.}\
  \bibnamefont {Bonn}}, \bibinfo {author} {\bibfnamefont {W.~N.}\ \bibnamefont
  {Hardy}}, \bibinfo {author} {\bibfnamefont {B.}~\bibnamefont {Keimer}},
  \bibinfo {author} {\bibfnamefont {G.~A.}\ \bibnamefont {Sawatzky}}, \ and\
  \bibinfo {author} {\bibfnamefont {D.~G.}\ \bibnamefont {Hawthorn}},\ }\href
  {\doibase 10.1103/PhysRevLett.109.167001} {\bibfield  {journal} {\bibinfo
  {journal} {Phys. Rev. Lett.}\ }\textbf {\bibinfo {volume} {109}},\ \bibinfo
  {pages} {167001} (\bibinfo {year} {2012})}\BibitemShut {NoStop}%
\bibitem [{\citenamefont {Croft}\ \emph {et~al.}(2014)\citenamefont {Croft},
  \citenamefont {Lester}, \citenamefont {Senn}, \citenamefont {Bombardi},\ and\
  \citenamefont {Hayden}}]{Croft2014}%
  \BibitemOpen
  \bibfield  {author} {\bibinfo {author} {\bibfnamefont {T.~P.}\ \bibnamefont
  {Croft}}, \bibinfo {author} {\bibfnamefont {C.}~\bibnamefont {Lester}},
  \bibinfo {author} {\bibfnamefont {M.~S.}\ \bibnamefont {Senn}}, \bibinfo
  {author} {\bibfnamefont {A.}~\bibnamefont {Bombardi}}, \ and\ \bibinfo
  {author} {\bibfnamefont {S.~M.}\ \bibnamefont {Hayden}},\ }\href
  {http://link.aps.org/doi/10.1103/PhysRevB.89.224513} {\bibfield  {journal}
  {\bibinfo  {journal} {Phys. Rev. B}\ }\textbf {\bibinfo {volume} {89}},\
  \bibinfo {pages} {224513} (\bibinfo {year} {2014})}\BibitemShut {NoStop}%
\bibitem [{\citenamefont {Tabis}\ \emph {et~al.}(2014)\citenamefont {Tabis},
  \citenamefont {Li}, \citenamefont {Tacon}, \citenamefont {Braicovich},
  \citenamefont {Kreyssig}, \citenamefont {Minola}, \citenamefont {Dellea},
  \citenamefont {Weschke}, \citenamefont {Veit}, \citenamefont {Ramazanoglu},
  \citenamefont {Goldman}, \citenamefont {Schmitt}, \citenamefont
  {Ghiringhelli}, \citenamefont {Bari{\v s}i{\'c}}, \citenamefont {Chan},
  \citenamefont {Dorow}, \citenamefont {Yu}, \citenamefont {Zhao},
  \citenamefont {Keimer},\ and\ \citenamefont {Greven}}]{Tabis2014}%
  \BibitemOpen
  \bibfield  {author} {\bibinfo {author} {\bibfnamefont {W.}~\bibnamefont
  {Tabis}}, \bibinfo {author} {\bibfnamefont {Y.}~\bibnamefont {Li}}, \bibinfo
  {author} {\bibfnamefont {M.~L.}\ \bibnamefont {Tacon}}, \bibinfo {author}
  {\bibfnamefont {L.}~\bibnamefont {Braicovich}}, \bibinfo {author}
  {\bibfnamefont {A.}~\bibnamefont {Kreyssig}}, \bibinfo {author}
  {\bibfnamefont {M.}~\bibnamefont {Minola}}, \bibinfo {author} {\bibfnamefont
  {G.}~\bibnamefont {Dellea}}, \bibinfo {author} {\bibfnamefont
  {E.}~\bibnamefont {Weschke}}, \bibinfo {author} {\bibfnamefont {M.~J.}\
  \bibnamefont {Veit}}, \bibinfo {author} {\bibfnamefont {M.}~\bibnamefont
  {Ramazanoglu}}, \bibinfo {author} {\bibfnamefont {A.~I.}\ \bibnamefont
  {Goldman}}, \bibinfo {author} {\bibfnamefont {T.}~\bibnamefont {Schmitt}},
  \bibinfo {author} {\bibfnamefont {G.}~\bibnamefont {Ghiringhelli}}, \bibinfo
  {author} {\bibfnamefont {N.}~\bibnamefont {Bari{\v s}i{\'c}}}, \bibinfo
  {author} {\bibfnamefont {M.~K.}\ \bibnamefont {Chan}}, \bibinfo {author}
  {\bibfnamefont {C.~J.}\ \bibnamefont {Dorow}}, \bibinfo {author}
  {\bibfnamefont {G.}~\bibnamefont {Yu}}, \bibinfo {author} {\bibfnamefont
  {X.}~\bibnamefont {Zhao}}, \bibinfo {author} {\bibfnamefont {B.}~\bibnamefont
  {Keimer}}, \ and\ \bibinfo {author} {\bibfnamefont {M.}~\bibnamefont
  {Greven}},\ }\href {http://dx.doi.org/10.1038/ncomms6875} {\bibfield
  {journal} {\bibinfo  {journal} {Nat. Commun.}\ }\textbf {\bibinfo {volume}
  {5}},\ \bibinfo {pages} {5875} (\bibinfo {year} {2014})}\BibitemShut
  {NoStop}%
\bibitem [{\citenamefont {Comin}\ \emph {et~al.}(2014)\citenamefont {Comin},
  \citenamefont {Frano}, \citenamefont {Yee}, \citenamefont {Yoshida},
  \citenamefont {Eisaki}, \citenamefont {Schierle}, \citenamefont {Weschke},
  \citenamefont {Sutarto}, \citenamefont {He}, \citenamefont {Soumyanarayanan},
  \citenamefont {He}, \citenamefont {Le~Tacon}, \citenamefont {Elfimov},
  \citenamefont {Hoffman}, \citenamefont {Sawatzky}, \citenamefont {Keimer},\
  and\ \citenamefont {Damascelli}}]{Comin2014}%
  \BibitemOpen
  \bibfield  {author} {\bibinfo {author} {\bibfnamefont {R.}~\bibnamefont
  {Comin}}, \bibinfo {author} {\bibfnamefont {A.}~\bibnamefont {Frano}},
  \bibinfo {author} {\bibfnamefont {M.~M.}\ \bibnamefont {Yee}}, \bibinfo
  {author} {\bibfnamefont {Y.}~\bibnamefont {Yoshida}}, \bibinfo {author}
  {\bibfnamefont {H.}~\bibnamefont {Eisaki}}, \bibinfo {author} {\bibfnamefont
  {E.}~\bibnamefont {Schierle}}, \bibinfo {author} {\bibfnamefont
  {E.}~\bibnamefont {Weschke}}, \bibinfo {author} {\bibfnamefont
  {R.}~\bibnamefont {Sutarto}}, \bibinfo {author} {\bibfnamefont
  {F.}~\bibnamefont {He}}, \bibinfo {author} {\bibfnamefont {A.}~\bibnamefont
  {Soumyanarayanan}}, \bibinfo {author} {\bibfnamefont {Y.}~\bibnamefont {He}},
  \bibinfo {author} {\bibfnamefont {M.}~\bibnamefont {Le~Tacon}}, \bibinfo
  {author} {\bibfnamefont {I.~S.}\ \bibnamefont {Elfimov}}, \bibinfo {author}
  {\bibfnamefont {J.~E.}\ \bibnamefont {Hoffman}}, \bibinfo {author}
  {\bibfnamefont {G.~A.}\ \bibnamefont {Sawatzky}}, \bibinfo {author}
  {\bibfnamefont {B.}~\bibnamefont {Keimer}}, \ and\ \bibinfo {author}
  {\bibfnamefont {A.}~\bibnamefont {Damascelli}},\ }\href
  {http://www.sciencemag.org/content/343/6169/390.abstract} {\bibfield
  {journal} {\bibinfo  {journal} {Science}\ }\textbf {\bibinfo {volume}
  {343}},\ \bibinfo {pages} {390} (\bibinfo {year} {2014})}\BibitemShut
  {NoStop}%
\bibitem [{\citenamefont {da~Silva~Neto}\ \emph {et~al.}(2014)\citenamefont
  {da~Silva~Neto}, \citenamefont {Aynajian}, \citenamefont {Frano},
  \citenamefont {Comin}, \citenamefont {Schierle}, \citenamefont {Weschke},
  \citenamefont {Gyenis}, \citenamefont {Wen}, \citenamefont {Schneeloch},
  \citenamefont {Xu}, \citenamefont {Ono}, \citenamefont {Gu}, \citenamefont
  {Le~Tacon},\ and\ \citenamefont {Yazdani}}]{daSilvaNeto2014}%
  \BibitemOpen
  \bibfield  {author} {\bibinfo {author} {\bibfnamefont {E.~H.}\ \bibnamefont
  {da~Silva~Neto}}, \bibinfo {author} {\bibfnamefont {P.}~\bibnamefont
  {Aynajian}}, \bibinfo {author} {\bibfnamefont {A.}~\bibnamefont {Frano}},
  \bibinfo {author} {\bibfnamefont {R.}~\bibnamefont {Comin}}, \bibinfo
  {author} {\bibfnamefont {E.}~\bibnamefont {Schierle}}, \bibinfo {author}
  {\bibfnamefont {E.}~\bibnamefont {Weschke}}, \bibinfo {author} {\bibfnamefont
  {A.}~\bibnamefont {Gyenis}}, \bibinfo {author} {\bibfnamefont
  {J.}~\bibnamefont {Wen}}, \bibinfo {author} {\bibfnamefont {J.}~\bibnamefont
  {Schneeloch}}, \bibinfo {author} {\bibfnamefont {Z.}~\bibnamefont {Xu}},
  \bibinfo {author} {\bibfnamefont {S.}~\bibnamefont {Ono}}, \bibinfo {author}
  {\bibfnamefont {G.}~\bibnamefont {Gu}}, \bibinfo {author} {\bibfnamefont
  {M.}~\bibnamefont {Le~Tacon}}, \ and\ \bibinfo {author} {\bibfnamefont
  {A.}~\bibnamefont {Yazdani}},\ }\href {\doibase 10.1126/science.1243479}
  {\bibfield  {journal} {\bibinfo  {journal} {Science}\ }\textbf {\bibinfo
  {volume} {343}},\ \bibinfo {pages} {393} (\bibinfo {year}
  {2014})}\BibitemShut {NoStop}%
\bibitem [{\citenamefont {Tranquada}\ \emph {et~al.}(1995)\citenamefont
  {Tranquada}, \citenamefont {Sternlieb}, \citenamefont {Axe}, \citenamefont
  {Nakamura},\ and\ \citenamefont {Uchida}}]{Tranquada1995}%
  \BibitemOpen
  \bibfield  {author} {\bibinfo {author} {\bibfnamefont {J.~M.}\ \bibnamefont
  {Tranquada}}, \bibinfo {author} {\bibfnamefont {B.~J.}\ \bibnamefont
  {Sternlieb}}, \bibinfo {author} {\bibfnamefont {J.~D.}\ \bibnamefont {Axe}},
  \bibinfo {author} {\bibfnamefont {Y.}~\bibnamefont {Nakamura}}, \ and\
  \bibinfo {author} {\bibfnamefont {S.}~\bibnamefont {Uchida}},\ }\href
  {http://dx.doi.org/10.1038/375561a0} {\bibfield  {journal} {\bibinfo
  {journal} {Nature}\ }\textbf {\bibinfo {volume} {375}},\ \bibinfo {pages}
  {561} (\bibinfo {year} {1995})}\BibitemShut {NoStop}%
\bibitem [{\citenamefont {H\"ucker}\ \emph {et~al.}(2014)\citenamefont
  {H\"ucker}, \citenamefont {Christensen}, \citenamefont {Holmes},
  \citenamefont {Blackburn}, \citenamefont {Forgan}, \citenamefont {Liang},
  \citenamefont {Bonn}, \citenamefont {Hardy}, \citenamefont {Gutowski},
  \citenamefont {Zimmermann}, \citenamefont {Hayden},\ and\ \citenamefont
  {Chang}}]{Hucker2014}%
  \BibitemOpen
  \bibfield  {author} {\bibinfo {author} {\bibfnamefont {M.}~\bibnamefont
  {H\"ucker}}, \bibinfo {author} {\bibfnamefont {N.~B.}\ \bibnamefont
  {Christensen}}, \bibinfo {author} {\bibfnamefont {A.~T.}\ \bibnamefont
  {Holmes}}, \bibinfo {author} {\bibfnamefont {E.}~\bibnamefont {Blackburn}},
  \bibinfo {author} {\bibfnamefont {E.~M.}\ \bibnamefont {Forgan}}, \bibinfo
  {author} {\bibfnamefont {R.}~\bibnamefont {Liang}}, \bibinfo {author}
  {\bibfnamefont {D.~A.}\ \bibnamefont {Bonn}}, \bibinfo {author}
  {\bibfnamefont {W.~N.}\ \bibnamefont {Hardy}}, \bibinfo {author}
  {\bibfnamefont {O.}~\bibnamefont {Gutowski}}, \bibinfo {author}
  {\bibfnamefont {M.~v.}\ \bibnamefont {Zimmermann}}, \bibinfo {author}
  {\bibfnamefont {S.~M.}\ \bibnamefont {Hayden}}, \ and\ \bibinfo {author}
  {\bibfnamefont {J.}~\bibnamefont {Chang}},\ }\href
  {http://link.aps.org/doi/10.1103/PhysRevB.90.054514} {\bibfield  {journal}
  {\bibinfo  {journal} {Phys. Rev. B}\ }\textbf {\bibinfo {volume} {90}},\
  \bibinfo {pages} {054514} (\bibinfo {year} {2014})}\BibitemShut {NoStop}%
\bibitem [{\citenamefont {Blanco-Canosa}\ \emph {et~al.}(2014)\citenamefont
  {Blanco-Canosa}, \citenamefont {Frano}, \citenamefont {Schierle},
  \citenamefont {Porras}, \citenamefont {Loew}, \citenamefont {Minola},
  \citenamefont {Bluschke}, \citenamefont {Weschke}, \citenamefont {Keimer},\
  and\ \citenamefont {Le~Tacon}}]{Blanco-Canosa2014}%
  \BibitemOpen
  \bibfield  {author} {\bibinfo {author} {\bibfnamefont {S.}~\bibnamefont
  {Blanco-Canosa}}, \bibinfo {author} {\bibfnamefont {A.}~\bibnamefont
  {Frano}}, \bibinfo {author} {\bibfnamefont {E.}~\bibnamefont {Schierle}},
  \bibinfo {author} {\bibfnamefont {J.}~\bibnamefont {Porras}}, \bibinfo
  {author} {\bibfnamefont {T.}~\bibnamefont {Loew}}, \bibinfo {author}
  {\bibfnamefont {M.}~\bibnamefont {Minola}}, \bibinfo {author} {\bibfnamefont
  {M.}~\bibnamefont {Bluschke}}, \bibinfo {author} {\bibfnamefont
  {E.}~\bibnamefont {Weschke}}, \bibinfo {author} {\bibfnamefont
  {B.}~\bibnamefont {Keimer}}, \ and\ \bibinfo {author} {\bibfnamefont
  {M.}~\bibnamefont {Le~Tacon}},\ }\href {\doibase 10.1103/PhysRevB.90.054513}
  {\bibfield  {journal} {\bibinfo  {journal} {Phys. Rev. B}\ }\textbf {\bibinfo
  {volume} {90}},\ \bibinfo {pages} {054513} (\bibinfo {year}
  {2014})}\BibitemShut {NoStop}%
\bibitem [{\citenamefont {Wu}\ \emph {et~al.}(2013)\citenamefont {Wu},
  \citenamefont {Mayaffre}, \citenamefont {Kr\"amer}, \citenamefont
  {Horvati\`c}, \citenamefont {Berthier}, \citenamefont {Kuhns}, \citenamefont
  {Reyes}, \citenamefont {Liang}, \citenamefont {Hardy}, \citenamefont {Bonn},\
  and\ \citenamefont {Julien}}]{Wu2013}%
  \BibitemOpen
  \bibfield  {author} {\bibinfo {author} {\bibfnamefont {T.}~\bibnamefont
  {Wu}}, \bibinfo {author} {\bibfnamefont {H.}~\bibnamefont {Mayaffre}},
  \bibinfo {author} {\bibfnamefont {S.}~\bibnamefont {Kr\"amer}}, \bibinfo
  {author} {\bibfnamefont {M.}~\bibnamefont {Horvati\`c}}, \bibinfo {author}
  {\bibfnamefont {C.}~\bibnamefont {Berthier}}, \bibinfo {author}
  {\bibfnamefont {P.~L.}\ \bibnamefont {Kuhns}}, \bibinfo {author}
  {\bibfnamefont {A.~P.}\ \bibnamefont {Reyes}}, \bibinfo {author}
  {\bibfnamefont {R.}~\bibnamefont {Liang}}, \bibinfo {author} {\bibfnamefont
  {W.~N.}\ \bibnamefont {Hardy}}, \bibinfo {author} {\bibfnamefont {D.~A.}\
  \bibnamefont {Bonn}}, \ and\ \bibinfo {author} {\bibfnamefont {M.-H.}\
  \bibnamefont {Julien}},\ }\href {http://dx.doi.org/10.1038/ncomms3113}
  {\bibfield  {journal} {\bibinfo  {journal} {Nat. Commun.}\ }\textbf {\bibinfo
  {volume} {4}},\ \bibinfo {pages} {2113} (\bibinfo {year} {2013})}\BibitemShut
  {NoStop}%
\bibitem [{\citenamefont {Doiron-Leyraud}\ \emph {et~al.}(2007)\citenamefont
  {Doiron-Leyraud}, \citenamefont {Proust}, \citenamefont {LeBoeuf},
  \citenamefont {Levallois}, \citenamefont {Bonnemaison}, \citenamefont
  {Liang}, \citenamefont {Bonn}, \citenamefont {Hardy},\ and\ \citenamefont
  {Taillefer}}]{Doiron-Leyraud2007}%
  \BibitemOpen
  \bibfield  {author} {\bibinfo {author} {\bibfnamefont {N.}~\bibnamefont
  {Doiron-Leyraud}}, \bibinfo {author} {\bibfnamefont {C.}~\bibnamefont
  {Proust}}, \bibinfo {author} {\bibfnamefont {D.}~\bibnamefont {LeBoeuf}},
  \bibinfo {author} {\bibfnamefont {J.}~\bibnamefont {Levallois}}, \bibinfo
  {author} {\bibfnamefont {J.-B.}\ \bibnamefont {Bonnemaison}}, \bibinfo
  {author} {\bibfnamefont {R.}~\bibnamefont {Liang}}, \bibinfo {author}
  {\bibfnamefont {D.~A.}\ \bibnamefont {Bonn}}, \bibinfo {author}
  {\bibfnamefont {W.~N.}\ \bibnamefont {Hardy}}, \ and\ \bibinfo {author}
  {\bibfnamefont {L.}~\bibnamefont {Taillefer}},\ }\href
  {http://dx.doi.org/10.1038/nature05872} {\bibfield  {journal} {\bibinfo
  {journal} {Nature}\ }\textbf {\bibinfo {volume} {447}},\ \bibinfo {pages}
  {565} (\bibinfo {year} {2007})}\BibitemShut {NoStop}%
\bibitem [{\citenamefont {Doiron-Leyraud}\ \emph {et~al.}(2015)\citenamefont
  {Doiron-Leyraud}, \citenamefont {Badoux}, \citenamefont {Ren{\'e}~de Cotret},
  \citenamefont {Lepault}, \citenamefont {LeBoeuf}, \citenamefont
  {Lalibert{\'e}}, \citenamefont {Hassinger}, \citenamefont {Ramshaw},
  \citenamefont {Bonn}, \citenamefont {Hardy}, \citenamefont {Liang},
  \citenamefont {Park}, \citenamefont {Vignolles}, \citenamefont {Vignolle},
  \citenamefont {Taillefer},\ and\ \citenamefont
  {Proust}}]{Doiron-Leyraud2015}%
  \BibitemOpen
  \bibfield  {author} {\bibinfo {author} {\bibfnamefont {N.}~\bibnamefont
  {Doiron-Leyraud}}, \bibinfo {author} {\bibfnamefont {S.}~\bibnamefont
  {Badoux}}, \bibinfo {author} {\bibfnamefont {S.}~\bibnamefont {Ren{\'e}~de
  Cotret}}, \bibinfo {author} {\bibfnamefont {S.}~\bibnamefont {Lepault}},
  \bibinfo {author} {\bibfnamefont {D.}~\bibnamefont {LeBoeuf}}, \bibinfo
  {author} {\bibfnamefont {F.}~\bibnamefont {Lalibert{\'e}}}, \bibinfo {author}
  {\bibfnamefont {E.}~\bibnamefont {Hassinger}}, \bibinfo {author}
  {\bibfnamefont {B.~J.}\ \bibnamefont {Ramshaw}}, \bibinfo {author}
  {\bibfnamefont {D.~A.}\ \bibnamefont {Bonn}}, \bibinfo {author}
  {\bibfnamefont {W.~N.}\ \bibnamefont {Hardy}}, \bibinfo {author}
  {\bibfnamefont {R.}~\bibnamefont {Liang}}, \bibinfo {author} {\bibfnamefont
  {J.~H.~.}\ \bibnamefont {Park}}, \bibinfo {author} {\bibfnamefont
  {D.}~\bibnamefont {Vignolles}}, \bibinfo {author} {\bibfnamefont
  {B.}~\bibnamefont {Vignolle}}, \bibinfo {author} {\bibfnamefont
  {L.}~\bibnamefont {Taillefer}}, \ and\ \bibinfo {author} {\bibfnamefont
  {C.}~\bibnamefont {Proust}},\ }\href
  {https://www.nature.com/articles/ncomms7034} {\bibfield  {journal} {\bibinfo
  {journal} {Nat. Commun.}\ }\textbf {\bibinfo {volume} {6}},\ \bibinfo {pages}
  {6034} (\bibinfo {year} {2015})}\BibitemShut {NoStop}%
\bibitem [{\citenamefont {LeBoeuf}\ \emph {et~al.}(2007)\citenamefont
  {LeBoeuf}, \citenamefont {Doiron-Leyraud}, \citenamefont {Levallois},
  \citenamefont {Daou}, \citenamefont {Bonnemaison}, \citenamefont {Hussey},
  \citenamefont {Balicas}, \citenamefont {Ramshaw}, \citenamefont {Liang},
  \citenamefont {Bonn}, \citenamefont {Hardy}, \citenamefont {Adachi},
  \citenamefont {Proust},\ and\ \citenamefont {Taillefer}}]{LeBoeuf2007}%
  \BibitemOpen
  \bibfield  {author} {\bibinfo {author} {\bibfnamefont {D.}~\bibnamefont
  {LeBoeuf}}, \bibinfo {author} {\bibfnamefont {N.}~\bibnamefont
  {Doiron-Leyraud}}, \bibinfo {author} {\bibfnamefont {J.}~\bibnamefont
  {Levallois}}, \bibinfo {author} {\bibfnamefont {R.}~\bibnamefont {Daou}},
  \bibinfo {author} {\bibfnamefont {J.-B.}\ \bibnamefont {Bonnemaison}},
  \bibinfo {author} {\bibfnamefont {N.~E.}\ \bibnamefont {Hussey}}, \bibinfo
  {author} {\bibfnamefont {L.}~\bibnamefont {Balicas}}, \bibinfo {author}
  {\bibfnamefont {B.~J.}\ \bibnamefont {Ramshaw}}, \bibinfo {author}
  {\bibfnamefont {R.}~\bibnamefont {Liang}}, \bibinfo {author} {\bibfnamefont
  {D.~A.}\ \bibnamefont {Bonn}}, \bibinfo {author} {\bibfnamefont {W.~N.}\
  \bibnamefont {Hardy}}, \bibinfo {author} {\bibfnamefont {S.}~\bibnamefont
  {Adachi}}, \bibinfo {author} {\bibfnamefont {C.}~\bibnamefont {Proust}}, \
  and\ \bibinfo {author} {\bibfnamefont {L.}~\bibnamefont {Taillefer}},\ }\href
  {http://dx.doi.org/10.1038/nature06332} {\bibfield  {journal} {\bibinfo
  {journal} {Nature}\ }\textbf {\bibinfo {volume} {450}},\ \bibinfo {pages}
  {533} (\bibinfo {year} {2007})}\BibitemShut {NoStop}%
\bibitem [{\citenamefont {LeBoeuf}\ \emph {et~al.}(2011)\citenamefont
  {LeBoeuf}, \citenamefont {Doiron-Leyraud}, \citenamefont {Vignolle},
  \citenamefont {Sutherland}, \citenamefont {Ramshaw}, \citenamefont
  {Levallois}, \citenamefont {Daou}, \citenamefont {Lalibert\'e}, \citenamefont
  {Cyr-Choini\`ere}, \citenamefont {Chang}, \citenamefont {Jo}, \citenamefont
  {Balicas}, \citenamefont {Liang}, \citenamefont {Bonn}, \citenamefont
  {Hardy}, \citenamefont {Proust},\ and\ \citenamefont
  {Taillefer}}]{LeBoeuf2011}%
  \BibitemOpen
  \bibfield  {author} {\bibinfo {author} {\bibfnamefont {D.}~\bibnamefont
  {LeBoeuf}}, \bibinfo {author} {\bibfnamefont {N.}~\bibnamefont
  {Doiron-Leyraud}}, \bibinfo {author} {\bibfnamefont {B.}~\bibnamefont
  {Vignolle}}, \bibinfo {author} {\bibfnamefont {M.}~\bibnamefont
  {Sutherland}}, \bibinfo {author} {\bibfnamefont {B.~J.}\ \bibnamefont
  {Ramshaw}}, \bibinfo {author} {\bibfnamefont {J.}~\bibnamefont {Levallois}},
  \bibinfo {author} {\bibfnamefont {R.}~\bibnamefont {Daou}}, \bibinfo {author}
  {\bibfnamefont {F.}~\bibnamefont {Lalibert\'e}}, \bibinfo {author}
  {\bibfnamefont {O.}~\bibnamefont {Cyr-Choini\`ere}}, \bibinfo {author}
  {\bibfnamefont {J.}~\bibnamefont {Chang}}, \bibinfo {author} {\bibfnamefont
  {Y.~J.}\ \bibnamefont {Jo}}, \bibinfo {author} {\bibfnamefont
  {L.}~\bibnamefont {Balicas}}, \bibinfo {author} {\bibfnamefont
  {R.}~\bibnamefont {Liang}}, \bibinfo {author} {\bibfnamefont {D.~A.}\
  \bibnamefont {Bonn}}, \bibinfo {author} {\bibfnamefont {W.~N.}\ \bibnamefont
  {Hardy}}, \bibinfo {author} {\bibfnamefont {C.}~\bibnamefont {Proust}}, \
  and\ \bibinfo {author} {\bibfnamefont {L.}~\bibnamefont {Taillefer}},\ }\href
  {\doibase 10.1103/PhysRevB.83.054506} {\bibfield  {journal} {\bibinfo
  {journal} {Phys. Rev. B}\ }\textbf {\bibinfo {volume} {83}},\ \bibinfo
  {pages} {054506} (\bibinfo {year} {2011})}\BibitemShut {NoStop}%
\bibitem [{\citenamefont {Fink}\ \emph {et~al.}(2011)\citenamefont {Fink},
  \citenamefont {Soltwisch}, \citenamefont {Geck}, \citenamefont {Schierle},
  \citenamefont {Weschke},\ and\ \citenamefont {B\"uchner}}]{Fink2011}%
  \BibitemOpen
  \bibfield  {author} {\bibinfo {author} {\bibfnamefont {J.}~\bibnamefont
  {Fink}}, \bibinfo {author} {\bibfnamefont {V.}~\bibnamefont {Soltwisch}},
  \bibinfo {author} {\bibfnamefont {J.}~\bibnamefont {Geck}}, \bibinfo {author}
  {\bibfnamefont {E.}~\bibnamefont {Schierle}}, \bibinfo {author}
  {\bibfnamefont {E.}~\bibnamefont {Weschke}}, \ and\ \bibinfo {author}
  {\bibfnamefont {B.}~\bibnamefont {B\"uchner}},\ }\href
  {https://journals.aps.org/prb/abstract/10.1103/PhysRevB.83.092503} {\bibfield
   {journal} {\bibinfo  {journal} {Phys. Rev. B}\ }\textbf {\bibinfo {volume}
  {83}},\ \bibinfo {pages} {092503} (\bibinfo {year} {2011})}\BibitemShut
  {NoStop}%
\bibitem [{\citenamefont {Lalibert\'e}\ \emph {et~al.}(2011)\citenamefont
  {Lalibert\'e}, \citenamefont {Chang}, \citenamefont {Doiron-Leyraud},
  \citenamefont {Hassinger}, \citenamefont {Daou}, \citenamefont {Rondeau},
  \citenamefont {Ramshaw}, \citenamefont {Liang}, \citenamefont {Bonn},
  \citenamefont {Hardy}, \citenamefont {Pyon}, \citenamefont {Takayama},
  \citenamefont {Takagi}, \citenamefont {Sheikin}, \citenamefont {Malone},
  \citenamefont {Proust}, \citenamefont {Behnia},\ and\ \citenamefont
  {Taillefer}}]{Laliberte2011}%
  \BibitemOpen
  \bibfield  {author} {\bibinfo {author} {\bibfnamefont {F.}~\bibnamefont
  {Lalibert\'e}}, \bibinfo {author} {\bibfnamefont {J.}~\bibnamefont {Chang}},
  \bibinfo {author} {\bibfnamefont {N.}~\bibnamefont {Doiron-Leyraud}},
  \bibinfo {author} {\bibfnamefont {E.}~\bibnamefont {Hassinger}}, \bibinfo
  {author} {\bibfnamefont {R.}~\bibnamefont {Daou}}, \bibinfo {author}
  {\bibfnamefont {M.}~\bibnamefont {Rondeau}}, \bibinfo {author} {\bibfnamefont
  {B.}~\bibnamefont {Ramshaw}}, \bibinfo {author} {\bibfnamefont
  {R.}~\bibnamefont {Liang}}, \bibinfo {author} {\bibfnamefont
  {D.}~\bibnamefont {Bonn}}, \bibinfo {author} {\bibfnamefont {W.}~\bibnamefont
  {Hardy}}, \bibinfo {author} {\bibfnamefont {S.}~\bibnamefont {Pyon}},
  \bibinfo {author} {\bibfnamefont {T.}~\bibnamefont {Takayama}}, \bibinfo
  {author} {\bibfnamefont {H.}~\bibnamefont {Takagi}}, \bibinfo {author}
  {\bibfnamefont {I.}~\bibnamefont {Sheikin}}, \bibinfo {author} {\bibfnamefont
  {L.}~\bibnamefont {Malone}}, \bibinfo {author} {\bibfnamefont
  {C.}~\bibnamefont {Proust}}, \bibinfo {author} {\bibfnamefont
  {K.}~\bibnamefont {Behnia}}, \ and\ \bibinfo {author} {\bibfnamefont
  {L.}~\bibnamefont {Taillefer}},\ }\href
  {http://dx.doi.org/10.1038/ncomms1440} {\bibfield  {journal} {\bibinfo
  {journal} {Nat. Commun.}\ }\textbf {\bibinfo {volume} {2}},\ \bibinfo {pages}
  {432} (\bibinfo {year} {2011})}\BibitemShut {NoStop}%
\bibitem [{\citenamefont {Liang}\ \emph {et~al.}(2006)\citenamefont {Liang},
  \citenamefont {Bonn},\ and\ \citenamefont {Hardy}}]{Liang2006}%
  \BibitemOpen
  \bibfield  {author} {\bibinfo {author} {\bibfnamefont {R.}~\bibnamefont
  {Liang}}, \bibinfo {author} {\bibfnamefont {D.~A.}\ \bibnamefont {Bonn}}, \
  and\ \bibinfo {author} {\bibfnamefont {W.~N.}\ \bibnamefont {Hardy}},\ }\href
  {\doibase 10.1103/PhysRevB.73.180505} {\bibfield  {journal} {\bibinfo
  {journal} {Phys. Rev. B}\ }\textbf {\bibinfo {volume} {73}},\ \bibinfo
  {pages} {180505} (\bibinfo {year} {2006})}\BibitemShut {NoStop}%
\bibitem [{\citenamefont {Ando}\ \emph {et~al.}(2004)\citenamefont {Ando},
  \citenamefont {Komiya}, \citenamefont {Segawa}, \citenamefont {Ono},\ and\
  \citenamefont {Kurita}}]{Ando2004}%
  \BibitemOpen
  \bibfield  {author} {\bibinfo {author} {\bibfnamefont {Y.}~\bibnamefont
  {Ando}}, \bibinfo {author} {\bibfnamefont {S.}~\bibnamefont {Komiya}},
  \bibinfo {author} {\bibfnamefont {K.}~\bibnamefont {Segawa}}, \bibinfo
  {author} {\bibfnamefont {S.}~\bibnamefont {Ono}}, \ and\ \bibinfo {author}
  {\bibfnamefont {Y.}~\bibnamefont {Kurita}},\ }\href
  {http://link.aps.org/doi/10.1103/PhysRevLett.93.267001} {\bibfield  {journal}
  {\bibinfo  {journal} {Phys. Rev. Lett.}\ }\textbf {\bibinfo {volume} {93}},\
  \bibinfo {pages} {267001} (\bibinfo {year} {2004})}\BibitemShut {NoStop}%
\bibitem [{\citenamefont {Daou}\ \emph {et~al.}(2010)\citenamefont {Daou},
  \citenamefont {Chang}, \citenamefont {LeBoeuf}, \citenamefont
  {Cyr-Choini\`ere}, \citenamefont {Lalibert\'e}, \citenamefont
  {Doiron-Leyraud}, \citenamefont {Ramshaw}, \citenamefont {Liang},
  \citenamefont {Bonn}, \citenamefont {Hardy},\ and\ \citenamefont
  {Taillefer}}]{Daou2010}%
  \BibitemOpen
  \bibfield  {author} {\bibinfo {author} {\bibfnamefont {R.}~\bibnamefont
  {Daou}}, \bibinfo {author} {\bibfnamefont {J.}~\bibnamefont {Chang}},
  \bibinfo {author} {\bibfnamefont {D.}~\bibnamefont {LeBoeuf}}, \bibinfo
  {author} {\bibfnamefont {O.}~\bibnamefont {Cyr-Choini\`ere}}, \bibinfo
  {author} {\bibfnamefont {F.}~\bibnamefont {Lalibert\'e}}, \bibinfo {author}
  {\bibfnamefont {N.}~\bibnamefont {Doiron-Leyraud}}, \bibinfo {author}
  {\bibfnamefont {B.~J.}\ \bibnamefont {Ramshaw}}, \bibinfo {author}
  {\bibfnamefont {R.}~\bibnamefont {Liang}}, \bibinfo {author} {\bibfnamefont
  {D.~A.}\ \bibnamefont {Bonn}}, \bibinfo {author} {\bibfnamefont {W.~N.}\
  \bibnamefont {Hardy}}, \ and\ \bibinfo {author} {\bibfnamefont
  {L.}~\bibnamefont {Taillefer}},\ }\href
  {http://dx.doi.org/10.1038/nature08716} {\bibfield  {journal} {\bibinfo
  {journal} {Nature}\ }\textbf {\bibinfo {volume} {463}},\ \bibinfo {pages}
  {519} (\bibinfo {year} {2010})}\BibitemShut {NoStop}%
\bibitem [{\citenamefont {Nakayama}\ \emph {et~al.}(2014)\citenamefont
  {Nakayama}, \citenamefont {Onda}, \citenamefont {Yamada}, \citenamefont
  {Fujihisa}, \citenamefont {Sakata}, \citenamefont {Nakamoto}, \citenamefont
  {Shimizu}, \citenamefont {Nakano}, \citenamefont {Ohmura}, \citenamefont
  {Ishikawa},\ and\ \citenamefont {Yamada}}]{Nakayama2014}%
  \BibitemOpen
  \bibfield  {author} {\bibinfo {author} {\bibfnamefont {A.}~\bibnamefont
  {Nakayama}}, \bibinfo {author} {\bibfnamefont {Y.}~\bibnamefont {Onda}},
  \bibinfo {author} {\bibfnamefont {S.}~\bibnamefont {Yamada}}, \bibinfo
  {author} {\bibfnamefont {H.}~\bibnamefont {Fujihisa}}, \bibinfo {author}
  {\bibfnamefont {M.}~\bibnamefont {Sakata}}, \bibinfo {author} {\bibfnamefont
  {Y.}~\bibnamefont {Nakamoto}}, \bibinfo {author} {\bibfnamefont
  {K.}~\bibnamefont {Shimizu}}, \bibinfo {author} {\bibfnamefont
  {S.}~\bibnamefont {Nakano}}, \bibinfo {author} {\bibfnamefont
  {A.}~\bibnamefont {Ohmura}}, \bibinfo {author} {\bibfnamefont
  {F.}~\bibnamefont {Ishikawa}}, \ and\ \bibinfo {author} {\bibfnamefont
  {Y.}~\bibnamefont {Yamada}},\ }\href {\doibase 10.7566/JPSJ.83.093601}
  {\bibfield  {journal} {\bibinfo  {journal} {J. Phys. Soc. Jpn.}\ }\textbf
  {\bibinfo {volume} {83}},\ \bibinfo {pages} {093601} (\bibinfo {year}
  {2014})}\BibitemShut {NoStop}%
\bibitem [{\citenamefont {Grissonnanche}\ \emph {et~al.}(2014)\citenamefont
  {Grissonnanche}, \citenamefont {Cyr-Choini\`ere}, \citenamefont
  {Lalibert\'e}, \citenamefont {Ren\'e~de Cotret}, \citenamefont
  {Juneau-Fecteau}, \citenamefont {Dufour-Beaus\'ejour}, \citenamefont
  {Delage}, \citenamefont {LeBoeuf}, \citenamefont {Chang}, \citenamefont
  {Ramshaw}, \citenamefont {Bonn}, \citenamefont {Hardy}, \citenamefont
  {Liang}, \citenamefont {Adachi}, \citenamefont {Hussey}, \citenamefont
  {Vignolle}, \citenamefont {Proust}, \citenamefont {Sutherland}, \citenamefont
  {Kramer}, \citenamefont {Park}, \citenamefont {Graf}, \citenamefont
  {Doiron-Leyraud},\ and\ \citenamefont {Taillefer}}]{Grissonnanche2014}%
  \BibitemOpen
  \bibfield  {author} {\bibinfo {author} {\bibfnamefont {G.}~\bibnamefont
  {Grissonnanche}}, \bibinfo {author} {\bibfnamefont {O.}~\bibnamefont
  {Cyr-Choini\`ere}}, \bibinfo {author} {\bibfnamefont {F.}~\bibnamefont
  {Lalibert\'e}}, \bibinfo {author} {\bibfnamefont {S.}~\bibnamefont {Ren\'e~de
  Cotret}}, \bibinfo {author} {\bibfnamefont {A.}~\bibnamefont
  {Juneau-Fecteau}}, \bibinfo {author} {\bibfnamefont {S.}~\bibnamefont
  {Dufour-Beaus\'ejour}}, \bibinfo {author} {\bibfnamefont {M.-E.}\
  \bibnamefont {Delage}}, \bibinfo {author} {\bibfnamefont {D.}~\bibnamefont
  {LeBoeuf}}, \bibinfo {author} {\bibfnamefont {J.}~\bibnamefont {Chang}},
  \bibinfo {author} {\bibfnamefont {B.~J.}\ \bibnamefont {Ramshaw}}, \bibinfo
  {author} {\bibfnamefont {D.~A.}\ \bibnamefont {Bonn}}, \bibinfo {author}
  {\bibfnamefont {W.~N.}\ \bibnamefont {Hardy}}, \bibinfo {author}
  {\bibfnamefont {R.}~\bibnamefont {Liang}}, \bibinfo {author} {\bibfnamefont
  {S.}~\bibnamefont {Adachi}}, \bibinfo {author} {\bibfnamefont {N.~E.}\
  \bibnamefont {Hussey}}, \bibinfo {author} {\bibfnamefont {B.}~\bibnamefont
  {Vignolle}}, \bibinfo {author} {\bibfnamefont {C.}~\bibnamefont {Proust}},
  \bibinfo {author} {\bibfnamefont {M.}~\bibnamefont {Sutherland}}, \bibinfo
  {author} {\bibfnamefont {S.}~\bibnamefont {Kramer}}, \bibinfo {author}
  {\bibfnamefont {J.-H.}\ \bibnamefont {Park}}, \bibinfo {author}
  {\bibfnamefont {D.}~\bibnamefont {Graf}}, \bibinfo {author} {\bibfnamefont
  {N.}~\bibnamefont {Doiron-Leyraud}}, \ and\ \bibinfo {author} {\bibfnamefont
  {L.}~\bibnamefont {Taillefer}},\ }\href
  {http://dx.doi.org/10.1038/ncomms4280} {\bibfield  {journal} {\bibinfo
  {journal} {Nat. Commun.}\ }\textbf {\bibinfo {volume} {5}},\ \bibinfo {pages}
  {3280} (\bibinfo {year} {2014})}\BibitemShut {NoStop}%
\bibitem [{\citenamefont {Souliou}\ \emph {et~al.}(2018)\citenamefont
  {Souliou}, \citenamefont {Gretarsson}, \citenamefont {Garbarino},
  \citenamefont {Bosak}, \citenamefont {Porras}, \citenamefont {Loew},
  \citenamefont {Keimer},\ and\ \citenamefont {Le~Tacon}}]{Souliou2018}%
  \BibitemOpen
  \bibfield  {author} {\bibinfo {author} {\bibfnamefont {S.~M.}\ \bibnamefont
  {Souliou}}, \bibinfo {author} {\bibfnamefont {H.}~\bibnamefont {Gretarsson}},
  \bibinfo {author} {\bibfnamefont {G.}~\bibnamefont {Garbarino}}, \bibinfo
  {author} {\bibfnamefont {A.}~\bibnamefont {Bosak}}, \bibinfo {author}
  {\bibfnamefont {J.}~\bibnamefont {Porras}}, \bibinfo {author} {\bibfnamefont
  {T.}~\bibnamefont {Loew}}, \bibinfo {author} {\bibfnamefont {B.}~\bibnamefont
  {Keimer}}, \ and\ \bibinfo {author} {\bibfnamefont {M.}~\bibnamefont
  {Le~Tacon}},\ }\href {\doibase 10.1103/PhysRevB.97.020503} {\bibfield
  {journal} {\bibinfo  {journal} {Phys. Rev. B}\ }\textbf {\bibinfo {volume}
  {97}},\ \bibinfo {pages} {020503} (\bibinfo {year} {2018})}\BibitemShut
  {NoStop}%
\bibitem [{\citenamefont {Huang}\ \emph {et~al.}(2018)\citenamefont {Huang},
  \citenamefont {Jang}, \citenamefont {Fujita}, \citenamefont {Nishizaki},
  \citenamefont {Lin}, \citenamefont {Wang}, \citenamefont {Ying},
  \citenamefont {Smith}, \citenamefont {Kenney-Benson}, \citenamefont {Shen},
  \citenamefont {Mao}, \citenamefont {Kao}, \citenamefont {Liu},\ and\
  \citenamefont {Lee}}]{Huang2018}%
  \BibitemOpen
  \bibfield  {author} {\bibinfo {author} {\bibfnamefont {H.}~\bibnamefont
  {Huang}}, \bibinfo {author} {\bibfnamefont {H.}~\bibnamefont {Jang}},
  \bibinfo {author} {\bibfnamefont {M.}~\bibnamefont {Fujita}}, \bibinfo
  {author} {\bibfnamefont {T.}~\bibnamefont {Nishizaki}}, \bibinfo {author}
  {\bibfnamefont {Y.}~\bibnamefont {Lin}}, \bibinfo {author} {\bibfnamefont
  {J.}~\bibnamefont {Wang}}, \bibinfo {author} {\bibfnamefont {J.}~\bibnamefont
  {Ying}}, \bibinfo {author} {\bibfnamefont {J.~S.}\ \bibnamefont {Smith}},
  \bibinfo {author} {\bibfnamefont {C.}~\bibnamefont {Kenney-Benson}}, \bibinfo
  {author} {\bibfnamefont {G.}~\bibnamefont {Shen}}, \bibinfo {author}
  {\bibfnamefont {W.~L.}\ \bibnamefont {Mao}}, \bibinfo {author} {\bibfnamefont
  {C.-C.}\ \bibnamefont {Kao}}, \bibinfo {author} {\bibfnamefont {Y.-J.}\
  \bibnamefont {Liu}}, \ and\ \bibinfo {author} {\bibfnamefont {J.-S.}\
  \bibnamefont {Lee}},\ }\href {\doibase 10.1103/PhysRevB.97.174508} {\bibfield
   {journal} {\bibinfo  {journal} {Phys. Rev. B}\ }\textbf {\bibinfo {volume}
  {97}},\ \bibinfo {pages} {174508} (\bibinfo {year} {2018})}\BibitemShut
  {NoStop}%
\bibitem [{\citenamefont {Liang}\ \emph {et~al.}(2012)\citenamefont {Liang},
  \citenamefont {Bonn},\ and\ \citenamefont {Hardy}}]{Liang2012}%
  \BibitemOpen
  \bibfield  {author} {\bibinfo {author} {\bibfnamefont {R.}~\bibnamefont
  {Liang}}, \bibinfo {author} {\bibfnamefont {D.~A.}\ \bibnamefont {Bonn}}, \
  and\ \bibinfo {author} {\bibfnamefont {W.~N.}\ \bibnamefont {Hardy}},\ }\href
  {\doibase 10.1080/14786435.2012.669065} {\bibfield  {journal} {\bibinfo
  {journal} {Phil. Mag.}\ }\textbf {\bibinfo {volume} {92}},\ \bibinfo {pages}
  {2563} (\bibinfo {year} {2012})}\BibitemShut {NoStop}%
\bibitem [{\citenamefont {Sadewasser}\ \emph {et~al.}(1997)\citenamefont
  {Sadewasser}, \citenamefont {Wang}, \citenamefont {Schilling}, \citenamefont
  {Zheng}, \citenamefont {Paulikas},\ and\ \citenamefont
  {Veal}}]{Sadewasser1997}%
  \BibitemOpen
  \bibfield  {author} {\bibinfo {author} {\bibfnamefont {S.}~\bibnamefont
  {Sadewasser}}, \bibinfo {author} {\bibfnamefont {Y.}~\bibnamefont {Wang}},
  \bibinfo {author} {\bibfnamefont {J.~S.}\ \bibnamefont {Schilling}}, \bibinfo
  {author} {\bibfnamefont {H.}~\bibnamefont {Zheng}}, \bibinfo {author}
  {\bibfnamefont {A.~P.}\ \bibnamefont {Paulikas}}, \ and\ \bibinfo {author}
  {\bibfnamefont {B.~W.}\ \bibnamefont {Veal}},\ }\href
  {https://journals.aps.org/prb/abstract/10.1103/PhysRevB.56.14168} {\bibfield
  {journal} {\bibinfo  {journal} {Phys. Rev. B}\ }\textbf {\bibinfo {volume}
  {56}},\ \bibinfo {pages} {14168} (\bibinfo {year} {1997})}\BibitemShut
  {NoStop}%
\bibitem [{\citenamefont {Sadewasser}\ \emph {et~al.}(2000)\citenamefont
  {Sadewasser}, \citenamefont {Schilling}, \citenamefont {Paulikas},\ and\
  \citenamefont {Veal}}]{Sadewasser2000}%
  \BibitemOpen
  \bibfield  {author} {\bibinfo {author} {\bibfnamefont {S.}~\bibnamefont
  {Sadewasser}}, \bibinfo {author} {\bibfnamefont {J.~S.}\ \bibnamefont
  {Schilling}}, \bibinfo {author} {\bibfnamefont {A.~P.}\ \bibnamefont
  {Paulikas}}, \ and\ \bibinfo {author} {\bibfnamefont {B.~W.}\ \bibnamefont
  {Veal}},\ }\href
  {https://journals.aps.org/prb/abstract/10.1103/PhysRevB.61.741} {\bibfield
  {journal} {\bibinfo  {journal} {Phys. Rev. B}\ }\textbf {\bibinfo {volume}
  {61}},\ \bibinfo {pages} {741} (\bibinfo {year} {2000})}\BibitemShut
  {NoStop}%
\bibitem [{\citenamefont {Cyr-Choini\`ere}\ \emph {et~al.}(2018)\citenamefont
  {Cyr-Choini\`ere}, \citenamefont {Daou}, \citenamefont {Lalibert\'e},
  \citenamefont {Collignon}, \citenamefont {Badoux}, \citenamefont {LeBoeuf},
  \citenamefont {Chang}, \citenamefont {Ramshaw}, \citenamefont {Bonn},
  \citenamefont {Hardy}, \citenamefont {Liang}, \citenamefont {Yan},
  \citenamefont {Cheng}, \citenamefont {Zhou}, \citenamefont {Goodenough},
  \citenamefont {Pyon}, \citenamefont {Takayama}, \citenamefont {Takagi},
  \citenamefont {Doiron-Leyraud},\ and\ \citenamefont
  {Taillefer}}]{Cyr-Choiniere2018}%
  \BibitemOpen
  \bibfield  {author} {\bibinfo {author} {\bibfnamefont {O.}~\bibnamefont
  {Cyr-Choini\`ere}}, \bibinfo {author} {\bibfnamefont {R.}~\bibnamefont
  {Daou}}, \bibinfo {author} {\bibfnamefont {F.}~\bibnamefont {Lalibert\'e}},
  \bibinfo {author} {\bibfnamefont {C.}~\bibnamefont {Collignon}}, \bibinfo
  {author} {\bibfnamefont {S.}~\bibnamefont {Badoux}}, \bibinfo {author}
  {\bibfnamefont {D.}~\bibnamefont {LeBoeuf}}, \bibinfo {author} {\bibfnamefont
  {J.}~\bibnamefont {Chang}}, \bibinfo {author} {\bibfnamefont {B.~J.}\
  \bibnamefont {Ramshaw}}, \bibinfo {author} {\bibfnamefont {D.~A.}\
  \bibnamefont {Bonn}}, \bibinfo {author} {\bibfnamefont {W.~N.}\ \bibnamefont
  {Hardy}}, \bibinfo {author} {\bibfnamefont {R.}~\bibnamefont {Liang}},
  \bibinfo {author} {\bibfnamefont {J.-Q.}\ \bibnamefont {Yan}}, \bibinfo
  {author} {\bibfnamefont {J.-G.}\ \bibnamefont {Cheng}}, \bibinfo {author}
  {\bibfnamefont {J.-S.}\ \bibnamefont {Zhou}}, \bibinfo {author}
  {\bibfnamefont {J.~B.}\ \bibnamefont {Goodenough}}, \bibinfo {author}
  {\bibfnamefont {S.}~\bibnamefont {Pyon}}, \bibinfo {author} {\bibfnamefont
  {T.}~\bibnamefont {Takayama}}, \bibinfo {author} {\bibfnamefont
  {H.}~\bibnamefont {Takagi}}, \bibinfo {author} {\bibfnamefont
  {N.}~\bibnamefont {Doiron-Leyraud}}, \ and\ \bibinfo {author} {\bibfnamefont
  {L.}~\bibnamefont {Taillefer}},\ }\href {\doibase 10.1103/PhysRevB.97.064502}
  {\bibfield  {journal} {\bibinfo  {journal} {Phys. Rev. B}\ }\textbf {\bibinfo
  {volume} {97}},\ \bibinfo {pages} {064502} (\bibinfo {year}
  {2018})}\BibitemShut {NoStop}%
\bibitem [{\citenamefont {Collignon}\ \emph {et~al.}(2017)\citenamefont
  {Collignon}, \citenamefont {Badoux}, \citenamefont {Afshar}, \citenamefont
  {Michon}, \citenamefont {Lalibert{\'e}}, \citenamefont {Cyr-Choini{\`e}re},
  \citenamefont {Zhou}, \citenamefont {Licciardello}, \citenamefont {Wiedmann},
  \citenamefont {Doiron-Leyraud},\ and\ \citenamefont
  {Taillefer}}]{Collignon2017}%
  \BibitemOpen
  \bibfield  {author} {\bibinfo {author} {\bibfnamefont {C.}~\bibnamefont
  {Collignon}}, \bibinfo {author} {\bibfnamefont {S.}~\bibnamefont {Badoux}},
  \bibinfo {author} {\bibfnamefont {S.~A.~A.}\ \bibnamefont {Afshar}}, \bibinfo
  {author} {\bibfnamefont {B.}~\bibnamefont {Michon}}, \bibinfo {author}
  {\bibfnamefont {F.}~\bibnamefont {Lalibert{\'e}}}, \bibinfo {author}
  {\bibfnamefont {O.}~\bibnamefont {Cyr-Choini{\`e}re}}, \bibinfo {author}
  {\bibfnamefont {J.-S.}\ \bibnamefont {Zhou}}, \bibinfo {author}
  {\bibfnamefont {S.}~\bibnamefont {Licciardello}}, \bibinfo {author}
  {\bibfnamefont {S.}~\bibnamefont {Wiedmann}}, \bibinfo {author}
  {\bibfnamefont {N.}~\bibnamefont {Doiron-Leyraud}}, \ and\ \bibinfo {author}
  {\bibfnamefont {L.}~\bibnamefont {Taillefer}},\ }\href {\doibase
  10.1103/PhysRevB.95.224517} {\bibfield  {journal} {\bibinfo  {journal} {Phys.
  Rev. B}\ }\textbf {\bibinfo {volume} {95}},\ \bibinfo {pages} {224517}
  (\bibinfo {year} {2017})}\BibitemShut {NoStop}%
\bibitem [{\citenamefont {Matt}\ \emph {et~al.}(2015)\citenamefont {Matt},
  \citenamefont {Fatuzzo}, \citenamefont {Sassa}, \citenamefont {M\aa{}nsson},
  \citenamefont {Fatale}, \citenamefont {Bitetta}, \citenamefont {Shi},
  \citenamefont {Pailh\`es}, \citenamefont {Berntsen}, \citenamefont
  {Kurosawa}, \citenamefont {Oda}, \citenamefont {Momono}, \citenamefont
  {Lipscombe}, \citenamefont {Hayden}, \citenamefont {Yan}, \citenamefont
  {Zhou}, \citenamefont {Goodenough}, \citenamefont {Pyon}, \citenamefont
  {Takayama}, \citenamefont {Takagi}, \citenamefont {Patthey}, \citenamefont
  {Bendounan}, \citenamefont {Razzoli}, \citenamefont {Shi}, \citenamefont
  {Plumb}, \citenamefont {Radovic}, \citenamefont {Grioni}, \citenamefont
  {Mesot}, \citenamefont {Tjernberg},\ and\ \citenamefont {Chang}}]{Matt2015}%
  \BibitemOpen
  \bibfield  {author} {\bibinfo {author} {\bibfnamefont {C.~E.}\ \bibnamefont
  {Matt}}, \bibinfo {author} {\bibfnamefont {C.~G.}\ \bibnamefont {Fatuzzo}},
  \bibinfo {author} {\bibfnamefont {Y.}~\bibnamefont {Sassa}}, \bibinfo
  {author} {\bibfnamefont {M.}~\bibnamefont {M\aa{}nsson}}, \bibinfo {author}
  {\bibfnamefont {S.}~\bibnamefont {Fatale}}, \bibinfo {author} {\bibfnamefont
  {V.}~\bibnamefont {Bitetta}}, \bibinfo {author} {\bibfnamefont
  {X.}~\bibnamefont {Shi}}, \bibinfo {author} {\bibfnamefont {S.}~\bibnamefont
  {Pailh\`es}}, \bibinfo {author} {\bibfnamefont {M.~H.}\ \bibnamefont
  {Berntsen}}, \bibinfo {author} {\bibfnamefont {T.}~\bibnamefont {Kurosawa}},
  \bibinfo {author} {\bibfnamefont {M.}~\bibnamefont {Oda}}, \bibinfo {author}
  {\bibfnamefont {N.}~\bibnamefont {Momono}}, \bibinfo {author} {\bibfnamefont
  {O.~J.}\ \bibnamefont {Lipscombe}}, \bibinfo {author} {\bibfnamefont {S.~M.}\
  \bibnamefont {Hayden}}, \bibinfo {author} {\bibfnamefont {J.-Q.}\
  \bibnamefont {Yan}}, \bibinfo {author} {\bibfnamefont {J.-S.}\ \bibnamefont
  {Zhou}}, \bibinfo {author} {\bibfnamefont {J.~B.}\ \bibnamefont
  {Goodenough}}, \bibinfo {author} {\bibfnamefont {S.}~\bibnamefont {Pyon}},
  \bibinfo {author} {\bibfnamefont {T.}~\bibnamefont {Takayama}}, \bibinfo
  {author} {\bibfnamefont {H.}~\bibnamefont {Takagi}}, \bibinfo {author}
  {\bibfnamefont {L.}~\bibnamefont {Patthey}}, \bibinfo {author} {\bibfnamefont
  {A.}~\bibnamefont {Bendounan}}, \bibinfo {author} {\bibfnamefont
  {E.}~\bibnamefont {Razzoli}}, \bibinfo {author} {\bibfnamefont
  {M.}~\bibnamefont {Shi}}, \bibinfo {author} {\bibfnamefont {N.~C.}\
  \bibnamefont {Plumb}}, \bibinfo {author} {\bibfnamefont {M.}~\bibnamefont
  {Radovic}}, \bibinfo {author} {\bibfnamefont {M.}~\bibnamefont {Grioni}},
  \bibinfo {author} {\bibfnamefont {J.}~\bibnamefont {Mesot}}, \bibinfo
  {author} {\bibfnamefont {O.}~\bibnamefont {Tjernberg}}, \ and\ \bibinfo
  {author} {\bibfnamefont {J.}~\bibnamefont {Chang}},\ }\href {\doibase
  10.1103/PhysRevB.92.134524} {\bibfield  {journal} {\bibinfo  {journal} {Phys.
  Rev. B}\ }\textbf {\bibinfo {volume} {92}},\ \bibinfo {pages} {134524}
  (\bibinfo {year} {2015})}\BibitemShut {NoStop}%
\bibitem [{\citenamefont {Doiron-Leyraud}\ \emph {et~al.}(2017)\citenamefont
  {Doiron-Leyraud}, \citenamefont {Cyr-Choini{\`e}re}, \citenamefont {Badoux},
  \citenamefont {Ataei}, \citenamefont {Collignon}, \citenamefont {Gourgout},
  \citenamefont {Dufour-Beaus{\'e}jour}, \citenamefont {Tafti}, \citenamefont
  {Lalibert{\'e}}, \citenamefont {Boulanger}, \citenamefont {Matusiak},
  \citenamefont {Graf}, \citenamefont {Kim}, \citenamefont {Zhou},
  \citenamefont {Momono}, \citenamefont {Kurosawa}, \citenamefont {Takagi},\
  and\ \citenamefont {Taillefer}}]{Doiron-Leyraud2017}%
  \BibitemOpen
  \bibfield  {author} {\bibinfo {author} {\bibfnamefont {N.}~\bibnamefont
  {Doiron-Leyraud}}, \bibinfo {author} {\bibfnamefont {O.}~\bibnamefont
  {Cyr-Choini{\`e}re}}, \bibinfo {author} {\bibfnamefont {S.}~\bibnamefont
  {Badoux}}, \bibinfo {author} {\bibfnamefont {A.}~\bibnamefont {Ataei}},
  \bibinfo {author} {\bibfnamefont {C.}~\bibnamefont {Collignon}}, \bibinfo
  {author} {\bibfnamefont {A.}~\bibnamefont {Gourgout}}, \bibinfo {author}
  {\bibfnamefont {S.}~\bibnamefont {Dufour-Beaus{\'e}jour}}, \bibinfo {author}
  {\bibfnamefont {F.~F.}\ \bibnamefont {Tafti}}, \bibinfo {author}
  {\bibfnamefont {F.}~\bibnamefont {Lalibert{\'e}}}, \bibinfo {author}
  {\bibfnamefont {M.-E.}\ \bibnamefont {Boulanger}}, \bibinfo {author}
  {\bibfnamefont {M.}~\bibnamefont {Matusiak}}, \bibinfo {author}
  {\bibfnamefont {D.}~\bibnamefont {Graf}}, \bibinfo {author} {\bibfnamefont
  {M.}~\bibnamefont {Kim}}, \bibinfo {author} {\bibfnamefont {J.-S.}\
  \bibnamefont {Zhou}}, \bibinfo {author} {\bibfnamefont {N.}~\bibnamefont
  {Momono}}, \bibinfo {author} {\bibfnamefont {T.}~\bibnamefont {Kurosawa}},
  \bibinfo {author} {\bibfnamefont {H.}~\bibnamefont {Takagi}}, \ and\ \bibinfo
  {author} {\bibfnamefont {L.}~\bibnamefont {Taillefer}},\ }\href {\doibase
  10.1038/s41467-017-02122-x} {\bibfield  {journal} {\bibinfo  {journal} {Nat.
  Commun.}\ }\textbf {\bibinfo {volume} {8}},\ \bibinfo {pages} {2044}
  (\bibinfo {year} {2017})}\BibitemShut {NoStop}%
\bibitem [{\citenamefont {Cvitani\'{c}}\ \emph {et~al.}(2014)\citenamefont
  {Cvitani\'{c}}, \citenamefont {Pelc}, \citenamefont {Po\v{z}ek},
  \citenamefont {Amit},\ and\ \citenamefont {Keren}}]{Cvitanic2014}%
  \BibitemOpen
  \bibfield  {author} {\bibinfo {author} {\bibfnamefont {T.}~\bibnamefont
  {Cvitani\'{c}}}, \bibinfo {author} {\bibfnamefont {D.}~\bibnamefont {Pelc}},
  \bibinfo {author} {\bibfnamefont {M.}~\bibnamefont {Po\v{z}ek}}, \bibinfo
  {author} {\bibfnamefont {E.}~\bibnamefont {Amit}}, \ and\ \bibinfo {author}
  {\bibfnamefont {A.}~\bibnamefont {Keren}},\ }\href
  {http://link.aps.org/doi/10.1103/PhysRevB.90.054508} {\bibfield  {journal}
  {\bibinfo  {journal} {Phys. Rev. B}\ }\textbf {\bibinfo {volume} {90}},\
  \bibinfo {pages} {054508} (\bibinfo {year} {2014})}\BibitemShut {NoStop}%
\bibitem [{\citenamefont {Yelland}\ \emph {et~al.}(2008)\citenamefont
  {Yelland}, \citenamefont {Singleton}, \citenamefont {Mielke}, \citenamefont
  {Harrison}, \citenamefont {Balakirev}, \citenamefont {Dabrowski},\ and\
  \citenamefont {Cooper}}]{Yelland2008}%
  \BibitemOpen
  \bibfield  {author} {\bibinfo {author} {\bibfnamefont {E.~A.}\ \bibnamefont
  {Yelland}}, \bibinfo {author} {\bibfnamefont {J.}~\bibnamefont {Singleton}},
  \bibinfo {author} {\bibfnamefont {C.~H.}\ \bibnamefont {Mielke}}, \bibinfo
  {author} {\bibfnamefont {N.}~\bibnamefont {Harrison}}, \bibinfo {author}
  {\bibfnamefont {F.~F.}\ \bibnamefont {Balakirev}}, \bibinfo {author}
  {\bibfnamefont {B.}~\bibnamefont {Dabrowski}}, \ and\ \bibinfo {author}
  {\bibfnamefont {J.~R.}\ \bibnamefont {Cooper}},\ }\href {\doibase
  10.1103/PhysRevLett.100.047003} {\bibfield  {journal} {\bibinfo  {journal}
  {Phys. Rev. Lett.}\ }\textbf {\bibinfo {volume} {100}},\ \bibinfo {pages}
  {047003} (\bibinfo {year} {2008})}\BibitemShut {NoStop}%
\bibitem [{\citenamefont {Bangura}\ \emph {et~al.}(2008)\citenamefont
  {Bangura}, \citenamefont {Fletcher}, \citenamefont {Carrington},
  \citenamefont {Levallois}, \citenamefont {Nardone}, \citenamefont {Vignolle},
  \citenamefont {Heard}, \citenamefont {Doiron-Leyraud}, \citenamefont
  {LeBoeuf}, \citenamefont {Taillefer}, \citenamefont {Adachi}, \citenamefont
  {Proust},\ and\ \citenamefont {Hussey}}]{Bangura2008}%
  \BibitemOpen
  \bibfield  {author} {\bibinfo {author} {\bibfnamefont {A.~F.}\ \bibnamefont
  {Bangura}}, \bibinfo {author} {\bibfnamefont {J.~D.}\ \bibnamefont
  {Fletcher}}, \bibinfo {author} {\bibfnamefont {A.}~\bibnamefont
  {Carrington}}, \bibinfo {author} {\bibfnamefont {J.}~\bibnamefont
  {Levallois}}, \bibinfo {author} {\bibfnamefont {M.}~\bibnamefont {Nardone}},
  \bibinfo {author} {\bibfnamefont {B.}~\bibnamefont {Vignolle}}, \bibinfo
  {author} {\bibfnamefont {P.~J.}\ \bibnamefont {Heard}}, \bibinfo {author}
  {\bibfnamefont {N.}~\bibnamefont {Doiron-Leyraud}}, \bibinfo {author}
  {\bibfnamefont {D.}~\bibnamefont {LeBoeuf}}, \bibinfo {author} {\bibfnamefont
  {L.}~\bibnamefont {Taillefer}}, \bibinfo {author} {\bibfnamefont
  {S.}~\bibnamefont {Adachi}}, \bibinfo {author} {\bibfnamefont
  {C.}~\bibnamefont {Proust}}, \ and\ \bibinfo {author} {\bibfnamefont {N.~E.}\
  \bibnamefont {Hussey}},\ }\href {\doibase 10.1103/PhysRevLett.100.047004}
  {\bibfield  {journal} {\bibinfo  {journal} {Phys. Rev. Lett.}\ }\textbf
  {\bibinfo {volume} {100}},\ \bibinfo {pages} {047004} (\bibinfo {year}
  {2008})}\BibitemShut {NoStop}%
\bibitem [{\citenamefont {Badoux}\ \emph {et~al.}(2016)\citenamefont {Badoux},
  \citenamefont {Tabis}, \citenamefont {Lalibert\'e}, \citenamefont
  {Grissonnanche}, \citenamefont {Vignolle}, \citenamefont {Vignolles},
  \citenamefont {B{\'e}ard}, \citenamefont {Bonn}, \citenamefont {Hardy},
  \citenamefont {Liang}, \citenamefont {Doiron-Leyraud},\ and\ \citenamefont
  {Proust}}]{Badoux2016}%
  \BibitemOpen
  \bibfield  {author} {\bibinfo {author} {\bibfnamefont {S.}~\bibnamefont
  {Badoux}}, \bibinfo {author} {\bibfnamefont {W.}~\bibnamefont {Tabis}},
  \bibinfo {author} {\bibfnamefont {F.}~\bibnamefont {Lalibert\'e}}, \bibinfo
  {author} {\bibfnamefont {G.}~\bibnamefont {Grissonnanche}}, \bibinfo {author}
  {\bibfnamefont {B.}~\bibnamefont {Vignolle}}, \bibinfo {author}
  {\bibfnamefont {D.}~\bibnamefont {Vignolles}}, \bibinfo {author}
  {\bibfnamefont {J.}~\bibnamefont {B{\'e}ard}}, \bibinfo {author}
  {\bibfnamefont {D.~A.}\ \bibnamefont {Bonn}}, \bibinfo {author}
  {\bibfnamefont {W.~N.}\ \bibnamefont {Hardy}}, \bibinfo {author}
  {\bibfnamefont {R.}~\bibnamefont {Liang}}, \bibinfo {author} {\bibfnamefont
  {L.}~\bibnamefont {Doiron-Leyraud}, \bibfnamefont {N.~Taillefer}}, \ and\
  \bibinfo {author} {\bibfnamefont {C.}~\bibnamefont {Proust}},\ }\href
  {http://dx.doi.org/10.1038/nature16983} {\bibfield  {journal} {\bibinfo
  {journal} {Nature}\ }\textbf {\bibinfo {volume} {7593}},\ \bibinfo {pages}
  {210} (\bibinfo {year} {2016})}\BibitemShut {NoStop}%
\bibitem [{\citenamefont {Putzke}\ \emph {et~al.}(2018)\citenamefont {Putzke},
  \citenamefont {Ayres}, \citenamefont {Buhot}, \citenamefont {Licciardello},
  \citenamefont {Hussey}, \citenamefont {Friedemann},\ and\ \citenamefont
  {Carrington}}]{Putzke2018}%
  \BibitemOpen
  \bibfield  {author} {\bibinfo {author} {\bibfnamefont {C.}~\bibnamefont
  {Putzke}}, \bibinfo {author} {\bibfnamefont {J.}~\bibnamefont {Ayres}},
  \bibinfo {author} {\bibfnamefont {J.}~\bibnamefont {Buhot}}, \bibinfo
  {author} {\bibfnamefont {S.}~\bibnamefont {Licciardello}}, \bibinfo {author}
  {\bibfnamefont {N.~E.}\ \bibnamefont {Hussey}}, \bibinfo {author}
  {\bibfnamefont {S.}~\bibnamefont {Friedemann}}, \ and\ \bibinfo {author}
  {\bibfnamefont {A.}~\bibnamefont {Carrington}},\ }\href {\doibase
  10.1103/PhysRevLett.120.117002} {\bibfield  {journal} {\bibinfo  {journal}
  {Phys. Rev. Lett.}\ }\textbf {\bibinfo {volume} {120}},\ \bibinfo {pages}
  {117002} (\bibinfo {year} {2018})}\BibitemShut {NoStop}%
\bibitem [{\citenamefont {Almasan}\ \emph {et~al.}(1992)\citenamefont
  {Almasan}, \citenamefont {Han}, \citenamefont {Lee}, \citenamefont {Paulius},
  \citenamefont {Maple}, \citenamefont {Veal}, \citenamefont {Downey},
  \citenamefont {Paulikas}, \citenamefont {Fisk},\ and\ \citenamefont
  {Schirber}}]{Almasan1992}%
  \BibitemOpen
  \bibfield  {author} {\bibinfo {author} {\bibfnamefont {C.~C.}\ \bibnamefont
  {Almasan}}, \bibinfo {author} {\bibfnamefont {S.~H.}\ \bibnamefont {Han}},
  \bibinfo {author} {\bibfnamefont {B.~W.}\ \bibnamefont {Lee}}, \bibinfo
  {author} {\bibfnamefont {L.~M.}\ \bibnamefont {Paulius}}, \bibinfo {author}
  {\bibfnamefont {M.~B.}\ \bibnamefont {Maple}}, \bibinfo {author}
  {\bibfnamefont {B.~W.}\ \bibnamefont {Veal}}, \bibinfo {author}
  {\bibfnamefont {J.~W.}\ \bibnamefont {Downey}}, \bibinfo {author}
  {\bibfnamefont {A.~P.}\ \bibnamefont {Paulikas}}, \bibinfo {author}
  {\bibfnamefont {Z.}~\bibnamefont {Fisk}}, \ and\ \bibinfo {author}
  {\bibfnamefont {J.~E.}\ \bibnamefont {Schirber}},\ }\href
  {http://journals.aps.org/prl/abstract/10.1103/PhysRevLett.69.680} {\bibfield
  {journal} {\bibinfo  {journal} {Phys. Rev. Lett.}\ }\textbf {\bibinfo
  {volume} {69}},\ \bibinfo {pages} {680} (\bibinfo {year} {1992})}\BibitemShut
  {NoStop}%
\bibitem [{\citenamefont {Neumeier}\ and\ \citenamefont
  {Zimmermann}(1993)}]{Neumeier1993}%
  \BibitemOpen
  \bibfield  {author} {\bibinfo {author} {\bibfnamefont {J.~J.}\ \bibnamefont
  {Neumeier}}\ and\ \bibinfo {author} {\bibfnamefont {H.~A.}\ \bibnamefont
  {Zimmermann}},\ }\href
  {https://journals.aps.org/prb/abstract/10.1103/PhysRevB.47.8385} {\bibfield
  {journal} {\bibinfo  {journal} {Phys. Rev. B}\ }\textbf {\bibinfo {volume}
  {47}},\ \bibinfo {pages} {8385} (\bibinfo {year} {1993})}\BibitemShut
  {NoStop}%
\bibitem [{\citenamefont {Yamada}\ and\ \citenamefont
  {Ido}(1992)}]{Yamada1992}%
  \BibitemOpen
  \bibfield  {author} {\bibinfo {author} {\bibfnamefont {N.}~\bibnamefont
  {Yamada}}\ and\ \bibinfo {author} {\bibfnamefont {M.}~\bibnamefont {Ido}},\
  }\href {http://www.sciencedirect.com/science/article/pii/092145349290029C}
  {\bibfield  {journal} {\bibinfo  {journal} {Physica C}\ }\textbf {\bibinfo
  {volume} {203}},\ \bibinfo {pages} {240 } (\bibinfo {year}
  {1992})}\BibitemShut {NoStop}%
\bibitem [{\citenamefont {Chu}\ \emph {et~al.}(1993)\citenamefont {Chu},
  \citenamefont {Gao}, \citenamefont {Chen}, \citenamefont {Huang},
  \citenamefont {Meng},\ and\ \citenamefont {Xue}}]{Chu1993}%
  \BibitemOpen
  \bibfield  {author} {\bibinfo {author} {\bibfnamefont {C.~W.}\ \bibnamefont
  {Chu}}, \bibinfo {author} {\bibfnamefont {L.}~\bibnamefont {Gao}}, \bibinfo
  {author} {\bibfnamefont {F.}~\bibnamefont {Chen}}, \bibinfo {author}
  {\bibfnamefont {Z.~J.}\ \bibnamefont {Huang}}, \bibinfo {author}
  {\bibfnamefont {R.~L.}\ \bibnamefont {Meng}}, \ and\ \bibinfo {author}
  {\bibfnamefont {Y.~Y.}\ \bibnamefont {Xue}},\ }\href
  {https://www.nature.com/nature/journal/v365/n6444/abs/365323a0.html}
  {\bibfield  {journal} {\bibinfo  {journal} {Nature}\ }\textbf {\bibinfo
  {volume} {365}},\ \bibinfo {pages} {323} (\bibinfo {year}
  {1993})}\BibitemShut {NoStop}%
\bibitem [{\citenamefont {Crawford}\ \emph {et~al.}(2005)\citenamefont
  {Crawford}, \citenamefont {Harlow}, \citenamefont {Deemyad}, \citenamefont
  {Tissen}, \citenamefont {Schilling}, \citenamefont {McCarron}, \citenamefont
  {Tozer}, \citenamefont {Cox}, \citenamefont {Ichikawa}, \citenamefont
  {Uchida},\ and\ \citenamefont {Huang}}]{Crawford2005}%
  \BibitemOpen
  \bibfield  {author} {\bibinfo {author} {\bibfnamefont {M.~K.}\ \bibnamefont
  {Crawford}}, \bibinfo {author} {\bibfnamefont {R.~L.}\ \bibnamefont
  {Harlow}}, \bibinfo {author} {\bibfnamefont {S.}~\bibnamefont {Deemyad}},
  \bibinfo {author} {\bibfnamefont {V.}~\bibnamefont {Tissen}}, \bibinfo
  {author} {\bibfnamefont {J.~S.}\ \bibnamefont {Schilling}}, \bibinfo {author}
  {\bibfnamefont {E.~M.}\ \bibnamefont {McCarron}}, \bibinfo {author}
  {\bibfnamefont {S.~W.}\ \bibnamefont {Tozer}}, \bibinfo {author}
  {\bibfnamefont {D.~E.}\ \bibnamefont {Cox}}, \bibinfo {author} {\bibfnamefont
  {N.}~\bibnamefont {Ichikawa}}, \bibinfo {author} {\bibfnamefont
  {S.}~\bibnamefont {Uchida}}, \ and\ \bibinfo {author} {\bibfnamefont
  {Q.}~\bibnamefont {Huang}},\ }\href {\doibase 10.1103/PhysRevB.71.104513}
  {\bibfield  {journal} {\bibinfo  {journal} {Phys. Rev. B}\ }\textbf {\bibinfo
  {volume} {71}},\ \bibinfo {pages} {104513} (\bibinfo {year}
  {2005})}\BibitemShut {NoStop}%
\bibitem [{\citenamefont {H\"ucker}\ \emph {et~al.}(2010)\citenamefont
  {H\"ucker}, \citenamefont {v.~Zimmermann}, \citenamefont {Debessai},
  \citenamefont {Schilling}, \citenamefont {Tranquada},\ and\ \citenamefont
  {Gu}}]{Hucker2010}%
  \BibitemOpen
  \bibfield  {author} {\bibinfo {author} {\bibfnamefont {M.}~\bibnamefont
  {H\"ucker}}, \bibinfo {author} {\bibfnamefont {M.}~\bibnamefont
  {v.~Zimmermann}}, \bibinfo {author} {\bibfnamefont {M.}~\bibnamefont
  {Debessai}}, \bibinfo {author} {\bibfnamefont {J.~S.}\ \bibnamefont
  {Schilling}}, \bibinfo {author} {\bibfnamefont {J.~M.}\ \bibnamefont
  {Tranquada}}, \ and\ \bibinfo {author} {\bibfnamefont {G.~D.}\ \bibnamefont
  {Gu}},\ }\href {\doibase 10.1103/PhysRevLett.104.057004} {\bibfield
  {journal} {\bibinfo  {journal} {Phys. Rev. Lett.}\ }\textbf {\bibinfo
  {volume} {104}},\ \bibinfo {pages} {057004} (\bibinfo {year}
  {2010})}\BibitemShut {NoStop}%
\bibitem [{\citenamefont {Benischke}\ \emph {et~al.}(1992)\citenamefont
  {Benischke}, \citenamefont {Weber}, \citenamefont {Fietz}, \citenamefont
  {Metzger}, \citenamefont {Grube}, \citenamefont {Wolf},\ and\ \citenamefont
  {W{\"u}hl}}]{Benischke1992}%
  \BibitemOpen
  \bibfield  {author} {\bibinfo {author} {\bibfnamefont {R.}~\bibnamefont
  {Benischke}}, \bibinfo {author} {\bibfnamefont {T.}~\bibnamefont {Weber}},
  \bibinfo {author} {\bibfnamefont {W.~H.}\ \bibnamefont {Fietz}}, \bibinfo
  {author} {\bibfnamefont {J.}~\bibnamefont {Metzger}}, \bibinfo {author}
  {\bibfnamefont {K.}~\bibnamefont {Grube}}, \bibinfo {author} {\bibfnamefont
  {T.}~\bibnamefont {Wolf}}, \ and\ \bibinfo {author} {\bibfnamefont
  {H.}~\bibnamefont {W{\"u}hl}},\ }\href
  {http://www.sciencedirect.com/science/article/pii/092145349290036C}
  {\bibfield  {journal} {\bibinfo  {journal} {Physica C}\ }\textbf {\bibinfo
  {volume} {203}},\ \bibinfo {pages} {293 } (\bibinfo {year}
  {1992})}\BibitemShut {NoStop}%
\bibitem [{\citenamefont {Kraut}\ \emph {et~al.}(1993)\citenamefont {Kraut},
  \citenamefont {Meingast}, \citenamefont {Br{\"a}uchle}, \citenamefont
  {Claus}, \citenamefont {Erb}, \citenamefont {M{\"u}ller-Vogt},\ and\
  \citenamefont {W{\"u}hl}}]{Kraut1993}%
  \BibitemOpen
  \bibfield  {author} {\bibinfo {author} {\bibfnamefont {O.}~\bibnamefont
  {Kraut}}, \bibinfo {author} {\bibfnamefont {C.}~\bibnamefont {Meingast}},
  \bibinfo {author} {\bibfnamefont {G.}~\bibnamefont {Br{\"a}uchle}}, \bibinfo
  {author} {\bibfnamefont {H.}~\bibnamefont {Claus}}, \bibinfo {author}
  {\bibfnamefont {A.}~\bibnamefont {Erb}}, \bibinfo {author} {\bibfnamefont
  {G.}~\bibnamefont {M{\"u}ller-Vogt}}, \ and\ \bibinfo {author} {\bibfnamefont
  {H.}~\bibnamefont {W{\"u}hl}},\ }\href
  {http://www.sciencedirect.com/science/article/pii/092145349390180X}
  {\bibfield  {journal} {\bibinfo  {journal} {Physica C}\ }\textbf {\bibinfo
  {volume} {205}},\ \bibinfo {pages} {139} (\bibinfo {year}
  {1993})}\BibitemShut {NoStop}%
\bibitem [{\citenamefont {Alireza}\ \emph {et~al.}(2017)\citenamefont
  {Alireza}, \citenamefont {Zhang}, \citenamefont {Guo}, \citenamefont
  {Porras}, \citenamefont {Loew}, \citenamefont {Hsu}, \citenamefont
  {Lonzarich}, \citenamefont {Le~Tacon}, \citenamefont {Keimer},\ and\
  \citenamefont {Sebastian}}]{Alireza2017}%
  \BibitemOpen
  \bibfield  {author} {\bibinfo {author} {\bibfnamefont {P.~L.}\ \bibnamefont
  {Alireza}}, \bibinfo {author} {\bibfnamefont {G.~H.}\ \bibnamefont {Zhang}},
  \bibinfo {author} {\bibfnamefont {W.}~\bibnamefont {Guo}}, \bibinfo {author}
  {\bibfnamefont {J.}~\bibnamefont {Porras}}, \bibinfo {author} {\bibfnamefont
  {T.}~\bibnamefont {Loew}}, \bibinfo {author} {\bibfnamefont {Y.-T.}\
  \bibnamefont {Hsu}}, \bibinfo {author} {\bibfnamefont {G.~G.}\ \bibnamefont
  {Lonzarich}}, \bibinfo {author} {\bibfnamefont {M.}~\bibnamefont {Le~Tacon}},
  \bibinfo {author} {\bibfnamefont {B.}~\bibnamefont {Keimer}}, \ and\ \bibinfo
  {author} {\bibfnamefont {S.~E.}\ \bibnamefont {Sebastian}},\ }\href {\doibase
  10.1103/PhysRevB.95.100505} {\bibfield  {journal} {\bibinfo  {journal} {Phys.
  Rev. B}\ }\textbf {\bibinfo {volume} {95}},\ \bibinfo {pages} {100505}
  (\bibinfo {year} {2017})}\BibitemShut {NoStop}%
\bibitem [{\citenamefont {Haug}\ \emph {et~al.}(2010)\citenamefont {Haug},
  \citenamefont {Hinkov}, \citenamefont {Sidis}, \citenamefont {Bourges},
  \citenamefont {Christensen}, \citenamefont {Ivanov}, \citenamefont {Keller},
  \citenamefont {Lin},\ and\ \citenamefont {Keimer}}]{Haug2010}%
  \BibitemOpen
  \bibfield  {author} {\bibinfo {author} {\bibfnamefont {D.}~\bibnamefont
  {Haug}}, \bibinfo {author} {\bibfnamefont {V.}~\bibnamefont {Hinkov}},
  \bibinfo {author} {\bibfnamefont {Y.}~\bibnamefont {Sidis}}, \bibinfo
  {author} {\bibfnamefont {P.}~\bibnamefont {Bourges}}, \bibinfo {author}
  {\bibfnamefont {N.~B.}\ \bibnamefont {Christensen}}, \bibinfo {author}
  {\bibfnamefont {A.}~\bibnamefont {Ivanov}}, \bibinfo {author} {\bibfnamefont
  {T.}~\bibnamefont {Keller}}, \bibinfo {author} {\bibfnamefont {C.~T.}\
  \bibnamefont {Lin}}, \ and\ \bibinfo {author} {\bibfnamefont
  {B.}~\bibnamefont {Keimer}},\ }\href
  {http://stacks.iop.org/1367-2630/12/i=10/a=105006} {\bibfield  {journal}
  {\bibinfo  {journal} {New J. Phys.}\ }\textbf {\bibinfo {volume} {12}},\
  \bibinfo {pages} {105006} (\bibinfo {year} {2010})}\BibitemShut {NoStop}%
\bibitem [{\citenamefont {Blanco-Canosa}\ \emph {et~al.}(2013)\citenamefont
  {Blanco-Canosa}, \citenamefont {Frano}, \citenamefont {Loew}, \citenamefont
  {Lu}, \citenamefont {Porras}, \citenamefont {Ghiringhelli}, \citenamefont
  {Minola}, \citenamefont {Mazzoli}, \citenamefont {Braicovich}, \citenamefont
  {Schierle}, \citenamefont {Weschke}, \citenamefont {Le~Tacon},\ and\
  \citenamefont {Keimer}}]{Blanco-Canosa2013}%
  \BibitemOpen
  \bibfield  {author} {\bibinfo {author} {\bibfnamefont {S.}~\bibnamefont
  {Blanco-Canosa}}, \bibinfo {author} {\bibfnamefont {A.}~\bibnamefont
  {Frano}}, \bibinfo {author} {\bibfnamefont {T.}~\bibnamefont {Loew}},
  \bibinfo {author} {\bibfnamefont {Y.}~\bibnamefont {Lu}}, \bibinfo {author}
  {\bibfnamefont {J.}~\bibnamefont {Porras}}, \bibinfo {author} {\bibfnamefont
  {G.}~\bibnamefont {Ghiringhelli}}, \bibinfo {author} {\bibfnamefont
  {M.}~\bibnamefont {Minola}}, \bibinfo {author} {\bibfnamefont
  {C.}~\bibnamefont {Mazzoli}}, \bibinfo {author} {\bibfnamefont
  {L.}~\bibnamefont {Braicovich}}, \bibinfo {author} {\bibfnamefont
  {E.}~\bibnamefont {Schierle}}, \bibinfo {author} {\bibfnamefont
  {E.}~\bibnamefont {Weschke}}, \bibinfo {author} {\bibfnamefont
  {M.}~\bibnamefont {Le~Tacon}}, \ and\ \bibinfo {author} {\bibfnamefont
  {B.}~\bibnamefont {Keimer}},\ }\href
  {http://link.aps.org/doi/10.1103/PhysRevLett.110.187001} {\bibfield
  {journal} {\bibinfo  {journal} {Phys. Rev. Lett.}\ }\textbf {\bibinfo
  {volume} {110}},\ \bibinfo {pages} {187001} (\bibinfo {year}
  {2013})}\BibitemShut {NoStop}%
\bibitem [{\citenamefont {Guguchia}\ \emph {et~al.}(2013)\citenamefont
  {Guguchia}, \citenamefont {Maisuradze}, \citenamefont {Ghambashidze},
  \citenamefont {Khasanov}, \citenamefont {Shengelaya},\ and\ \citenamefont
  {Keller}}]{Guguchia2013}%
  \BibitemOpen
  \bibfield  {author} {\bibinfo {author} {\bibfnamefont {Z.}~\bibnamefont
  {Guguchia}}, \bibinfo {author} {\bibfnamefont {A.}~\bibnamefont
  {Maisuradze}}, \bibinfo {author} {\bibfnamefont {G.}~\bibnamefont
  {Ghambashidze}}, \bibinfo {author} {\bibfnamefont {R.}~\bibnamefont
  {Khasanov}}, \bibinfo {author} {\bibfnamefont {A.}~\bibnamefont
  {Shengelaya}}, \ and\ \bibinfo {author} {\bibfnamefont {H.}~\bibnamefont
  {Keller}},\ }\href {http://stacks.iop.org/1367-2630/15/i=9/a=093005}
  {\bibfield  {journal} {\bibinfo  {journal} {New J. Phys.}\ }\textbf {\bibinfo
  {volume} {15}},\ \bibinfo {pages} {093005} (\bibinfo {year}
  {2013})}\BibitemShut {NoStop}%
\bibitem [{\citenamefont {Georges}\ \emph {et~al.}(1996)\citenamefont
  {Georges}, \citenamefont {Kotliar}, \citenamefont {Krauth},\ and\
  \citenamefont {Rozenberg}}]{Georges1996}%
  \BibitemOpen
  \bibfield  {author} {\bibinfo {author} {\bibfnamefont {A.}~\bibnamefont
  {Georges}}, \bibinfo {author} {\bibfnamefont {G.}~\bibnamefont {Kotliar}},
  \bibinfo {author} {\bibfnamefont {W.}~\bibnamefont {Krauth}}, \ and\ \bibinfo
  {author} {\bibfnamefont {M.~J.}\ \bibnamefont {Rozenberg}},\ }\href
  {https://journals.aps.org/rmp/abstract/10.1103/RevModPhys.68.13} {\bibfield
  {journal} {\bibinfo  {journal} {Rev. Mod. Phys.}\ }\textbf {\bibinfo {volume}
  {68}},\ \bibinfo {pages} {13} (\bibinfo {year} {1996})}\BibitemShut {NoStop}%
\bibitem [{\citenamefont {Tremblay}(2013)}]{Tremblay2013}%
  \BibitemOpen
  \bibfield  {author} {\bibinfo {author} {\bibfnamefont {A.-M.~S.}\
  \bibnamefont {Tremblay}},\ }\enquote {\bibinfo {title} {Strongly correlated
  superconductivity},}\ in\ \href
  {https://www.cond-mat.de/events/correl13/manuscripts/correl13.pdf} {\emph
  {\bibinfo {booktitle} {Emergent Phenomena in Correlated Matter Modeling and
  Simulation}}},\ Vol.~\bibinfo {volume} {3},\ \bibinfo {editor} {edited by\
  \bibinfo {editor} {\bibfnamefont {E.}~\bibnamefont {Pavarini}}, \bibinfo
  {editor} {\bibfnamefont {E.}~\bibnamefont {Koch}}, \ and\ \bibinfo {editor}
  {\bibfnamefont {U.}~\bibnamefont {Schollw{\"o}ck}}}\ (\bibinfo  {publisher}
  {Forschungszentrum J{\"u}lich},\ \bibinfo {year} {2013})\ Chap.~\bibinfo
  {chapter} {10}\BibitemShut {NoStop}%
\bibitem [{\citenamefont {Gull}\ \emph {et~al.}(2013)\citenamefont {Gull},
  \citenamefont {Parcollet},\ and\ \citenamefont {Millis}}]{Gull2013}%
  \BibitemOpen
  \bibfield  {author} {\bibinfo {author} {\bibfnamefont {E.}~\bibnamefont
  {Gull}}, \bibinfo {author} {\bibfnamefont {O.}~\bibnamefont {Parcollet}}, \
  and\ \bibinfo {author} {\bibfnamefont {A.~J.}\ \bibnamefont {Millis}},\
  }\href {\doibase 10.1103/PhysRevLett.110.216405} {\bibfield  {journal}
  {\bibinfo  {journal} {Phys. Rev. Lett.}\ }\textbf {\bibinfo {volume} {110}},\
  \bibinfo {pages} {216405} (\bibinfo {year} {2013})}\BibitemShut {NoStop}%
\bibitem [{\citenamefont {Werner}\ \emph {et~al.}(2009)\citenamefont {Werner},
  \citenamefont {Gull}, \citenamefont {Parcollet},\ and\ \citenamefont
  {Millis}}]{Werner2009}%
  \BibitemOpen
  \bibfield  {author} {\bibinfo {author} {\bibfnamefont {P.}~\bibnamefont
  {Werner}}, \bibinfo {author} {\bibfnamefont {E.}~\bibnamefont {Gull}},
  \bibinfo {author} {\bibfnamefont {O.}~\bibnamefont {Parcollet}}, \ and\
  \bibinfo {author} {\bibfnamefont {A.~J.}\ \bibnamefont {Millis}},\ }\href
  {\doibase 10.1103/PhysRevB.80.045120} {\bibfield  {journal} {\bibinfo
  {journal} {Phys. Rev. B}\ }\textbf {\bibinfo {volume} {80}},\ \bibinfo
  {pages} {045120} (\bibinfo {year} {2009})}\BibitemShut {NoStop}%
\bibitem [{\citenamefont {Sordi}\ \emph {et~al.}(2012)\citenamefont {Sordi},
  \citenamefont {S\'emon}, \citenamefont {Haule},\ and\ \citenamefont
  {Tremblay}}]{Sordi2012}%
  \BibitemOpen
  \bibfield  {author} {\bibinfo {author} {\bibfnamefont {G.}~\bibnamefont
  {Sordi}}, \bibinfo {author} {\bibfnamefont {P.}~\bibnamefont {S\'emon}},
  \bibinfo {author} {\bibfnamefont {K.}~\bibnamefont {Haule}}, \ and\ \bibinfo
  {author} {\bibfnamefont {A.-M.}\ \bibnamefont {Tremblay}},\ }\href
  {http://dx.doi.org/10.1038/srep00547} {\bibfield  {journal} {\bibinfo
  {journal} {Sci. Rep.}\ }\textbf {\bibinfo {volume} {2}},\ \bibinfo {pages}
  {547} (\bibinfo {year} {2012})}\BibitemShut {NoStop}%
\bibitem [{\citenamefont {Cyr-Choini\`ere}\ \emph {et~al.}(2017)\citenamefont
  {Cyr-Choini\`ere}, \citenamefont {Badoux}, \citenamefont {Grissonnanche},
  \citenamefont {Michon}, \citenamefont {Afshar}, \citenamefont {Fortier},
  \citenamefont {LeBoeuf}, \citenamefont {Graf}, \citenamefont {Day},
  \citenamefont {Bonn}, \citenamefont {Hardy}, \citenamefont {Liang},
  \citenamefont {Doiron-Leyraud},\ and\ \citenamefont
  {Taillefer}}]{Cyr-Choiniere2017}%
  \BibitemOpen
  \bibfield  {author} {\bibinfo {author} {\bibfnamefont {O.}~\bibnamefont
  {Cyr-Choini\`ere}}, \bibinfo {author} {\bibfnamefont {S.}~\bibnamefont
  {Badoux}}, \bibinfo {author} {\bibfnamefont {G.}~\bibnamefont
  {Grissonnanche}}, \bibinfo {author} {\bibfnamefont {B.}~\bibnamefont
  {Michon}}, \bibinfo {author} {\bibfnamefont {S.~A.~A.}\ \bibnamefont
  {Afshar}}, \bibinfo {author} {\bibfnamefont {S.}~\bibnamefont {Fortier}},
  \bibinfo {author} {\bibfnamefont {D.}~\bibnamefont {LeBoeuf}}, \bibinfo
  {author} {\bibfnamefont {D.}~\bibnamefont {Graf}}, \bibinfo {author}
  {\bibfnamefont {J.}~\bibnamefont {Day}}, \bibinfo {author} {\bibfnamefont
  {D.~A.}\ \bibnamefont {Bonn}}, \bibinfo {author} {\bibfnamefont {W.~N.}\
  \bibnamefont {Hardy}}, \bibinfo {author} {\bibfnamefont {R.}~\bibnamefont
  {Liang}}, \bibinfo {author} {\bibfnamefont {N.}~\bibnamefont
  {Doiron-Leyraud}}, \ and\ \bibinfo {author} {\bibfnamefont {L.}~\bibnamefont
  {Taillefer}},\ }\href {https://link.aps.org/doi/10.1103/PhysRevX.7.031042}
  {\bibfield  {journal} {\bibinfo  {journal} {Phys. Rev. X}\ }\textbf {\bibinfo
  {volume} {7}},\ \bibinfo {pages} {031042} (\bibinfo {year}
  {2017})}\BibitemShut {NoStop}%
\bibitem [{\citenamefont {LeBoeuf}\ \emph {et~al.}(2013)\citenamefont
  {LeBoeuf}, \citenamefont {Kramer}, \citenamefont {Hardy}, \citenamefont
  {Liang}, \citenamefont {Bonn},\ and\ \citenamefont {Proust}}]{LeBoeuf2013}%
  \BibitemOpen
  \bibfield  {author} {\bibinfo {author} {\bibfnamefont {D.}~\bibnamefont
  {LeBoeuf}}, \bibinfo {author} {\bibfnamefont {S.}~\bibnamefont {Kramer}},
  \bibinfo {author} {\bibfnamefont {W.~N.}\ \bibnamefont {Hardy}}, \bibinfo
  {author} {\bibfnamefont {R.}~\bibnamefont {Liang}}, \bibinfo {author}
  {\bibfnamefont {D.~A.}\ \bibnamefont {Bonn}}, \ and\ \bibinfo {author}
  {\bibfnamefont {C.}~\bibnamefont {Proust}},\ }\href
  {http://dx.doi.org/10.1038/nphys2502} {\bibfield  {journal} {\bibinfo
  {journal} {Nat. Phys.}\ }\textbf {\bibinfo {volume} {9}},\ \bibinfo {pages}
  {79} (\bibinfo {year} {2013})}\BibitemShut {NoStop}%
\bibitem [{\citenamefont {Gerber}\ \emph {et~al.}(2015)\citenamefont {Gerber},
  \citenamefont {Jang}, \citenamefont {Nojiri}, \citenamefont {Matsuzawa},
  \citenamefont {Yasumura}, \citenamefont {Bonn}, \citenamefont {Liang},
  \citenamefont {Hardy}, \citenamefont {Islam}, \citenamefont {Mehta},
  \citenamefont {Song}, \citenamefont {Sikorski}, \citenamefont {Stefanescu},
  \citenamefont {Feng}, \citenamefont {Kivelson}, \citenamefont {Devereaux},
  \citenamefont {Shen}, \citenamefont {Kao}, \citenamefont {Lee}, \citenamefont
  {Zhu},\ and\ \citenamefont {Lee}}]{Gerber2015}%
  \BibitemOpen
  \bibfield  {author} {\bibinfo {author} {\bibfnamefont {S.}~\bibnamefont
  {Gerber}}, \bibinfo {author} {\bibfnamefont {H.}~\bibnamefont {Jang}},
  \bibinfo {author} {\bibfnamefont {H.}~\bibnamefont {Nojiri}}, \bibinfo
  {author} {\bibfnamefont {S.}~\bibnamefont {Matsuzawa}}, \bibinfo {author}
  {\bibfnamefont {H.}~\bibnamefont {Yasumura}}, \bibinfo {author}
  {\bibfnamefont {D.~A.}\ \bibnamefont {Bonn}}, \bibinfo {author}
  {\bibfnamefont {R.}~\bibnamefont {Liang}}, \bibinfo {author} {\bibfnamefont
  {W.~N.}\ \bibnamefont {Hardy}}, \bibinfo {author} {\bibfnamefont
  {Z.}~\bibnamefont {Islam}}, \bibinfo {author} {\bibfnamefont
  {A.}~\bibnamefont {Mehta}}, \bibinfo {author} {\bibfnamefont
  {S.}~\bibnamefont {Song}}, \bibinfo {author} {\bibfnamefont {M.}~\bibnamefont
  {Sikorski}}, \bibinfo {author} {\bibfnamefont {D.}~\bibnamefont
  {Stefanescu}}, \bibinfo {author} {\bibfnamefont {Y.}~\bibnamefont {Feng}},
  \bibinfo {author} {\bibfnamefont {S.~A.}\ \bibnamefont {Kivelson}}, \bibinfo
  {author} {\bibfnamefont {T.~P.}\ \bibnamefont {Devereaux}}, \bibinfo {author}
  {\bibfnamefont {Z.-X.}\ \bibnamefont {Shen}}, \bibinfo {author}
  {\bibfnamefont {C.-C.}\ \bibnamefont {Kao}}, \bibinfo {author} {\bibfnamefont
  {W.-S.}\ \bibnamefont {Lee}}, \bibinfo {author} {\bibfnamefont
  {D.}~\bibnamefont {Zhu}}, \ and\ \bibinfo {author} {\bibfnamefont {J.-S.}\
  \bibnamefont {Lee}},\ }\href {\doibase 10.1126/science.aac6257} {\bibfield
  {journal} {\bibinfo  {journal} {Science}\ }\textbf {\bibinfo {volume}
  {350}},\ \bibinfo {pages} {949} (\bibinfo {year} {2015})}\BibitemShut
  {NoStop}%
\bibitem [{\citenamefont {Gao}\ \emph {et~al.}(1994)\citenamefont {Gao},
  \citenamefont {Xue}, \citenamefont {Chen}, \citenamefont {Xiong},
  \citenamefont {Meng}, \citenamefont {Ramirez}, \citenamefont {Chu},
  \citenamefont {Eggert},\ and\ \citenamefont {Mao}}]{Gao1994}%
  \BibitemOpen
  \bibfield  {author} {\bibinfo {author} {\bibfnamefont {L.}~\bibnamefont
  {Gao}}, \bibinfo {author} {\bibfnamefont {Y.~Y.}\ \bibnamefont {Xue}},
  \bibinfo {author} {\bibfnamefont {F.}~\bibnamefont {Chen}}, \bibinfo {author}
  {\bibfnamefont {Q.}~\bibnamefont {Xiong}}, \bibinfo {author} {\bibfnamefont
  {R.~L.}\ \bibnamefont {Meng}}, \bibinfo {author} {\bibfnamefont
  {D.}~\bibnamefont {Ramirez}}, \bibinfo {author} {\bibfnamefont {C.~W.}\
  \bibnamefont {Chu}}, \bibinfo {author} {\bibfnamefont {J.~H.}\ \bibnamefont
  {Eggert}}, \ and\ \bibinfo {author} {\bibfnamefont {H.~K.}\ \bibnamefont
  {Mao}},\ }\href {\doibase 10.1103/PhysRevB.50.4260} {\bibfield  {journal}
  {\bibinfo  {journal} {Phys. Rev. B}\ }\textbf {\bibinfo {volume} {50}},\
  \bibinfo {pages} {4260} (\bibinfo {year} {1994})}\BibitemShut {NoStop}%
\bibitem [{\citenamefont {Tissen}\ \emph {et~al.}(1999)\citenamefont {Tissen},
  \citenamefont {Wang}, \citenamefont {Paulikas}, \citenamefont {Veal},\ and\
  \citenamefont {Schilling}}]{Tissen1999}%
  \BibitemOpen
  \bibfield  {author} {\bibinfo {author} {\bibfnamefont {V.~G.}\ \bibnamefont
  {Tissen}}, \bibinfo {author} {\bibfnamefont {Y.}~\bibnamefont {Wang}},
  \bibinfo {author} {\bibfnamefont {A.~P.}\ \bibnamefont {Paulikas}}, \bibinfo
  {author} {\bibfnamefont {B.~W.}\ \bibnamefont {Veal}}, \ and\ \bibinfo
  {author} {\bibfnamefont {J.~S.}\ \bibnamefont {Schilling}},\ }\href
  {http://www.sciencedirect.com/science/article/pii/S0921453499002610}
  {\bibfield  {journal} {\bibinfo  {journal} {Physica C}\ }\textbf {\bibinfo
  {volume} {316}},\ \bibinfo {pages} {21 } (\bibinfo {year}
  {1999})}\BibitemShut {NoStop}%
\bibitem [{\citenamefont {Fietz}\ \emph {et~al.}(1996)\citenamefont {Fietz},
  \citenamefont {Quenzel}, \citenamefont {Ludwig}, \citenamefont {Grube},
  \citenamefont {S.~I.~Schlachter}, \citenamefont {Wolf}, \citenamefont {Erb},
  \citenamefont {Kl{\"a}ser},\ and\ \citenamefont
  {M{\"u}ller-Vogtller-Vogt}}]{Fietz1996}%
  \BibitemOpen
  \bibfield  {author} {\bibinfo {author} {\bibfnamefont {W.~H.}\ \bibnamefont
  {Fietz}}, \bibinfo {author} {\bibfnamefont {R.}~\bibnamefont {Quenzel}},
  \bibinfo {author} {\bibfnamefont {H.~A.}\ \bibnamefont {Ludwig}}, \bibinfo
  {author} {\bibfnamefont {K.}~\bibnamefont {Grube}}, \bibinfo {author}
  {\bibfnamefont {F.~W.~H.}\ \bibnamefont {S.~I.~Schlachter}}, \bibinfo
  {author} {\bibfnamefont {T.}~\bibnamefont {Wolf}}, \bibinfo {author}
  {\bibfnamefont {A.}~\bibnamefont {Erb}}, \bibinfo {author} {\bibfnamefont
  {M.}~\bibnamefont {Kl{\"a}ser}}, \ and\ \bibinfo {author} {\bibfnamefont
  {G.}~\bibnamefont {M{\"u}ller-Vogtller-Vogt}},\ }\href
  {http://www.sciencedirect.com/science/article/pii/S0921453496004807}
  {\bibfield  {journal} {\bibinfo  {journal} {Physica C}\ }\textbf {\bibinfo
  {volume} {270}},\ \bibinfo {pages} {258 } (\bibinfo {year}
  {1996})}\BibitemShut {NoStop}%
\bibitem [{\citenamefont {Klotz}\ \emph {et~al.}(1991)\citenamefont {Klotz},
  \citenamefont {Reith},\ and\ \citenamefont {Schilling}}]{Klotz1991}%
  \BibitemOpen
  \bibfield  {author} {\bibinfo {author} {\bibfnamefont {S.}~\bibnamefont
  {Klotz}}, \bibinfo {author} {\bibfnamefont {W.}~\bibnamefont {Reith}}, \ and\
  \bibinfo {author} {\bibfnamefont {J.}~\bibnamefont {Schilling}},\ }\href
  {http://www.sciencedirect.com/science/article/pii/092145349190208G}
  {\bibfield  {journal} {\bibinfo  {journal} {Physica C}\ }\textbf {\bibinfo
  {volume} {172}},\ \bibinfo {pages} {423 } (\bibinfo {year}
  {1991})}\BibitemShut {NoStop}%
\bibitem [{\citenamefont {Tozer}\ \emph {et~al.}(1993)\citenamefont {Tozer},
  \citenamefont {Koston},\ and\ \citenamefont {McCarron~III}}]{Tozer1993}%
  \BibitemOpen
  \bibfield  {author} {\bibinfo {author} {\bibfnamefont {S.~W.}\ \bibnamefont
  {Tozer}}, \bibinfo {author} {\bibfnamefont {J.~L.}\ \bibnamefont {Koston}}, \
  and\ \bibinfo {author} {\bibfnamefont {E.~M.}\ \bibnamefont {McCarron~III}},\
  }\href {https://journals.aps.org/prb/abstract/10.1103/PhysRevB.47.8089}
  {\bibfield  {journal} {\bibinfo  {journal} {Phys. Rev. B}\ }\textbf {\bibinfo
  {volume} {47}},\ \bibinfo {pages} {8089} (\bibinfo {year}
  {1993})}\BibitemShut {NoStop}%
\bibitem [{\citenamefont {Fietz}\ \emph {et~al.}(1994)\citenamefont {Fietz},
  \citenamefont {Quenzel}, \citenamefont {Grube}, \citenamefont {Metzger},
  \citenamefont {Weber},\ and\ \citenamefont {Ludwig}}]{Fietz1994}%
  \BibitemOpen
  \bibfield  {author} {\bibinfo {author} {\bibfnamefont {W.~H.}\ \bibnamefont
  {Fietz}}, \bibinfo {author} {\bibfnamefont {R.}~\bibnamefont {Quenzel}},
  \bibinfo {author} {\bibfnamefont {K.}~\bibnamefont {Grube}}, \bibinfo
  {author} {\bibfnamefont {J.}~\bibnamefont {Metzger}}, \bibinfo {author}
  {\bibfnamefont {T.}~\bibnamefont {Weber}}, \ and\ \bibinfo {author}
  {\bibfnamefont {H.~A.}\ \bibnamefont {Ludwig}},\ }\href
  {http://www.sciencedirect.com/science/article/pii/0921453494921148}
  {\bibfield  {journal} {\bibinfo  {journal} {Physica C}\ }\textbf {\bibinfo
  {volume} {235-240 Part 3}},\ \bibinfo {pages} {1785 } (\bibinfo {year}
  {1994})}\BibitemShut {NoStop}%
\bibitem [{\citenamefont {Yoshida}\ \emph {et~al.}(1999)\citenamefont
  {Yoshida}, \citenamefont {Rykov}, \citenamefont {Tajima},\ and\ \citenamefont
  {Terasaki}}]{Yoshida1999}%
  \BibitemOpen
  \bibfield  {author} {\bibinfo {author} {\bibfnamefont {K.}~\bibnamefont
  {Yoshida}}, \bibinfo {author} {\bibfnamefont {A.~I.}\ \bibnamefont {Rykov}},
  \bibinfo {author} {\bibfnamefont {S.}~\bibnamefont {Tajima}}, \ and\ \bibinfo
  {author} {\bibfnamefont {I.}~\bibnamefont {Terasaki}},\ }\href
  {https://journals.aps.org/prb/abstract/10.1103/PhysRevB.60.R15035} {\bibfield
   {journal} {\bibinfo  {journal} {Phys. Rev. B}\ }\textbf {\bibinfo {volume}
  {60}},\ \bibinfo {pages} {R15035} (\bibinfo {year} {1999})}\BibitemShut
  {NoStop}%
\bibitem [{\citenamefont {Lortz}\ \emph {et~al.}(2006)\citenamefont {Lortz},
  \citenamefont {Tomita}, \citenamefont {Wang}, \citenamefont {Junod},
  \citenamefont {Schilling}, \citenamefont {Masui},\ and\ \citenamefont
  {Tajima}}]{Lortz2006}%
  \BibitemOpen
  \bibfield  {author} {\bibinfo {author} {\bibfnamefont {R.}~\bibnamefont
  {Lortz}}, \bibinfo {author} {\bibfnamefont {T.}~\bibnamefont {Tomita}},
  \bibinfo {author} {\bibfnamefont {Y.}~\bibnamefont {Wang}}, \bibinfo {author}
  {\bibfnamefont {A.}~\bibnamefont {Junod}}, \bibinfo {author} {\bibfnamefont
  {J.}~\bibnamefont {Schilling}}, \bibinfo {author} {\bibfnamefont
  {T.}~\bibnamefont {Masui}}, \ and\ \bibinfo {author} {\bibfnamefont
  {S.}~\bibnamefont {Tajima}},\ }\href
  {http://www.sciencedirect.com/science/article/pii/S0921453405009032}
  {\bibfield  {journal} {\bibinfo  {journal} {Physica C}\ }\textbf {\bibinfo
  {volume} {434}},\ \bibinfo {pages} {194 } (\bibinfo {year}
  {2006})}\BibitemShut {NoStop}%
\end{thebibliography}

%



\end{document}